\documentclass[twocolumn,amsmath,amssymb,prb,superscriptaddress,a4paper,floatfix,showkeys]{revtex4}
\usepackage{graphicx}
\usepackage{mathptmx}      
\DeclareRobustCommand\openone{\leavevmode\hbox{\small1\normalsize\kern-.33em1}}

\newcommand\OTOCLED{OTOCLED}
\newcommand\CNTUH{CNTUH}
\begin{document}

\title{Extended Boole-Bell inequalities applicable to quantum theory}

\author{Hans De Raedt}
\email{h.a.de.raedt@rug.nl}
\affiliation{%
Department of Applied Physics,
Zernike Institute of Advanced Materials,
University of Groningen, Nijenborgh 4, NL-9747 AG Groningen, The Netherlands
}%
\author{Karl Hess}
\email{k-hess@illinois.edu}           
\affiliation{%
Beckman Institute, Department of Electrical Engineering and Department
of Physics, University of Illinois, Urbana, Il 61801, USA
}%
\author{Kristel Michielsen}
\email{k.michielsen@fz-juelich.de}           
\affiliation{%
Institute for Advanced Simulation, J\"ulich Supercomputing Centre,
Research Centre J\"ulich, D-52425 J\"ulich, Germany
}%
\date{\today}

\begin{abstract}
We address the basic meaning of apparent contradictions of quantum
theory and probability frameworks as expressed by
Bell's inequalities. We show that these contradictions have their
origin in the incomplete considerations of
the premises of the derivation of the inequalities.
A careful consideration of past work, including that of
Boole and Vorob'ev, has lead us to the formulation of extended
Boole-Bell inequalities that are binding for both classical and
quantum models.
The Einstein-Podolsky-Rosen-Bohm gedanken experiment
and a macroscopic quantum coherence experiment proposed by Leggett
and Garg are both shown to obey the extended Boole-Bell
inequalities. These examples as well as additional
discussions also provide reasons for apparent violations of these
inequalities.
\end{abstract}

\maketitle
\tableofcontents

\section{Introduction}\label{Introduction}
\label{Notation}

The foundations of quantum theory and quantum information theory
encompass central questions that connect the ontology of two valued
``elements of reality'' to epistemic propositions about the possible
correlations between data related to these two valued elements. It
is usually maintained that the concepts of realism, macroscopic
realism, Einstein locality and contextuality need to be revised to
explain certain correlations of measurements related to the work of
Einstein, Podolsky and Rosen (EPR) \cite{EPR35}. In this paper, we
offer explanations of the problems surrounding models of EPR
experiments that do not touch the very basic concepts of realism and
locality but instead find a satisfactory resolution by a careful
amalgamate of the contributions of Boole~\cite{BO1862},
Vorob'ev~\cite{VORO62} and Bell~\cite{BELL64,BELL93}.

We start on the purely mathematical side by noting that the
inequalities of Boole~\cite{BO1862} impose restrictions on the
correlations of certain sets of three or more two-valued integer
variables. Then, we show that elementary algebra suffices to prove
inequalities that have the same structure as those of Boole and
impose restrictions on the values of nonnegative functions of
triples, quadruples etc. of two-valued variables.
These inequalities are also similar to those of Bell~\cite{BELL64,BELL93}
but the proof of the former requires fewer assumptions.
Finally, starting from the
commonly accepted postulates of quantum theory we present a rigorous
derivation of inequalities for quantum theory equivalent to those of
Boole, again by using only linear algebra and the properties of
non negative functions of three or more two-valued variables.
Although the conditions to prove all of these inequalities are
different to those in Boole's or Bell's work, the inequalities
themselves have the same structure as those of Boole and Bell.
Because of this similarity we refer to them as the extended
Boole-Bell inequalities (EBBI).

Our proofs of the EBBI do not require metaphysical assumptions but include the
inequalities of Bell and apply to quantum theory as well. Should the
EBBI be violated, the logical implication is that one or more of the
necessary conditions to prove these inequalities are not satisfied.
As these conditions do not refer to concepts such as locality or
macroscopic realism, no revision of these concepts is necessitated
by Bell's work.
Furthermore, it follows from our work that, given Bell's
premises, the Bell inequalities cannot be violated, not even by
influences at a distance.

Many aspects of all of this have been discussed in the literature by
de la Pe\~na {\sl et al.}~\cite{PENA72},
Fine~\cite{FINE74,FINE82,FINE82a,FINE82b,FINE96},
Pitowsky~\cite{PITO94}, Hess and Philipp~\cite{HESS01,HESS05},
Khrennikov~\cite{KHRE99,KHRE07,KHRE09,KHRE10}, and many other
authors~\cite{MUYN86,JAYN89,BROD93,SICA99,BAER99,ACCA05,KRAC05,SANT05,LOUB05,MORG06,ADEN07,BARU08,NIEU09,MATZ09,KARL09}.
The number of papers indicating dissent with Bell and his followers
represents a rousing chorus and is still increasing.

The structure of the paper is as follows. We add two introductory
subsections that explain the main points of statistics and classical
probability theory that need to be carefully considered when
discussing EPR experiments. In Section~\ref{Boole}, we discuss
general, conceptual aspects of the works of Boole~\cite{BO1862},
Kolmogorov-Vorob'ev~\cite{VORO62} and Bell~\cite{BELL64,BELL93} and
of their mutual relationships. Section~\ref{Boole} also presents a
derivation of Boole's conditions of possible
experience~\cite{BO1862} which differs from Boole's.
In Section~\ref{functions} we demonstrate by elementary arithmetics that real
non negative functions of dichotomic variables satisfy inequalities
that are of the same form as the Boole inequalities.
Section~\ref{QuantumTheory} extends the results of Section~\ref{functions}
to quantum theory. We use only commonly accepted postulates of
quantum theory to prove that a quantum system describing triples of
two-valued dynamical variables can never violate EBBI.
Although the quantum theoretical description of experiments
that measure two or more observables may involve non-commuting operators,
we show that this does not affect the derivation and application of EBBI
for the type of experiments we consider in this paper.
In Section~\ref{Leggett}, we
consider the interaction of the spins of three neutrons with the
magnetic moment of a SQUID (Superconducting Quantum Interference
Device), a two-state system~\cite{LEGG85}, at given time intervals.
We present a rigorous proof that the quantum theoretical description
of this experiment results in two-particle averages that cannot
violate the EBBI, in contrast to statements made in
Ref.~\onlinecite{LEGG85}. Section~\ref{EPR} discusses two types of
Einstein-Podolsky-Rosen-Bohm (EPRB) experiments. For the original
EPRB experiment~\cite{BOHM51}, we show that the apparent violation
of the EBBI appears as a consequence of substituting the expression
obtained from a quantum model with two spins into inequalities, the
EBBI, that hold for systems of three spins only. Hence, no
conclusions can be drawn from this violation. We analyze
realizable extensions of the EPRB experiment~\cite{SICA99} in which
the EBBI are satisfied. In Section~\ref{apparent} we explain why
actual experiments frequently appear to violate Boole(Bell)-type
inequalities.
We demonstrate apparent violations for
a real-live situation involving doctors and patients,
for a local realist factorizable model
and for laboratory EPRB experiments.
A summary and conclusions are given in Section~\ref{conclusions}.

\subsection{Experiments: data and statistics}\label{statistics}

We consider experiments and observations that can be represented by
two-valued variables $S=+1,-1$. For example, in a coin tossing
experiment one may assign $S=+1$ to the observation of head and
$S=-1$ to the observation of tails. In a Stern-Gerlach experiment,
one may define the observation of a ``click'' on one detector as
corresponding to $S=+1$ and the observation of a ``click'' on the
other detector as corresponding to $S=-1$.

During one experimental run, that lasts for a certain period of
time, a large set of data may be gathered. Further post-measurement
data analysis requires that this data set is labeled accordingly.
Data labeling not only involves simply enumerating the observations
but also needs to keep track of the experimental conditions under
which the data are gathered. The detail of labeling determines the
questions that can be asked, the hypothesis that can be checked, the
correlations that can be calculated and so on. Furthermore, if
several runs are made, the labels should include a unique
identification of each run.

Adding labels according to the experimental conditions requires a
careful consideration of the conditions that might influence the
experimental outcomes during the time period of the measurements.
For example, in the coin tossing experiment it might be essential to
know how many coins are tossed during one run, but it might also be
important to know the location where the various players are tossing
the coins. In this case, the two-valued variables $S$ acquire three
labels, one label numbering the coin, one label representing the
location where the player tosses the coin and one label simply
numbering the tosses. Similarly, in an EPRB experiment
the variables $S$ should carry the
index (1 or 2) of the magnet, an
index representing the orientation of the relevant magnetic field,
and a time label for the detection of the event.
Note that even if the time label or any other
label as for example a temperature label or an earth magnetic field label
does not seem to be of direct
importance for the experimental outcomes, the time label might well
be essential for the data analysis procedure and hence the variables
$S$ should also be labeled accordingly. Later, during the
post-processing step, one can then test the hypothesis that one or
the other label may be irrelevant but the converse is impossible: If
we have discarded (willingly or unwillingly) one or more labels
during the data collection process, these labels cannot be recovered
and we may well draw conclusions that seem paradoxical.

In some experiments, we collect one data element at a time, in
others such as the EPR thought experiment we collect two. We will
consider experiments that produce $n$-tuples of two-valued data that
are collected by ``observers'' who may not be aware of all aspects
of certain dynamical processes that have created the data. It is
thus crucial to employ an exact nomenclature that describes the
handling of data.

The data of $n$-tuples collected by the observer are therefore
denoted by
\begin{eqnarray}
\Upsilon^{(n)}&\equiv&\{
(S_{1,\alpha},\ldots,S_{n,\alpha})|\alpha=1,\ldots,M\},
\label{data0}
\end{eqnarray}
where each $S_{i,\alpha}$ ($i=1,\ldots,n)$ may assume the values
$\pm1$ and $M$ denotes the number of $n$-tuples which may be regarded as
fixed. We limit the discussion to pairs ($n=2$), triples ($n=3$)
and, occasionally, quadruples ($n=4$). Data sets of different runs
of a given sequence of experiments are denoted by
${\widehat\Upsilon}^{(n)}\equiv\{( {\widehat
S}_{1,\alpha},\ldots,{\widehat S}_{n,\alpha} )|$ $\alpha=1,\ldots,M\}$,
and ${\widetilde\Upsilon}^{(n)}\equiv\{( {\widetilde
S}_{1,\alpha},\ldots,{\widetilde S}_{n,\alpha})|\alpha=1,\ldots,M\}$
for the second, and third
run, respectively.

As a first step in the analysis of the data, it is common practice
to extract new sets from the data set $\Upsilon^{(n)}$ by grouping
the data in different ways. The reduced data sets that are obtained
by removing some elements of each $n$-tuple are denoted as
\begin{eqnarray}
\Gamma^{(n)}_{i}&\equiv&\{ S_{i,\alpha}|\alpha=1,\ldots,M\},
\nonumber \\
\Gamma^{(n)}_{ij}&\equiv&\{
(S_{i,\alpha},S_{j,\alpha})|\alpha=1,\ldots,M\},
\nonumber \\
\Gamma^{(n)}_{ijk}&\equiv&\{
(S_{i,\alpha},S_{j,\alpha},S_{k,\alpha})|\alpha=1,\ldots,M\},
\nonumber \\
&\dots&, \label{data1}
\end{eqnarray}
where $1\le i<j<\ldots\le n$. Although the approach taken in this
paper readily extends to $n>3$, we confine the discussion to
experiments and their description in terms of no more than three
dichotomic variables, because no additional insight is gained by
considering $n>3$.

We illustrate the use of the notation by an example. Let $n=3$,
meaning that an experiment produces triples of data that we collect
to form the set $\Upsilon^{(3)}$. Suppose that we want to analyze
this data by extracting three data sets of pairs, namely
$\Gamma^{(3)}_{12}$, $\Gamma^{(3)}_{13}$, and $\Gamma^{(3)}_{23}$.
Without further knowledge about the conditions under which the
experiments are carried out, we have
\begin{eqnarray}
\Gamma^{(3)}_{ij}&\not=&\Upsilon^{(2)}\quad,\quad
(i,j)=(1,2),(1,3),(2,3) , \label{data2}
\end{eqnarray}
even though the symbols that appear in both sets are the same. In
other words, in general there is no justification, logical or
physical, to assume that the data in $\Gamma^{(3)}_{ij}$ and
$\Upsilon^{(2)}$ have the same properties. A similar notation is
used to label averages of (products of) the $S_{i,\alpha}$. For
instance, ${F}^{(3)}_{ij}$ and $F^{(2)}$ are used to denote the
average over $\alpha$ of all products of pairs
$(S_{i,\alpha},S_{j,\alpha})$ of the reduced data set
${\Gamma}^{(3)}_{ij}$ and of the set $\Upsilon^{(2)}$, respectively.
If the number of subscripts is equal to $n$ we may, without creating
ambiguities, omit all the subscripts. Thus, we have
${\Gamma}^{(2)}\equiv{\Gamma}^{(2)}_{12}$,
 ${F}^{(3)}\equiv{F}^{(3)}_{123}$, and so on.

In 1862, Boole showed that whatever process generates a data set
$\Upsilon^{(3)}$ of triples of variables $S=\pm 1$, the averages of
all products of pairs in a data set $\Gamma_{ij}^{(3)}$ with
$(i,j)=(1,2),(1,3),(2,3)$ have to satisfy the
inequalities~\cite{BO1862}
%
\begin{equation}
|F_{ij}^{(3)}\pm F_{ik}^{(3)}| \le 1\pm F_{jk}^{(3)}\, ,\,
(i,j,k)=(1,2,3),(3,1,2),(2,3,1), \label{intro1}
\end{equation}
where $F_{ij}^{(3)}$ denote the averages of
all products of pairs in the set of triples $(S_1,S_2,S_3)$ (see
Eq.~(\ref{bel4})).
To prove Boole's inequalities Eq.~(\ref{intro1}) it is essential
that all pairs are selected from one and the same set of triples~\cite{BO1862}.
If we select pairs from three different
sets of pairs of dichotomic variables, then Boole's inequalities
Eq.~(\ref{intro1}) cannot be derived and may be violated. Indeed, if
the original data are collected in three sets of pairs, that is if
the data sets are $\Upsilon ^{(2)}$, ${\widehat \Upsilon} ^{(2)}$,
${\widetilde \Upsilon} ^{(2)}$ instead of $\Upsilon^{(3)}$, then the
average of products of pairs in these three sets have to satisfy the
less restrictive inequalities
\begin{equation}
|F^{(2)}\pm {\widehat F}^{(2)}| \le 3\pm |{\widetilde F}^{(2)}|.
\label{intro2}
\end{equation}
If we then test the hypothesis that
$F^{(2)}=F_{12}^{(3)}$, ${\widehat F}^{(2)}=F_{13}^{(3)}$, and
${\widetilde F}^{(2)}=F_{23}^{(3)}$ and find that Boole's
inequalities Eq.~(\ref{intro1}) are violated we can only conclude
that this hypothesis was incorrect. Therefore, if the data collected
in an experiment result in pair correlations that violate the Boole
inequalities, one or more of the following conditions must be true:
\begin{enumerate}
\item{The pairs of two-valued data have not been selected properly, that
is the pairs have not been selected from one data set with triples
of two-valued data.}
\item{There is no one-to-one mapping of the experimental outcomes to the chosen two-valued variables (see Subsection~\ref{lbasis}).}
\item{The labeling of the data is deficient.}
\item{The data processing procedure violates one or more rules of integer arithmetic.}
\end{enumerate}
No other conclusion can be drawn from the apparent violation because
the only assumptions needed to derive Boole's inequalities are that
the variables $S$ take values $+1,-1$, that integer arithmetic holds
and that the pairs of variables $S$ are selected from a set
containing triples of variables $S$.

The Boole inequalities Eq.~(\ref{intro1}) can be used to test the
hypothesis that the process giving rise to the data generates at
least triples. A theoretical model that purports to describe this
process should account for the possibility that the correspondence
between the empirical averages and the averages calculated from the
model may be deficient. Therefore, it is important to see to what
extent one can generalize Boole's results to theories within the
context of a theoretical model itself, that is without making
specific hypotheses about the relation between the empirical data
and the model. This is of particular relevance to quantum theory as
the latter, by construction, does not make predictions about
individual events but about averages only~\cite{HOME97}.

\subsection{Logical basis of probability frameworks}\label{lbasis}

We introduce here some aspects of the works of Boole \cite{BO1862},
Kolmogorov-Vorob'ev \cite{VORO62}, Bell~\cite{BELL64,BELL93} and
others with
particular emphasis on the connection of
probability models to logical elements and at the same time to data
sets. In particular we discuss two questions that need to be agreed
upon when dealing with any given set of experimental data in a
probabilistic model for two-valued possible outcomes:

\begin{itemize}
\item[(i)]{Can the data be brought into a one-to-one correspondence with elements
$x_1$, $x_2$, $x_3$, $\ldots$ ($x_i=0,1$) or $S_1, S_2, S_3,\ldots$ ($S_i=\pm1$)
of a two-valued logic,
and do we thus have a one-to-one correspondence of logical elements
to data (\OTOCLED)? This correspondence must be based on sense
impressions related to the experiments and measurements.}
\item[(ii)]{Are the data justifiably grouped into $n$-tuples ($n \geq 2$) corresponding to a
specific hypothesis about the correlation of the experimental facts?
We call this the correlated $n$-tuple hypothesis (\CNTUH). For
example, if we investigate the consequences of a particular illness
in a large number of patients and we have the hypothesis that there
are three symptoms to the illness, we assign to each patient a
triple such as $(S_1 = +1, S_2 = -1, S_3 =+1)$ meaning the patient
was positive for symptom $1$ and $3$ and negative for $2$.}
\end{itemize}

The second question has been addressed in
Subsection~\ref{statistics} and we will concentrate mostly on the
first.

We investigate the correlations of pair outcomes such as $(S_1 = +1,
S_2= -1)$ that are consistent with possible experience and denote
the rules that we obtain for these pair correlations with Boole as
conditions of possible experience (COPE). Note that this name
(chosen by Boole) is somewhat misleading because the actual premises
that have COPE as a consequence contain the requirement of a
one-to-one correspondence with logical elements as well as a
hypothesis that $n$-tuples of these elements ``belong together'',
for instance because they correspond to symptoms of single patients.
This belonging together means that we give meaning or preference to
certain sets and we concatenate these sets by regarding them as a
logical ``indivisible whole''. In the case of Boole, the
indivisibility corresponds to the allocation of three symptoms to a
single patient and the corresponding use (see below) of elements of
logic grouped in triples~\cite{BO1862}. The work of
Kolmogorov-Vorob'ev deals also with such $n$-tuple groupings by use
of $n$ functions (random variables) on one common probability
space~\cite{VORO62}. Bell groups data into triples or quadruples by
letting each three or four of his functions representing the data
depend on the identical element of reality
$\lambda$~\cite{BELL64,BELL93}. Finally, we group below into
$n$-tuples by forming functions on sets of two, three or four
variables.

If COPE show an inconsistency with the data, then we may conclude
either that our view contained in (i) or (ii) or both must in some
way be inadequate or we may go further and conclude that the
concepts that form the basis for the language of (i) and (ii) such
as reality, macroscopic reality or locality are inadequate. For
example, the symptoms observed for a given patient might be
influenced by those of others at a distance which may make a
different grouping necessary.

As mentioned, it is one of the main results of this paper that the
inconsistencies of pair correlations of data of EPRB experiments and
other experiments related to quantum mechanics as indicated by
certain inequalities such as those of John Bell \cite{BELL93} are
the consequences of the inadequacies of (i) and/or (ii) in
describing the data instead of inadequacies of basic concepts such
as realism or macroscopic realism. Locality considerations also need
not be blamed for the inconsistencies although these have a special
standing: Influences at a distance can never be disproved. We show
our point by the fact that if (ii) is valid for $n$-tuple size $n
\geq 3$ then the inequalities of Boole, of Vorob'ev (and others) and
of Bell (that represent non-trivial restrictions for the
pair-correlations) are valid even if we relate the data only to
dichotomic variables and treat them as independent of their
connections to any logic. This means we deal then with the axioms of
integers to derive the inequalities and can then never find a
violation. If a violation is found then the hypothesis in (ii) that
lead to the grouping in $n$-tuples must be rejected.

To set the stage we discuss a number of examples. Boole
\cite{BO1862} introduced a system of elements of mathematical logic
(Boolean variables) such as $true$ and $false$ that can be brought
into a one-to-one correspondence with two numbers such as $x = 0, 1$
or $S = \pm 1$ and that follow the algebra of these integers. This
system is then linked to actual experimental outcomes. In
Kolmogorov's final form of probability theory one deals in a logical
fashion with the more general elementary events as well as random
variables (that can assume more than two values) and constructs a
sample space and probability space. The question of the truth
content of a proposition is thus reduced to the question of the
truth of the axioms of the probability framework that is used.
However, the concept of ``truth'' does not deal with the assertions
of a purely mathematical framework because by the word ``true'' we
invariably designate the one-to-one correspondence with a ``real''
observation or measurement of some object. It is therefore the
\OTOCLED\ that takes central stage. However, \OTOCLED\ occupies only
a paragraph in standard probability texts (see e.g.
Feller~\cite{FELL68}) and we therefore add an instructive example.

Consider a coin toss that can result in the outcomes heads and
tails. We may link these outcomes to the values that a two-valued
logical variable $x$ may assume. If we deal with more than one coin,
we need to introduce different variables because it is obvious that
different coin tosses can result in different outcomes while each
single coin can only show one outcome. Furthermore, the coins need
not be fair and may have different bias. Therefore different logical
elements $x, {\widehat x}, {\widetilde x},...$ need to be introduced
to describe the correspondence to the actual experiments. Things
become complicated if these coins contain some magnetic substance
and various magnets with different orientations influence the
different experiments. Then we may need to introduce a corresponding
different logical symbol for different coins as well as for
different magnet orientations e.g. use different subscripts such as
${\bf a}, {\bf b}, {\bf c}$ for different magnet orientations.
Furthermore there may be some other influences that co-determine the
toss outcomes. For example we may decide that we perform composite
experiments on three coins at a time and we need to include in
addition subtle changes in the earth magnetic field for each such
three-coin-experiment \CNTUH\ that we label by an index $\alpha = 1,
2,...,M$. Logical elements tracking all these differences are then
denoted by e.g. $x_{{\bf a},\alpha}, {\widehat x}_{{\bf b},\alpha},
{\widetilde x}_{{\bf c},\alpha}$. Thus, the one-to-one
correspondence of logical elements (or elementary events etc.) to
observations or measurements as well as ordering into $n$-tuples
requires the knowledge of all the intricacies of the actual vehicles
and apparatuses of the measurements. Only if we have this knowledge
and only if we can establish a one-to-one correspondence of logical
elements and actual experiments and measurements that accounts for
all important details, can we use the algebra of the logical
variables to describe the experiments.

While this knowledge may be available for idealized coins, it is in
general not available in physical experiments and is not available
by definition if we attempt to describe these experiments by
probability theory. This simply means that our introduction of
logical elements in groups of $n$-tuples and choice of
correspondence to actual experimental facts represents a ``theory''
that may or may not be sufficient to guarantee full consistency.
This fact becomes particularly important when we consider
correlations of different experiments or correlations in composite
(more than one coin) experiments. In the above mentioned experiment
that involves a changing magnetic field, the correlations between
all the data will be different if we use one coin, two coins, three
coins or more coins in any given composite experiment.
Generalizations of this simple example to physical experiments are
used below when discussing Boole's inequalities and in
Section~\ref{apparent}.

In general physical experiments (involving e.g. observers such as
Alice and Bob, a cat, a decaying radioactive substance and the
moon), one usually indicates possible differences in experimental
outcomes by the introduction of Einstein's space-time. The statement
``the moon shines while Bob cooks'' is not precise enough to express
an everlasting truth that can be linked to logical elements such as
the $x_{{\bf a},\alpha}$ above. In order to establish a generally
valid correspondence more precise coordinates need to be given such
as ``the moon shines while Bob cooks dinner given space-time
coordinates $r_x, r_y, r_z, t$''. The outcomes of measurements and
observations are then described by functions of space-time and we
need in general to introduce a different logical element
corresponding to each different function and to each different
space-time label. In the Kolmogorov framework such expansion of
correspondence is established, for example, by the introduction of a
time label of random variables for Stochastic Processes or for
Martingales; generalization to space-time being relatively
straightforward.

The question arises naturally if criteria can be established on
whether the characterization of experiments (performed by using some
``theory'' related to the data) and the chosen one-to-one
correspondence of these experiments to logical elements (or
Kolmogorov's elementary events) and to $n$-tuples of data (a
grouping that co-determines certain correlations) is sufficiently
detailed so that no contradictions between actual experiments and
the results of the used probability theory model can arise. Such
criteria were derived in Boole's work of 1862 in form of the
mentioned inequalities. The combinatorial-topological content of
these inequalities was not explored by Boole and was derived much
later (1962) by Vorob'ev \cite{VORO62}. Again a few years later,
John Bell~\cite{BELL64} unveiled the importance of inequalities that
were virtually identical to Boole's and based on \CNTUH; the
difference being the application to medical statistics by Boole and
to quantum mechanics by Bell. Key for the understanding of Bell's
work is that Bell does not seem to have been aware of the fact
(proven by Boole in 1862, see Section~\ref{Boole}) that the
assumption of (ii) on the basis of dichotomic variables is
sufficient to always validate the known Boole-Bell inequalities
independent of any action or influence at a distance.

\section{Boole's conditions of possible experience}\label{Boole}

Here we summarize the work of Boole~\cite{BO1862} related to his
topic ``conditions of possible experience'' (COPE). We first explain the
basic facts in terms of Boole's inequalities for logical variables.
Subsequently we connect these inequalities derived for logical
variables to actual experiments and corresponding data
and link these inequalities to the work of Vorob'ev~\cite{VORO62}.

\subsection{Boole inequalities}

Let us consider three Boolean variables $x_1=0,1$, $x_2=0,1$, and $x_3=0,1$ and let us use the short hand
notation ${\bar x_i}=1-x_i$ for $i=1,2,3$.
Obviously the following identity holds:
%
\begin{eqnarray}
1&=&
{\bar x}_1{\bar x}_2{\bar x}_3
+
{     x}_1{\bar x}_2{\bar x}_3
+
{\bar x}_1{     x}_2{\bar x}_3
+
{     x}_1{     x}_2{\bar x}_3
\nonumber \\
&&+
{\bar x}_1{\bar x}_2{     x}_3
+
{     x}_1{\bar x}_2{     x}_3
+
{\bar x}_1{     x}_2{     x}_3
+
{     x}_1{     x}_2{     x}_3
.
\label{boo0}
\end{eqnarray}
We want to pick pairs of contributions such that each pair can be written as
a product of two Boolean variables only.
A nontrivial condition on the Boolean variables appears when we group
terms such that there is no way that we can continue adding two contributions
and reduce the number of variables in a term.
For instance,
\begin{eqnarray}
1&=&
{\bar x}_1{\bar x}_2{\bar x}_3
+
\left(
{     x}_1{\bar x}_2{\bar x}_3
+
{     x}_1{\bar x}_2{     x}_3
\right)
+
\left(
{\bar x}_1{\bar x}_2{     x}_3
+
{\bar x}_1{     x}_2{     x}_3
\right)
\nonumber \\
&&+
\left(
{     x}_1{     x}_2{\bar x}_3
+
{\bar x}_1{     x}_2{\bar x}_3
\right)
+
{     x}_1{     x}_2{     x}_3
\nonumber \\
&=&
{\bar x}_1{\bar x}_2{\bar x}_3
+
{     x}_1{\bar x}_2
+
{\bar x}_1{     x}_3
+
{     x}_2{\bar x}_3
+
{     x}_1{     x}_2{     x}_3
.
\label{boo2}
\end{eqnarray}
We rewrite Eq.~(\ref{boo2}) as
\begin{eqnarray}
{     x}_1{\bar x}_2
+
{\bar x}_1{     x}_3
+
{     x}_2{\bar x}_3
&=&1-
{\bar x}_1{\bar x}_2{\bar x}_3
-
{     x}_1{     x}_2{     x}_3
,
\label{boo3}
\end{eqnarray}
and as the two right most terms in Eq.~(\ref{boo3}) are zero or one, we have
\begin{eqnarray}
{     x}_1{\bar x}_2
+
{\bar x}_1{     x}_3
+
{     x}_2{\bar x}_3
&\le&
1
.
\label{boo4}
\end{eqnarray}
Similar inequalities can be derived by grouping terms differently. Alternatively,
if we replace ${x}_1$ by ${\bar x}_1$ in Eq.~(\ref{boo4}), we obtain
another inequality. Replacing ${x}_2$ by ${\bar x}_2$ in these two inequalities,
we obtain two new ones and replacing ${x}_3$ by ${\bar x}_3$
in the resulting four inequalities, we finally
end up with eight different but very similar inequalities.

It is often convenient to work with variables $S=\pm1$ instead of $x=0,1$.
Thus, we substitute $S_i=2x_i-1$ for $i=1,2,3$ in Eq.~(\ref{boo4})
and obtain
\begin{eqnarray}
-S_1S_2-S_1S_3-S_2S_3 &\le&1
,\nonumber \\
+S_1S_2+S_1S_3-S_2S_3 &\le&1
,
\label{boo7}
\end{eqnarray}
where the second inequality has been obtained from the first
by substituting $S_1\rightarrow-S_1$.
Note that we can write Eq.~(\ref{boo7}) as $|S_1S_2+S_1S_3|\le 1+S_2S_3$.
This inequality is in essence already a Boole inequality for logical
variables~\cite{BO1862}.

\subsection{Boole's inequalities and experience}

We now turn to the connection of the above results to actual data
and experience. We first note, and this is crucial, that
Eqs.~(\ref{boo4}) and (\ref{boo7}) are derived from Eq.~(\ref{boo2})
that was based on logical triples while Eqs.~(\ref{boo4}) and
(\ref{boo7}) deal with pair products only. If we wish to make a
connection of the logic to actual data, we then need to establish a
one-to-one correspondence
 of the logical triples to data-triples (\OTOCLED) and we need to
cover the set of all data by the set of all such triples. If and
only if this one-to-one correspondence is correctly established,
does Boole relate his inequalities to ``experience'' (see
discussions in Section~\ref{games}). We assume that this has been
accomplished and correspondingly add a new label $\alpha$ to the
variables. Then, using the notation introduced in
Section~\ref{Notation}, the set of data is $\Upsilon^{(3)}=\{
(S_{1,\alpha},S_{2,\alpha},S_{3,\alpha})|\alpha=1,\ldots,M\}$ and
$n=3$.

The averages of $S_{i,\alpha}S_{j,\alpha}$ over all $\alpha$ define
the correlations
\begin{eqnarray}
F^{(3)}_{ij}&=&\frac{1}{M} \sum_{\alpha=1}^M S_{i,\alpha}S_{j,\alpha}=F^{(3)}_{ji}
.
\label{bel4}
\end{eqnarray}
where $1\le i < j\le 3$.
Note, and this is essential,  that $F^{(3)}_{ij}$ is calculated from the pairs
in the reduced data set $\Gamma^{(3)}_{ij}$, not from pairs
in some data set $\Upsilon^{(2)}$.

From inequalities Eq.~(\ref{boo7}), it then follows directly that we have
\begin{eqnarray}
|F^{(3)}_{12}\pm F^{(3)}_{13}|&\le&1\pm F^{(3)}_{23}
,
\label{bel5}
\end{eqnarray}
where the inequality with the minus signs follows from
the one with the plus signs by letting $S_3\rightarrow-S_3$.
By permutation of the labels 1, 2, and 3 we find
%
\begin{equation}
|F^{(3)}_{ij}\pm F^{(3)}_{ik}|\le1\pm F^{(3)}_{jk}
,\, (i,j,k)=(1,2,3),(3,1,2),(2,3,1)
,
\label{BOOLE0}
\end{equation}
which are exactly Boole's conditions of possible experience in terms
of the concurrencies $(1+S_{i,\alpha}S_{j,\alpha})/2$~\cite{BO1862}.
Note that Boole wrote his inequalities in terms of frequencies.
The inequalities
Eq.~(\ref{BOOLE0}) have the same structure as the inequalities derived
by Bell~\cite{BELL64,BELL93}. Under the conditions stated, namely that
$F^{(3)}_{ij}$ is calculated from triples of data
$(S_{1,\alpha}, S_{2,\alpha}, S_{3,\alpha})$,
a violation of Eq.~(\ref{BOOLE0}) is mathematically impossible.

It is easy to repeat the steps that lead to Eq.~(\ref{BOOLE0}) if
the data are grouped into quadruples, that is the data set is
$\Upsilon^{(4)}=\{
(S_{1,\alpha},S_{2,\alpha},S_{3,\alpha},S_{4,\alpha})|\alpha=1,\ldots,M\}$.
Then, the correlations $F^{(4)}_{ij}$ satisfy inequalities such as
\begin{eqnarray}
|F^{(4)}_{13}- F^{(4)}_{23}+F^{(4)}_{14}+F^{(4)}_{24}|&\le&2
,
\label{CHSH0}
\end{eqnarray}
which is reminiscent of the Clauser-Horn-Shimony-Holt (CHSH) inequality~\cite{CLAU69}. Again, a
violation of inequalities of the type Eq.~(\ref{CHSH0})
is logically and mathematically impossible if
$F_{ij}^{(4)}$ is calculated from quadruples of data $(S_{1\alpha},
S_{2\alpha}, S_{3\alpha}, S_{4\alpha})$. In the remainder of this
paper, we focus on data sets containing at most triples, the
extension to quadruples etc. bringing no new insights.

\subsection{A trap to avoid I}\label{trapI}

We emphasize again that it is essential to keep track of the fact
that the correlations $F^{(3)}_{ij}$ have been calculated from the
data set that contains triples $\Upsilon^{(3)}$ instead of from
another set $\Upsilon^{(2)}$ in which the data has been collected in
pairs. Of course, the sorting in triples may not correspond to the
physical process of data creation. In general, there is no reason to
expect that one of the three $\Gamma^{(3)}_{ij}$'s is related to
$\Upsilon^{(2)}$, even though both sets contain two-valued
variables. It could be, as in the examples of
Section~\ref{apparent}, that the pair correlations are different if
the measurements are taken in pairs instead of triples. If the
experiment yields the data sets $\Upsilon^{(2)}$,
$\widehat\Upsilon^{(2)}$, and $\widetilde\Upsilon^{(2)}$ containing
pairs only and if we have physical differences in the taking of
pair-data, then we may have to replace Eq.~(\ref{boo7}) by the
inequalities
\begin{eqnarray}
-3\le -S_{1,\alpha}S_{2,\alpha}-{\widehat S}_{1,\alpha}{\widehat S}_{2,\alpha}-{\widetilde S}_{1,\alpha}{\widetilde S}_{2,\alpha} &\le&3
,\nonumber \\
-3 \le +S_{1,\alpha}S_{2,\alpha}+{\widehat S}_{1,\alpha}{\widehat S}_{2,\alpha}-{\widetilde S}_{1,\alpha}{\widetilde S}_{2,\alpha} &\le&3
,
\label{boo7h1}
\end{eqnarray}
for $\alpha=1,\ldots,M$. A more detailed account of these
considerations that also relates to the EPR-experiments discussed in Section~\ref{apparent}.

We may now again calculate averages. However, a different inequality
applies for the averages of pairs that we denote by $F^{(2)}$. From
inequality Eq.~(\ref{boo7h1}) obtained for data sets $\Upsilon
^{(2)}$, $\widehat\Upsilon ^{(2)}$ and $\widetilde\Upsilon ^{(2)}$
composed of pairs, we get
\begin{eqnarray}
|F^{(2)}\pm {\widehat F}^{(2)}|&\le&3 - |{\widetilde F}^{(2)}| ,
\label{bel5h1}
\end{eqnarray}
which differs from Bell's inequality~\cite{BELL64,BELL93} but is the
correct Boole inequality if pairs instead of triples of dichotomic
variables match the experimental facts.

\subsection{Relation to Kolmogorov's probability theory}

Although we do not need to involve references to Kolmogorov for the
reasoning presented here, it may be useful for some readers to
rephrase the above in this language. Conditions of the type shown in
Eq.~(\ref{BOOLE0}) have been studied in great detail by Vorob'ev
\cite{VORO62} on the basis of Kolmogorov's probability theory.
Vorob'ev showed in essence by very general combinatorial and
topological arguments that the non-trivial restriction of
Eq.~(\ref{boo7}) to $\leq 1$ instead of the trivial $\leq 3$ is a
consequence of the cyclical arrangement of the variables that form a
closed loop: the choice of variables in the first two terms
determines the choice for the variables in the third term. Vorob'ev
has proven that any nontrivial restriction expressed by this type of
inequalities is a consequence of a combinatorial-topological
``cyclicity''. For the Kolmogorov definitions this means that
violation of such inequalities implies that functions corresponding
to $S_1, S_2, S_3$ can not be defined on one probability space i.e.
are not Kolmogorov random variables. If no cyclicity is involved,
the functions can be defined on a single given Kolmogorov
probability space and no nontrivial restriction is obtained.

\subsection{Summary}

Using elementary arithmetic only, we have shown that whatever
process generates data sets organized in triples
\begin{eqnarray}
\Upsilon^{(3)}&\equiv&\{ (S_{1,\alpha},S_{2,\alpha},S_{3,\alpha})|\alpha=1,\ldots,M\}
,
\label{bel10}
\end{eqnarray}
the correlations $F^{(3)}_{ij}$ have to satisfy Boole's inequalities
Eq.~(\ref{BOOLE0}). If they do not, the procedure to compute
$F^{(3)}_{ij}$ from the data $\Upsilon^{(3)}$ violates a basic rule
of integer arithmetic. If the data are collected and grouped into
pairs, then in general the correlations need only obey inequality
Eq.~(\ref{bel5h1}).

\section{Boole inequalities for non negative functions}\label{functions}

Groups of two-valued data, generated by actual experiments or just
by numerical algorithms have to comply with the inequalities of
Section~\ref{Boole}, independent of the details of the physical or
arithmetic processes that produce the data. Assuming that the
premisses for an inequality to hold are satisfied, which may include
a certain grouping of the data (\CNTUH) or a one-to-one
correspondence of two-valued variables to logical elements (\OTOCLED)
or both, a violation of this inequality is then tantamount to a
violation of the rules of integer arithmetic.

We now ask whether there exist inequalities, similar to those of
Section~\ref{Boole},
for certain theoretical models that describe the two-valued variables
that result in the data.
As it is not our
intention to address this question in its full generality, we will
confine the discussion to models based on Kolmogorov's axioms of
probability theory and/or on the axioms of quantum theory.

The Kolmogorov framework features a well-defined
relation between the elements $\omega$ of the sample space $\Omega$
(representing the set of all possible outcomes) and the actual data.
In our case of countable $\Omega$, Kolmogorov ``events" $F$ are just
subsets of $\Omega$. The probability that $F$ will occur in an
experiment yet to be performed is expressed by a real valued
positive function on $\Omega$, the probability measure. This allows
us to calculate mathematical expectations and correlations related
to the data~\cite{FELL68}. Combined with our focus on dichotomic variables, this
naturally leads us to the study of non negative functions of $n$
dichotomic variables as presented below.

The quantum theoretical description of a system containing $n$
two-state objects leads one to consider non negative functions of $n$
dichotomic variables, each variable corresponding to an eigenvalue
of each of the $n$ dynamical variables. As the detailed relationship
between quantum theory and non negative functions is of no importance
for the remainder of this section, we relegate the derivation of
this relationship to Section~\ref{QuantumTheory}.

In the remainder of this section, we derive Boole-like inequalities
for real, non negative functions of dichotomic variables using
elementary algebra only.

\subsection{Two variables}
It is not difficult to see that any real-valued function $f^{(2)}=f^{(2)}(S_1,S_2)$
of two dichotomic variables $S_1=\pm1$ and $S_2=\pm1$ can be written as
\begin{eqnarray}
f^{(2)}(S_1,S_2)&=&
\frac{E^{(2)}_0+S_1E^{(2)}_1+S_2E^{(2)}_2+S_1S_2E^{(2)}}{4}
,
\label{trap0}
\end{eqnarray}
where
\begin{eqnarray}
E^{(2)}_0&=&\sum_{S_1=\pm1}\sum_{S_2=\pm1}f^{(2)}(S_1,S_2)
,
\label{trap1a} \\
E^{(2)}_i&=&\sum_{S_1=\pm1}\sum_{S_2=\pm1}S_if^{(2)}(S_1,S_2)\quad,\quad i=1,2
,
\label{trap1b} \\
E^{(2)}&=&\sum_{S_1=\pm1}\sum_{S_2=\pm1}S_1S_2f^{(2)}(S_1,S_2)
.
\label{trap1c}
\end{eqnarray}

We ask for the constraints on the $E$'s that appear in
Eq.~(\ref{trap0}) for non negative function $f^{(2)}(S_1,S_2)$. If
$f^{(2)}(S_1,S_2)\ge0$, from Eq.~(\ref{trap1a}) we have
$E^{(2)}_0\ge0$ and from
\begin{eqnarray}
E^{(2)}_0+S_1S_2E^{(2)}&\ge& -S_1(E^{(2)}_1+S_1S_2E^{(2)}_2)
,
\label{trap2}
\end{eqnarray}
it follows that
\begin{eqnarray}
E^{(2)}_0\pm E^{(2)}&\ge& |E^{(2)}_1\pm E^{(2)}_2|
.
\label{trap3}
\end{eqnarray}
Writing $4f^{(2)}(S_1,S_2)=E^{(2)}_0+S_1S_2E^{(2)}+S_1(E^{(2)}_1+S_1S_2E^{(2)}_2)$,
it directly follows that if both $E^{(2)}_0\ge0$ and Eq.~(\ref{trap3}) hold, then
$f^{(2)}(S_1,S_2)$ is non negative.
Thus, we have proven

\medskip\noindent
{\bf {Theorem I:}}
For a real-valued function $f^{(2)}(S_1,S_2)$ that is a
function of two variables $S_1=\pm1$ and $S_2=\pm1$
to be non negative, it is necessary and sufficient that
the expansion coefficients defined by Eqs.~(\ref{trap1a}), (\ref{trap1b}), (\ref{trap1c})
satisfy the inequalities
\begin{eqnarray}
0&\le&E^{(2)}_0\quad,\quad |E^{(2)}_1\pm E^{(2)}_2|\le E^{(2)}_0\pm E^{(2)}
.
\label{trap4}
\end{eqnarray}

As we deal with functions of two variables only, it is
not a surprise that the inequalities Eq.~(\ref{trap4}) do not resemble Boole's
inequalities Eq.~(\ref{BOOLE0}).

\subsection{Three and more variables}

Next, we consider real functions of three dichotomic variables.
As in the case of two  dichotomic variables, one readily verifies
that any real function of three dichotomic variables can be written as
\begin{eqnarray}
f^{(3)}(S_1,S_2,S_3)&=&
\frac{E^{(3)}_0+S_1E^{(3)}_1+S_2E^{(3)}_2+S_3E^{(3)}_3}{8}
\nonumber \\
&&+
\frac{S_1S_2E^{(3)}_{12}+S_1S_3E^{(3)}_{13}
+S_2S_3E^{(3)}_{23}}{8}
\nonumber \\
&&
+
\frac{S_1S_2S_3E^{(3)}}{8}
,
\label{booe0}
\end{eqnarray}
where
\begin{eqnarray}
E^{(3)}_0&=&\sum_{S_1=\pm1}\sum_{S_2=\pm1}\sum_{S_3=\pm1}f^{(3)}(S_1,S_2,S_3)
,
\label{booe1a} \\
E^{(3)}_i&=&\sum_{S_1=\pm1}\sum_{S_2=\pm1}\sum_{S_3=\pm1}S_if^{(3)}(S_1,S_2,S_3),
\label{booe1b} \\
E^{(3)}_{ij}&=&\sum_{S_1=\pm1}\sum_{S_2=\pm1}\sum_{S_3=\pm1}S_iS_jf^{(3)}(S_1,S_2,S_3),
\label{booe1c} \\
E^{(3)}&=&\sum_{S_1=\pm1}\sum_{S_2=\pm1}\sum_{S_3=\pm1}S_1S_2S_3f^{(3)}(S_1,S_2,S_3)
,
\label{booe1d}
\end{eqnarray}
where $i=1,2,3$ and $(i,j)=(1,2),(1,3),(2,3)$.

We postulate now that all functions $f^{(n)}$ obey $f^{(n)} \geq 0$ for $n\ge1$.
In the Kolmogorov framework this would be a step toward
defining a ``probability measure'' that, of course, also needs to
include the proper definition of algebras that are certain systems
$F$ of subsets of the sample space $\Omega$ and that relate to the pair,
triple or quadruple measurements. The coefficients $E^{(3)}_{ij}$
that appear in Eq.~(\ref{booe0}) relate to the pair correlations of
the various variables $S_i$ and we ask ourselves the question
whether Boole-type inequalities can be derived for them and what
form these inequalities will assume. We formalize our results by

\medskip\noindent
{\bf {Theorem II:}}
The following statements hold:

\begin{itemize}
\item[]{II.1 If $f^{(3)}(S_1,S_2,S_3)$ is a real non negative function of
three variables $S_1=\pm1$, $S_2=\pm1$, and $S_3=\pm1$,
the inequalities
\begin{eqnarray}
|E^{(3)}_{ij}\pm E^{(3)}_{ik}|&\le&E^{(3)}_0\pm E^{(3)}_{jk}
,
\label{BOOLE1}
\end{eqnarray}
with $(i,j,k)=(1,2,3),(3,1,2),(2,3,1)$ hold.
}
\medskip
\item[]{II.2 Given four real numbers satisfying
$|E^{(3)}_{ij}| \le E^{(3)}_{0}$
for $(i,j)=(1,2),(1,3)$, $(2,3)$
and satisfying Eq.~(\ref{BOOLE1}),
there exists a real, non negative function
\linebreak
$f^{(3)}(S_1,S_2,S_3)$
of three variables $S_1=\pm1$, $S_2=\pm1$, and $S_3=\pm1$,
such that Eqs.~(\ref{booe1a}) and (\ref{booe1c}) hold.
}
\end{itemize}

\medskip\noindent
{\bf {Proof:}} To prove II.1, we first note that from
$f^{(3)}(S_1,S_2,S_3)\ge0$ and Eqs.~(\ref{booe0}) -- (\ref{booe1d}),
it follows that $0\le E^{(3)}_0$ and that $|E^{(3)}_1|\le
E^{(3)}_0$, $|E^{(3)}_2|\le E^{(3)}_0$, $|E^{(3)}_3|\le E^{(3)}_0$,
$|E^{(3)}_{12}|\le E^{(3)}_0$, $|E^{(3)}_{13}|\le E^{(3)}_0$,
$|E^{(3)}_{23}|\le E^{(3)}_0$, and $|E^{(3)}|\le E^{(3)}_0$. We now
ask ourselves whether the non negativity of $f^{(3)}(S_1,S_2,S_3)$
enforces more stringent conditions on the $E$'s. We follow the same
procedure as the one that lead to Eq.~(\ref{BOOLE0}). Let us rewrite
Eq.~(\ref{booe1a}) as
\begin{eqnarray}
E^{(3)}_0&=&f^{(3)}(-1,-1,-1)
\nonumber \\
&&+
\left[
f^{(3)}(+1,-1,-1)
+f^{(3)}(+1,-1,+1)
\right]
\nonumber \\
&&+\left[
f^{(3)}(-1,-1,+1)
+
f^{(3)}(-1,+1,+1)
\right]
\nonumber \\
&&+\left[
f^{(3)}(-1,+1,-1)
+
f^{(3)}(+1,+1,-1)
\right]
\nonumber \\
&&+f^{(3)}(+1,+1,+1)
.
\label{booe2}
\end{eqnarray}
From the representation Eq.~(\ref{booe0}), it follows that
\begin{eqnarray}
f^{(3)}(+1,-1,-1)&+&f^{(3)}(+1,-1,+1)=
\nonumber \\
&&\frac{E^{(3)}_0+E^{(3)}_1-E^{(3)}_2 - E^{(3)}_{12}}{4}
,
\nonumber \\
f^{(3)}(-1,-1,+1)&+&f^{(3)}(-1,+1,+1)=
\nonumber \\
&&\frac{E^{(3)}_0-E^{(3)}_1+E^{(3)}_3 - E^{(3)}_{13}}{4}
,
\nonumber \\
f^{(3)}(-1,+1,-1)&+&f^{(3)}(+1,+1,-1)=
\nonumber \\
&&\frac{E^{(3)}_0+E^{(3)}_2-E^{(3)}_3 - E^{(3)}_{23}}{4}
,
\label{booe3}
\end{eqnarray}
such that Eq.~(\ref{booe2}) reduces to
\begin{eqnarray}
E^{(3)}_0&-&f^{(3)}(-1,-1,-1)-f^{(3)}(+1,+1,+1)
=
\nonumber \\
&&\frac{3E^{(3)}_0 - E^{(3)}_{12}- E^{(3)}_{13}- E^{(3)}_{23}}{4}
.
\label{booe4}
\end{eqnarray}
Using $0\le f^{(3)}(S_1,S_2,S_3)$,
we find
\begin{eqnarray}
-3E^{(3)}_0&\le&- E^{(3)}_{12}- E^{(3)}_{13}- E^{(3)}_{23}\le E^{(3)}_0
,
\label{booe5}
\end{eqnarray}
where the lower bound trivially follows from $|E_{12}^{(3)}|\le E_0^{(3)}$,
$|E_{13}^{(3)}|\le E_0^{(3)}$ and $|E_{23}^{(3)}|\le E_0^{(3)}$.
Using different groupings in pairs, we find that $E^{(3)}_{12}$,
$E^{(3)}_{13}$, and $E^{(3)}_{23}$ are bounded by the inequalities
\begin{eqnarray}
-3E^{(3)}_0&\le&-S_1S_2E^{(3)}_{12}-S_1S_3 E^{(3)}_{13}-S_2S_3 E^{(3)}_{23}\le E^{(3)}_0
,
\label{boole6}
\end{eqnarray}
for any choice of $S_1=\pm1$, $S_2=\pm1$ and $S_3=\pm1$.
Alternatively, we have the upper bound
\begin{eqnarray}
|E^{(3)}_{ij}\pm E^{(3)}_{ik}|&\le&E^{(3)}_0\pm E^{(3)}_{jk}
,
\label{BOOLE1a}
\end{eqnarray}
where $(i,j,k)=(1,2,3),(3,1,2),(2,3,1)$.
Thus, we have proven that if a real non negative function $f^{(3)}$
of three dichotomic variables exists, then the correlations defined
by Eq.~(\ref{booe1c}) satisfy the inequalities Eq.~(\ref{BOOLE1}).
Notice that Eq.~(\ref{BOOLE1}) is necessary but not sufficient for
$f^{(3)}$ to be non negative (see also Theorem IV).

To prove II.2, we assume that we are given four real numbers that
satisfy the inequalities $|A_{ij}| \le A_0$ and $|A_{ij}\pm A_{ik}|\le A_0\pm A_{jk}$
for $(i,j,k)=(1,2,3),(3,1,2)$, $(2,3,1)$.
Then, the function $g^{(3)}$ defined by
\begin{equation}
g^{(3)}(S_1,S_2,S_3)=
\frac{A_0+S_1S_2A_{12}+S_1S_3A_{13}+S_2S_3A_{23}}{8}
,
\label{g0}
\end{equation}
is non negative, as is easily seen by writing
$8g^{(3)}(S_1,S_2,S_3)=S_1S_2(A_{12}+S_2S_3A_{13})$ $+A_0+S_2S_3A_{23}$
and using the assumptions that
$|A_{ij}| \le A_0$  for $(i,j)=(1,2),(1,3)$, $(2,3)$
and $|A_{ij}\pm A_{ik}|\le A_0\pm A_{jk}$
for $(i,j,k)=(1,2,3),(3,1,2),(2,3,1)$.
Setting $A_0=E^{(3)}_0$ and $ A_{ij}=E^{(3)}_{ij}$
for $(i,j)=(1,2),(1,3),(2,3)$ completes the proof.

Although the context and derivation of Eq.~(\ref{BOOLE1a}) is
different from that used by Boole~\cite{BO1862} or
Bell~\cite{BELL64,BELL93}, the similarity to the Boole and Bell
inequalities is striking. Therefore, we will refer to
inequalities that have the same structure as Eq.~(\ref{BOOLE1}) as
the extended Boole-Bell inequalities (EBBI).

As in Section~\ref{Boole}, the above theorem readily generalizes to
functions of $n>3$ dichotomic variables. This generalization brings
no new insight.

\subsection{A trap to avoid II}\label{trapII}

In analogy with Section~\ref{trapI}, we now consider the case of
three different real non negative functions of two dichotomic
variables. In the spirit of the notation introduced earlier, we
denote these functions by $f^{(2)}$, ${\widehat f}^{(2)}$, and
${\widetilde f}^{(2)}$, respectively. The corresponding averages are
then $E^{(2)}_{0},\ldots,E^{(2)}$, ${\widehat
E}^{(2)}_{0},\ldots,{\widehat E}^{(2)}$, and ${\widetilde
E}^{(2)}_{0},\ldots,{\widetilde E}^{(2)}$, respectively. In view of
the complete arbitrariness of $f^{(2)}$, ${\widehat f}^{(2)}$, and
${\widetilde f}^{(2)}$, there is no reason to expect that
one can derive inequalities such as $|E^{(2)}\pm
{\widehat E}^{(2)}|\le E^{(2)}_0\pm {\widetilde E}^{(2)}$. Some
inequalities can be obtained by introducing additional assumptions
about the three functions. For instance, we have

\medskip\noindent
{\bf {Theorem III:}} Let $f^{(2)}(S,S')$, ${\widehat f}^{(2)}(S,S')$,
${\widetilde f}^{(2)}(S,S')$ be real non negative functions of
two variables $S=\pm1$ and $S'=\pm1$ defined by
\begin{eqnarray}
f^{(2)}(S,S')&=&\frac{E^{(2)}_0+SS'E^{(2)}}{4}
,\quad
{\widehat f}^{(2)}(S,S')=\frac{E^{(2)}_0+SS'{\widehat E}^{(2)}}{4}
,\nonumber \\
{\widetilde f}^{(2)}(S,S')&=&\frac{E^{(2)}_0+SS'{\widetilde E}^{(2)}}{4}
,
\label{booe7b}
\end{eqnarray}
then the inequalities
\begin{eqnarray}
|E^{(2)}\pm {\widehat E}^{(2)}|&\le&3E^{(2)}_0 - |{\widetilde E}^{(2)}|
,
\nonumber \\
|E^{(2)}\pm {\widetilde E}^{(2)}|&\le&3E^{(2)}_0 - |{\widehat E}^{(2)}|
,\nonumber \\
|{\widetilde E}^{(2)}\pm {\widehat E}^{(2)}|&\le&3E^{(2)}_0 - |{E}^{(2)}|
,
\label{booe11}
\end{eqnarray}
are satisfied.

\medskip\noindent
{\bf {Proof:}}
The assumption that $f^{(2)}$, ${\widehat f}^{(2)}$, and
${\widetilde f}^{(2)}$ are non negative obviously implies that
$0\le E^{(2)}_0$,
$|E^{(2)}|\le E^{(2)}_0$,
$|{\widehat E}^{(2)}|\le E^{(2)}_0$, and
$|{\widetilde E}^{(2)}|\le E^{(2)}_0$.
We consider
\begin{eqnarray}
&&f^{(2)}(S_1,-S_2)+{\widehat f}^{(2)}(-S_1,S_3)+{\widetilde f}^{(2)}(S_2,-S_3)
\nonumber \\
&&=
\frac{3E^{(2)}_0-S_1S_2E^{(2)}-S_1S_3{\widehat E}^{(2)}-S_2S_3{\widetilde E}^{(2)}}{4}
,
\label{booe8}
\end{eqnarray}
from which it immediately follows that
\begin{eqnarray}
S_1S_2E^{(2)}+S_1S_3{\widehat E}^{(2)}+S_2S_3{\widetilde E}^{(2)}
\le 3E^{(2)}_0
.
\label{booe9}
\end{eqnarray}
On the other hand, from $|E^{(2)}|\le E^{(2)}_0$, $|{\widehat
E}^{(2)}|\le E^{(2)}_0$ and $|{\widetilde E}^{(2)}|\le E^{(2)}_0$ it
follows that
\begin{equation}
-3E^{(2)}_0\le S_1S_2E^{(2)}+S_1S_3{\widehat E}^{(2)}+S_2S_3{\widetilde E}^{(2)}\le 3E^{(2)}_0.
\label{booextra}
\end{equation}
Hence Eq.~(\ref{booe9}) does not impose additional
constraints on the $E^{(2)}$'s that appear in Eq.~(\ref{booe7b}).
Rewriting Eq.~(\ref{booextra}) as
\begin{eqnarray}
-S_1S_2 (E^{(2)}+S_2S_3{\widehat E}^{(2)})
\le 3E^{(2)}_0 +S_2S_3{\widetilde E}^{(2)}
,
\nonumber \\
S_1S_2 (E^{(2)}+S_2S_3{\widehat E}^{(2)})
\le 3E^{(2)}_0 -S_2S_3{\widetilde E}^{(2)}
,
\label{booe10}
\end{eqnarray}
and noting that $S_1=\pm1$, $S_2=\pm1$, and $S_3=\pm1$
are arbitrary and that it is allowed to interchange
the roles of
$E^{(2)}$,
${\widehat E}^{(2)}$, and
${\widetilde E}^{(2)}$,
Eq.~(\ref{booe11}) follows.
Obviously, the inequalities Eq.~(\ref{booe11}) are the equivalent of
the inequalities Eq.~(\ref{bel5h1}) that we obtained in the case
that data sets consist of pairs, collected by performing three
different experiments.

In view of the logical contradictions that may follow from the
assumption that correlations of two dichotomic variables computed
from data sets of pairs satisfy the same inequalities as the same
correlations computed from data sets of triples, it is of interest
to inquire under what circumstances we can derive inequalities akin
to Eq.~(\ref{BOOLE1}), with the superscript $(3)$ replaced by the
superscript $(2)$. We have

\medskip\noindent
{\bf {Theorem IV:}}
The following statements hold:
\begin{itemize}
\item[]{IV.1
The three functions of two dichotomic variables defined by
\begin{eqnarray}
f^{(2)}(S_1,S_2)&=&
\frac{E^{(2)}_0+S_1E^{(2)}_1+S_2E^{(2)}_2+S_1S_2E^{(2)}}{4}
,
\nonumber \\
{\widehat f}^{(2)}(S_1,S_3)&=&
\frac{{\widehat E}^{(2)}_0+S_1{\widehat E}^{(2)}_1+S_3{\widehat E}^{(2)}_2+S_1S_3{\widehat E}^{(2)}}{4}
,
\nonumber \\
{\widetilde f}^{(2)}(S_2,S_3)&=&
\frac{{\widetilde E}^{(2)}_0+S_2{\widetilde E}^{(2)}_1+S_3{\widetilde E}^{(2)}_2+S_2S_3{\widetilde E}^{(2)}}{4}
,
\label{five5}
\end{eqnarray}
can be derived from a common function $f^{(3)}(S_1,S_2,S_3)$ of
three dichotomic variables by using
\begin{eqnarray}
f^{(2)}(S_1,S_2)&=&
\sum_{S_3=\pm1} f^{(3)}(S_1,S_2,S_3)
,\nonumber \\
{\widehat f}^{(2)}(S_1,S_3)&=&
\sum_{S_2=\pm1} f^{(3)}(S_1,S_2,S_3)
,\nonumber \\
{\widetilde f}^{(2)}(S_2,S_3)&=&
\sum_{S_1=\pm1} f^{(3)}(S_1,S_2,S_3)
,
\label{five1}
\end{eqnarray}
if and only if
$E^{(2)}_0={\widehat E}^{(2)}_0={\widetilde E}^{(2)}_0$,
$E^{(2)}_1={\widehat E}^{(2)}_1$,
$E^{(2)}_2={\widetilde E}^{(2)}_1$, and
${\widehat E}^{(2)}_2={\widetilde E}^{(2)}_2$.
}
\medskip
\item[]{IV.2
If (1) the three functions Eq.~(\ref{five5}) are non negative
and (2)
$E^{(2)}_0={\widehat E}^{(2)}_0={\widetilde E}^{(2)}_0$,
$E^{(2)}_1={\widehat E}^{(2)}_1$,
$E^{(2)}_2={\widetilde E}^{(2)}_1$,
${\widehat E}^{(2)}_2={\widetilde E}^{(2)}_2$,
and (3) the inequalities
\begin{eqnarray}
|E^{(2)}\pm {\widehat E}^{(2)}|&\le& E^{(2)}_0\pm {\widetilde E}^{(2)},
\nonumber \\
|E^{(2)}\pm {\widetilde E}^{(2)}|&\le& E^{(2)}_0\pm {\widehat E}^{(2)},
\nonumber \\
|{\widetilde E}^{(2)}\pm {\widehat E}^{(2)}|&\le& E^{(2)}_0\pm E^{(2)},
\label{BOOLE1X}
\end{eqnarray}
are satisfied, then there exists a non negative
$f^{(3)}(S_1,S_2,S_3)$ such that Eq.~(\ref{five1})
holds~\cite{FINE82b}. }
\medskip
\item[]{IV.3
If $f^{(3)}(S_1,S_2,S_3)$ is a real non negative function of
three dichotomic variables,
the three functions defined by Eq.~(\ref{five1})
are non negative and the coefficients $E^{(2)}$, ${\widehat E}^{(2)}$
and ${\widetilde E}^{(2)}$ that appear in their representation
Eq.~(\ref{five5}) satisfy the
inequalities Eq.~(\ref{BOOLE1X})~\cite{FINE82b}.
}
\end{itemize}

\medskip\noindent
{\bf {Proof:}}
Statement IV.1 directly follows from representation Eq.~(\ref{booe0}),
the fact that changing the order of summations does not change the result,
and the definitions
$E^{(3)}_0\equiv E^{(2)}_0={\widehat E}^{(2)}_0={\widetilde E}^{(2)}_0$,
$E^{(3)}_1\equiv E^{(2)}_1={\widehat E}^{(2)}_1$,
$E^{(3)}_2\equiv E^{(2)}_2={\widetilde E}^{(2)}_1$,
$E^{(3)}_3\equiv {\widehat E}^{(2)}_2={\widetilde E}^{(2)}_2$,
${E}^{(3)}_{12}\equiv{ E}^{(2)}$, ${E}^{(3)}_{13}\equiv{\widehat E}^{(2)}$, and
${E}^{(3)}_{23}\equiv{\widetilde E}^{(2)}$.
To prove IV.2, we write Eq.~(\ref{booe0}) as
\begin{eqnarray}
f^{(3)}(S_1,S_2,S_3)
&=&
\frac{E^{(3)}_0+S_1E^{(3)}_1+S_2E^{(3)}_2+S_1S_2E^{(3)}_{12}}{16}
\nonumber \\
&&+\frac{E^{(3)}_0+S_1E^{(3)}_1+S_3E^{(3)}_3+S_1S_3E^{(3)}_{13}}{16}
\nonumber \\
&&+\frac{E^{(3)}_0+S_2E^{(3)}_2+S_3E^{(3)}_3+S_2S_3E^{(3)}_{23}}{16}
\nonumber \\
&&+\frac{S_1S_2E^{(3)}_{12}+S_1S_3E^{(3)}_{13}+S_2S_3E^{(3)}_{23}-E^{(3)}_0}{16}
\nonumber \\
&&+\frac{S_1S_2S_3E^{(3)}}{8}
\nonumber \\
&=&
\frac{{f}^{(2)}(S_1,S_2)+{\widehat f}^{(2)}(S_1,S_3)+{\widetilde f}^{(2)}(S_2,S_3)}{4}
\nonumber \\&&
+\frac{S_1S_2E^{(3)}_{12}+S_1S_3E^{(3)}_{13}+S_2S_3E^{(3)}_{23}-E^{(3)}_0}{16}
\nonumber \\
&&+\frac{S_1S_2S_3E^{(3)}}{8}
,
\label{five3}
\end{eqnarray}
which is non negative if
\begin{eqnarray}
|E^{(3)}|
&\le&
2{f}^{(2)}(S_1,S_2)+2{\widehat f}^{(2)}(S_1,S_3)+2{\widetilde f}^{(2)}(S_2,S_3)
\nonumber \\&&
+\frac{S_1S_2E^{(3)}_{12}+S_1S_3E^{(3)}_{13}+S_2S_3E^{(3)}_{23}-E^{(3)}_0}{2}
,
\label{five4}
\end{eqnarray}
for any choice of $S_1=\pm1$, $S_2=\pm1$, and $S_3=\pm1$. By
assumption, the first three terms in Eq.~(\ref{five4}) are
non negative. Hence, Eq.~(\ref{five4}) always admits a solution for
$E^{(3)}$ if
$S_1S_2E^{(3)}_{12}+S_1S_3E^{(3)}_{13}+S_2S_3E^{(3)}_{23}\ge
E^{(3)}_0$ which by comparison with Eq.~(\ref{boole6}) is nothing
but the condition that the EBBI Eq.~(\ref{BOOLE1}) are
satisfied. Using IV.1 we conclude that, under the conditions stated,
the EBBI Eq.~(\ref{BOOLE1}) can be written as
Eq.~(\ref{BOOLE1X}). Finally, to prove IV.3, we note that if
expression Eq.~(\ref{booe0}) is non negative, the three functions
defined by Eq.~(\ref{five1}), being the sum of non negative numbers,
are non negative and the proof follows if we put
$E^{(2)}_0=E^{(3)}_0$, $E^{(2)}=E^{(3)}_{12}$, ${\widehat
E}^{(2)}=E^{(3)}_{13}$, and ${\widetilde E}^{(2)}=E^{(3)}_{23}$.

Theorem IV shows that if and only if the non negative two-variable
functions $f^{(2)}(S_1,S_2)$, ${\widehat f}^{(2)}(S_1,S_3)$,
${\widetilde f}^{(2)}(S_2,S_3)$ can be derived from a common real
non negative function $f^{(3)}(S_1,S_2,S_3)$ of three variables
$S_1=\pm1$, $S_2=\pm1$, and $S_3=\pm1$, only then it is allowed to
replace in the EBBI Eq.~(\ref{BOOLE1}) the superscripts
$(3)$ by the superscripts $(2)$.

\subsection{Relation to Bell's work}\label{Bell}

For completeness, we show now that the above construction includes
the restricted class of probabilistic models that form the core of
Bell's work~\cite{BELL93}. To see the mathematical structure of
these models, it suffices to use elementary arithmetic and a minimum
of probability concepts. Bell~\cite{BELL93} considers models
that are defined by
\begin{eqnarray}
f^{(2)}(S,S')&=&
\int  { f}^{(1)}(S|\lambda) {\widehat f}^{(1)}(S'|\lambda) \mu(\lambda)\,d\lambda
\nonumber \\
&=&\frac{1+SE^{(2)}_1+S'E^{(2)}_2+SS'E^{(2)}}{4}
,
\nonumber \\
{\widehat f}^{(2)}(S,S')&=&
\int {f}^{(1)}(S|\lambda) {\widetilde f}^{(1)}(S'|\lambda) \mu(\lambda)\,d\lambda
\nonumber \\
&=&\frac{1+S{\widehat E}^{(2)}_1+S'{\widehat E}^{(2)}_2+SS'{\widehat E}^{(2)}}{4}
,
\nonumber \\
{\widetilde f}^{(2)}(S,S')&=&
\int {\widehat f}^{(1)}(S|\lambda) {\widetilde f}^{(1)}(S'|\lambda) \mu(\lambda)\,d\lambda
\nonumber \\
&=&\frac{1+S{\widetilde E}^{(2)}_1+S'{\widetilde E}^{(2)}_2+SS'{\widetilde E}^{(2)}}{4}
,
\nonumber \\
\label{bell0}
\end{eqnarray}
where
\begin{eqnarray}
f^{(1)}(S|\lambda)&=&\frac{1+SE^{(1)}(\lambda)}{2},
\nonumber \\
{\widehat f}^{(1)}(S|\lambda)&=&\frac{1+S{\widehat E}^{(1)}(\lambda)}{2},
\nonumber \\
{\widetilde f}^{(1)}(S|\lambda)&=&\frac{1+S{\widetilde E}^{(1)}(\lambda)}{2}
,
\label{bell1}
\end{eqnarray}
$\mu(\lambda)$ is a probability density, 
a non negative function, which satisfies $\int \mu(\lambda)\,d\lambda
=1$ and $0\le f^{(1)}(S|\lambda)\le 1$,
$0\le {\widetilde f}^{(1)}(S|\lambda)\le 1$, and
$0\le {\widehat f}^{(1)}(S|\lambda)\le 1$.
The variable $\lambda$ is an element of a set that does not
need to be defined in detail. In Bell's work, $\lambda$ represents
the ``elements of reality'' corresponding to entangled pairs as
introduced by EPR but this representation is of no
concern for what follows in this section. From Eqs.~(\ref{bell0}) -- (\ref{bell1}) it follows that
\begin{eqnarray}
E^{(2)}_1&=&{\widehat E}^{(2)}_1=\int {E}^{(1)}(\lambda)\mu(\lambda)\,d\lambda,
\nonumber \\
E^{(2)}_2&=&{\widetilde E}^{(2)}_1=\int  {\widehat E}^{(1)}(\lambda)\mu(\lambda)\,d\lambda,
\nonumber \\
{\widehat E}^{(2)}_2&=&{\widetilde E}^{(2)}_2=\int  {\widetilde E}^{(1)}(\lambda)\mu(\lambda)\,d\lambda,
\nonumber \\
E^{(2)}&=&\int{E}^{(1)}(\lambda) {\widehat E}^{(1)}(\lambda)\mu(\lambda)\,d\lambda
,
\label{bell3x}
\end{eqnarray}
and so on.
Obviously, $f^{(2)}(S,S')$,
${\widehat f}^{(2)}(S,S')$, and ${\widetilde f}^{(2)}(S,S')$, being
sums of non negative contributions, are probabilities too.

We can easily construct the non negative function $f^{(3)}$ from which all three functions Eq.~(\ref{bell0})
can be derived by summing over the appropriate variable, namely~\cite{FINE82a}
\begin{eqnarray}
f^{(3)}(S,S',S'')&=&
\int  { f}^{(1)}(S|\lambda){\widehat f}^{(1)}(S'|\lambda) {\widetilde  f}^{(1)}(S''|\lambda) \mu(\lambda)\,d\lambda
,
\nonumber \\
&=&
\frac{E^{(3)}_0+SE^{(3)}_1+S'E^{(3)}_2+S''E^{(3)}_3}{8}
\nonumber \\
&&+\frac{SS'E^{(3)}_{12}+SS''E^{(3)}_{13} +S'S''E^{(3)}_{23}}{8}
\nonumber \\
&&+\frac{SS'S''E^{(3)}}{8}
.
\label{bell3}
\end{eqnarray}
In particular, we have $E^{(2)}=E^{(3)}_{12}$, ${\widehat
E}^{(2)}=E^{(3)}_{13}$, and ${\widetilde E}^{(2)}=E^{(3)}_{23}$.
From representation Eq.~(\ref{bell3}) it follows that the class of
models defined by Eq.~(\ref{bell0}) satisfies the conditions of
Theorem IV, hence these models satisfy the EBBI
Eq.~(\ref{BOOLE1X}).

The fact that there exists a non negative function of three variables
(Eq.~(\ref{bell3})) from which the three functions of two variables
(Eq.~(\ref{bell0})) can be recovered by summing over one of the
variables suffices to prove that the results of Bell's work are a
special case of Theorem IV. In Bell's original derivation of his
inequalities, no such arguments appear. However, it is well-known
that Bell's assumptions to prove his inequalities are equivalent to
the statement that there exists a three-variable joint probability
that returns the probabilities of Bell~\cite{FINE82a,FINE82b}. No
additional (metaphysical) assumptions about the nature of the model,
other than the assignment of non negative real values to pairs and
triples are required to arrive at this conclusion.

The relation of Bell's work to Theorems II and IV shows the
mathematical solidity and strength of Bell's work. It also shows, however, the
Achilles heel of Bell's interpretations: Because $\lambda$ has a
physical interpretation representing an element of reality,
Eq.~(\ref{bell0}) implies that in the actual
experiments identical $\lambda$'s are
available for each of the data pairs $(1,2),(1,3),(2,3)$.
This means that all of Bell's derivations assume from the start that
ordering the data into triples as well as into pairs must be
appropriate and commensurate with the physics. This ``hidden''
assumption was never discussed by Bell and his followers
~\cite{BELL93} and has ``invaded'' the mathematics in an innocuous
way. Once it is made, however, the
inequalities Eq.~(\ref{BOOLE1}) apply and even influences at a distance cannot change
this. The implications of this fact are discussed throughout this
paper and examples of actual classical experiments illustrating our
point are given in Section~\ref{apparent}.

\subsection{Summary}
The assignment of the range of a real-valued non negative function to
triple sets of outcomes implies that the inequalities
Eq.~(\ref{BOOLE1}) hold. Conversely, if the inequalities
Eq.~(\ref{BOOLE1}) are violated the real-valued function
$f^{(3)}(S_1,S_2,S_3)$ of the three two-valued variables $S_1$,
$S_2$ and $S_3$ cannot be non negative. No non-trivial restrictions
can be derived for $E^{(2)}$, that is for pair sets of outcomes,
unless the non negative functions of two variables can be obtained
from one non negative function of three variables.

To fully understand all the implications of this result and the true
content of Bell's derivations we need to return to the nature of
correlations between data. In case of assigning a positive value to
triples of data we put a ``correlation-measure'' (the positive value
of the function) to the correlation of positive and negative values
for three variables while if we consider pairs the measure is
imposed on two variables only.

In terms of Boole's elements of logic this means that the elements
of logic corresponding to e.g. the realizations of the value of the
variable $S_1$ for two different pairs may be altogether different.
One pair could be measured at different times, for different earth
magnetic fields than the other. We refer the reader to the more
detailed explanations in Section~\ref{apparent}. If the
realizations of $S_1, S_2, S_3$ correspond to the same logical
elements no matter which of the three cyclically arranged pairs is
chosen, then the inequalities Eq.~(\ref{BOOLE1}) are valid
irrespective if we deal with pairs or triples.

In Kolmogorov's framework one needs to define a measure
on an algebra and we deal with single indivisible elements
$\omega^{(3)}$ of a sample space that actualize (bring their
outcomes into existence) a given triple. If, on the other hand we
deal with a pair then we need sample space elements $\omega^{(2)}$
to actualize a given pair. This means that we deal, in principle,
with different sample spaces $\Omega$ and with different Kolmogorov
probability spaces when considering models for triples or pairs.

Note that our approach above is more explicit in expressing
the relationship of the mathematics to the experiments by
designating different functions to different experimental groupings
and in this way dealing more explicitly with the correlations. The
second trademark of our approach above is that \OTOCLED\ is not
explicitly addressed and may be different for each different
grouping of data be it into pairs or triples. In this respect our
approach is similar to that of quantum theory that does
not deal with the single outcomes and \OTOCLED. We show below that
we can therefore compare our approach and quantum theory to address
questions of the validity of Boole-type inequalities for experiments
generating pairs and triples of data.

Last but not least we note that John Bell~\cite{BELL64} based his
famous theorem on two assumptions: (a) Bell assumed in his original
paper by the algebraic operations of his Eqs.~(14) -- (22) and the
additional assumption that his $\lambda$ represents elements of
reality a clear grouping into triples because he implies the
existence of identical elements of reality for each of the three
pairs. (b) By the same operations Bell assumed that he deals with
dichotomic variables that follow the algebra of integers. From our
work above it is then an immediate corollary that Bell's
inequalities cannot be violated; not even by influences at a
distance.

\section{Extended Boole-Bell inequalities for quantum phenomena}\label{QuantumTheory}

We now apply the method of Section~\ref{functions} to quantum
theory. The main result of this section is that a quantum
theoretical model can never violate the extended Boole-Bell
inequalities because these EBBI can be derived within the framework
of quantum theory itself. This result follows directly from the
mathematical structure of quantum theory, just as the results of
Sections~\ref{Boole} and \ref{functions} follow from the rules of
elementary algebra. The basic concepts sufficient to derive the EBBI
for quantum theory are~\cite{BALL03}

\medskip\noindent
{\bf {Postulate I: }} To each state of the quantum system
there corresponds a unique state operator $\rho$
which must be Hermitian, non negative and of unit trace.

\medskip\noindent
{\bf {Postulate II: }} To each dynamical variable there corresponds
a Hermitian operator whose eigenvalues are the possible values
of the dynamical variable.

\medskip\noindent
{\bf {Postulate III: }} The average value of a dynamical variable,
represented by the operator $X$, in the state represented by $\rho$, is
$\langle X \rangle=\mathbf{Tr} \rho X$.

\medskip\noindent
We focus on systems that are being characterized by variables that
assume two values only. According to Postulate II, this implies that
the dynamical variables in the corresponding quantum system can be
represented by $2\times2$ Hermitian matrices. It is tradition to
describe such systems by means of the Pauli-spin matrices.
Each Pauli spin matrix represents a dynamical variable describing
the projection of the magnetic moment of a spin-1/2 particle
to one of the three spatial directions.
The Hilbert space ${\cal H}$ of a system of $n$ of these spin-1/2 objects
is the direct product of the $n$ two-dimensional Hilbert spaces
${\cal H}_i$, that is ${\cal H}={\cal H}_1\otimes\ldots\otimes {\cal H}_n$.
In this and the following sections, we denote the Pauli-spin
matrices describing the spin components of the $i$th spin-1/2 particle by
$\sigma_i=(\sigma^x_i,\sigma^y_i,\sigma^z_i)$.
The symbol $\sigma_i$ is to be interpreted as
(1) a two-by-two matrix when it acts on the Hilbert space
${\cal H}_i$ and (2) as a shorthand for
$\openone\otimes\ldots\otimes\openone\otimes
\sigma_i
\otimes\openone\otimes\ldots\otimes\openone$
when it acts on the full Hilbert space ${\cal H}$.
The eigenvalues of $\sigma^z_i$ are
$+1$ and $-1$ and the corresponding eigenvectors are the spin-up
state $|\uparrow\rangle_i$ and the spin-down state
$|\downarrow\rangle_i$, respectively. It is convenient to label the
eigenvalues by a two-valued variable $S=\pm 1$ such that
$|+1\rangle_i=|\uparrow\rangle_i$ and
$|-1\rangle_i=|\downarrow\rangle_i$. Thus, we have
$\sigma^z_i|S\rangle_i=S|S\rangle_i$ and $\sigma^z_i|S_1\ldots
S_n\rangle=S_i|S_1 \ldots S_n\rangle$. The state of a system of $n$
of these spin-1/2 particles is represented by a $2^n\times 2^n$
non negative definite, normalized matrix $\rho^{(n)}$. In the
following we will call $\rho^{(n)}$
the density matrix~\cite{BALL03}.

In the subsections that follow, we consider two different types of experiments
that produce $n$-tuples of two-valued variables.
First, we discuss experiments in which these
measurements are performed on $n$ different spin-1/2 particles (Section~\ref{nspins}).
In this case, quantum theory gives a description of the
$n$ dynamical variables representing the spins of
the $n$ spin-1/2 particles in terms of Pauli matrices that always
commute and guarantees the existence of a non-negative
function $P^{(n)}(S_1,\ldots,S_n)$ of the $n$ two-valued variables $S_1,\ldots,S_n$.

Second, in Section~\ref{filter} we consider $n$ successive measurements
of the filtering type on the spin of one spin-1/2 particle.
The quantum theoretical description of this experiment involves
Pauli spin matrices that may not commute but nevertheless,
quantum theory guarantees the existence of a non-negative
function $P^{(n)}(S_1,\ldots,S_n)$ of the $n$ two-valued variables $S_1,\ldots,S_n$.

From Section~\ref{functions}, we already know that the proof of the
EBBI only requires the existence of a non-negative function
$P^{(n)}(S_1,\ldots,S_n)$ for $n>2$. Therefore, for the type of
experiments such as the ones described in Sections~\ref{nspins} and
~\ref{filter}, quantum theory guarantees that the EBBI can be
derived and cannot be violated even if the quantum theoretical
description involves non-commuting operators: The non-commutativity
of these operators does not enter the derivation of the EBBI
and is therefore superfluous. This also holds for the EPRB
experiment described in Section~\ref{commut}.

EPRB experiments involve measurements that are performed on $n=2$
spin-1/2 particles and the pairs of two-valued variables are
determined by means of Stern-Gerlach magnets that perform
filtering-type experiments on the spins of the two spin-1/2
particles. A generalized EPRB set-up involves $m>2$ such experiments
with different settings (orientations of the Stern-Gerlach magnets)
that are being performed in parallel, yielding $m$ pairs of
two-valued data. The (products of) spin matrices that describe the
result of the $m$ different experiments do not necessarily commute.
However, as explained in more detail in Sections~\ref{example} -- \ref{commut},
(non-)commutation is not a necessary condition for the apparent violation of the EBBI.

\subsection{Spin measurements on $n$ different spin-1/2 particles}\label{nspins}

In the case of experiments that involve measurements of the spins
of $n$ different spin-1/2 particles along particular directions,
the corresponding Pauli matrices trivially commute, that is
$[\sigma_i^x,\sigma_j^y]=[\sigma_i^x,\sigma_j^z]=[\sigma_i^y,\sigma_j^z]=0$ for all $i\not=j$.

We assume that the $n$-particle system is in an arbitrary quantum state
described by the density matrix
\begin{equation}
\rho^{(n)}=
\sum_{\{S^{\;\prime}\},\{S^{\;\prime\prime}\}} a(S^{\;\prime}_1 \ldots S^{\;\prime}_n; S^{\;\prime\prime}_1 \ldots S^{\;\prime\prime}_n)|S^{\;\prime}_1 \ldots S^{\;\prime}_n\rangle \langle S^{\;\prime\prime}_1 \ldots S^{\;\prime\prime}_n|
,
\label{rho3z}
\end{equation}
where, in general, the $2^n\times2^n$ coefficients $a(S^{\;\prime}_1
\ldots S^{\;\prime}_n; S^{\;\prime\prime}_1 \ldots
S^{\;\prime\prime}_n)$ are complex numbers, with values restricted
by the conditions $\rho^{(n)}=(\rho^{(n)})^\dagger$ and $\mathbf{Tr}
\rho^{(n)}=1$. The sum in Eq.~(\ref{rho3z}) runs over all
$2^n\times2^n$ possible values $S^{\;\prime}_1=\pm1, \ldots,
S^{\;\prime}_n=\pm1, S^{\;\prime\prime}_1=\pm1, \ldots,
S^{\;\prime\prime}_n=\pm1$.
We ask for the average value, as postulated by quantum
theory, for observing a given $n$-tuple of eigenvalues
$(S_1,\ldots,S_n)$ of the $2^n\times 2^n$ matrix
$\sigma^z_1\ldots\sigma^z_n$. The $2^n\times 2^n$ Hermitian matrix
$M$ that corresponds to this collection of $n$ dynamical variables
is represented by $M=|S_1,\ldots,S_n\rangle\langle
S_1,\ldots,S_n|$~\cite{BALL03}. Note that $M=M^2$ is a diagonal
matrix that has one nonzero element (a one) only. According to
Postulate III, the average $\langle M\rangle$ is given by
\begin{eqnarray}
\lefteqn{P^{(n)}(S_1,\ldots,S_n)\equiv \mathbf{Tr} \rho^{(n)} M}
\nonumber \\
\nonumber \\
&&=
\sum_{\{S^{\;\prime}\},\{S^{\;\prime\prime}\}} a(S^{\;\prime}_1 \ldots S^{\;\prime}_n; S^{\;\prime\prime}_1 \ldots S^{\;\prime\prime}_n)
\langle S_1\ldots S_n|S^{\;\prime}_1\ldots S^{\;\prime}_n\rangle
\nonumber \\
&&\hbox to 2cm{}\times\langle S^{\;\prime\prime}_1 \ldots S^{\;\prime\prime}_n|S_1\ldots S_n\rangle
\nonumber \\
\nonumber \\
&&=
\sum_{\{S\}} a(S_1 \ldots S_n; S_1 \ldots S_n)
\nonumber \\&&
=
\langle S_1\ldots S_n|\rho^{(n)}|S_1\ldots S_n\rangle
,
\label{rho3}
\end{eqnarray}
where our notation suggests that $P^{(n)}(S_1,\ldots,S_n)$ may be interpreted as
a probability in Kolmogorov's sense. As we now show, this is indeed the case.

First because of Postulate I, $P^{(n)}(S_1,\ldots,S_n)$ is
the diagonal element of a non negative definite matrix with maximum
eigenvalue less or equal than one. Therefore, we have
$0\le P^{(n)}(S_1,\ldots,S_n)\le 1$. Second, by construction, the $2^n$
matrices $|S_1,\ldots,S_n\rangle\langle S_1,\ldots,S_n|$ for
$S_1=\pm1,\ldots,S_n=\pm1$ are an orthonormal and a complete
resolution of the identity matrix ($\sum_{\{S_i=\pm1\}}
|S_1,\ldots,S_n\rangle\langle S_1,\ldots,S_n|=\openone$), hence
$\sum_{\{S_i=\pm1\}}P^{(n)}(S_1,\ldots,S_n)=\mathbf{Tr} \rho^{(n)}
=1$. To complete the proof, we need to consider more general
observations. Let us write $M^{\;\prime}$ for the matrix that corresponds to
the observation of the $n$-tuple of eigenvalues
$(S^{\;\prime}_1,\ldots,S^{\;\prime}_n)\not=(S_1,\ldots,S_n)$. Obviously, $MM'=M'M=0$
and from Postulate III, $\langle MM'\rangle =P^{(n)}(
(S_1,\ldots,S_n)$ $ \wedge (S^{\;\prime}_1,\ldots,S^{\;\prime}_n) ) =0$, where $\wedge $
denotes the logical ``and' operation. Likewise the average value, as
postulated by quantum theory, of observing the $n$-tuple of
eigenvalues $(S_1,\ldots,S_n)$ or (inclusive) $(S^{\;\prime}_1,\ldots,S^{\;\prime}_n)$
is given by $\langle M + M^{\;\prime}\rangle =P^{(n)}( (S_1,\ldots,S_n)$ $\vee
(S^{\;\prime}_1,\ldots,S^{\;\prime}_n))
=P^{(n)}(S_1,\ldots,S_n)+P^{(n)}(S^{\;\prime}_1,\ldots,S^{\;\prime}_n)$ where $\vee$
denotes the logical inclusive ``or'' operation. These results
trivially extend to observations that correspond to more than two
projectors, completing the proof that the sample space formed by the
$2^n$ elementary events $(S_1,\ldots,S_n)$ and the function Eq.~(\ref{rho3})
may therefore be regarded as a
joint probability in the Kolmogorov sense.
Alternatively, one could use the consistent history approach
to define the probabilities for the elementary events $(S_1,\ldots,S_n)$~\cite{OMNE99,GRIF02}.
Note that Eq.~(\ref{rho3}) does not entail a complete description
of the state of the quantum system with $n$ different spin-1/2 particles
because Eq.~(\ref{rho3}) relates to the diagonal elements of $\rho^{(n)}$ only.

Within quantum theory, Eq.~(\ref{rho3z}) gives the complete description of
the state of a system with $n$ different spin-1/2 particles.
From this state, we can extract all the complete descriptions
of systems with $k<n$ different spin-1/2 particles by performing partial traces
and find relations between $P^{(n)}(S_1,\ldots,S_n)$
and $P^{(k)}(S_1,\ldots,S_{k})$ for $k<n$.
In this case, all the $k$-tuples $(S_1,\ldots,S_k)$, $k=1,\ldots,n-1$
trivially form one common Kolmogorov sample space (see
the concrete examples of Sections~\ref{Leggett} and \ref{EPR})):
All $k$-tuples ($k<n$) are drawn from one master set of all $n$-tuples,
all for the same experiment with precisely the same preparation and
measurement procedure.
Evidently, it would then be a serious mistake to regard
this $P^{(k)}(S_1,\ldots,S_{k})$ for $k<n$ as the
probability to observe the $k$-tuples $(S_1,\ldots,S_{k})$
in a different system of $k$ spin-1/2 particles.
To make this mathematical precise, it is necessary
to add a label $n$ to the variables $S_i$ such that there
cannot be doubt as to from which experiment they have been obtained.
Then, in general we have
\begin{eqnarray}
P^{(k)}(S_1^{(k)},\ldots,S_{k}^{(k)})\not=P^{(k)}(S_1^{(n)},\ldots,S_{k}^{(n)})
\quad\mathrm{for}\quad k<n
.
\label{xxxx}
\end{eqnarray}
In particular, given $P^{(2)}(S_1^{(2)},S_2^{(2)})$, $P^{(2)}(S_1^{(2)},S_3^{(2)})$
and $P^{(2)}(S_2^{(2)},S_3)^{(2)}$ one may or may not be able
to construct a common Kolmogorov sample space
and find the $P^{(3)}(S_1^{(3)},S_2^{(3)},S_3^{(3)})$ from which
the two-particle probabilities are the marginals.
As we have already seen in Section~\ref{functions}, the necessary and sufficient
condition for this common Kolmogorov sample space to exist is that the EBBI are satisfied.
Clearly, this condition is independent of whether or not
the operators in the quantum theoretical model commute, see
Sections~\ref{example} -- \ref{commut} for more details.

Summarizing: For experiments that measure the spins of $n$ different spin-1/2
particles along particular directions, quantum theory gives a description
of the $n$ dynamical variables representing the spins of these
particles in terms of Pauli matrices that always commute and
guarantees the existence of a non-negative function
$P^{(n)}(S_1,\ldots,S_n)$ of the $n$ two-valued variables that correspond
to the eigenvalues of these matrices.
The formulation of quantum mechanics dictates the difference of the logical elements
in the different joint probabilities for different experiments.
Quantum mechanics gets around the awkward notation introduced above by
forbidding us to consider the single outcomes any further.
However, when we write down joint probabilities we need to consider very carefully
the different logical elements that determine the joint probabilities and we
need to present them mathematically as different objects.

\subsection{Filtering-type measurements on the spin of one spin-1/2 particle}\label{filter}

We consider an experiment in which we perform successive measurements of
the filtering-type on one spin-1/2 particle only and show that also
for this case, quantum theory guarantees the existence of $P^{(n)}(S_1,\ldots,S_n)$
as a probability on the sample space of elementary events $(S_1,\ldots,S_n)$.

\begin{figure*}[t]
\begin{center}
\includegraphics[width=14cm]{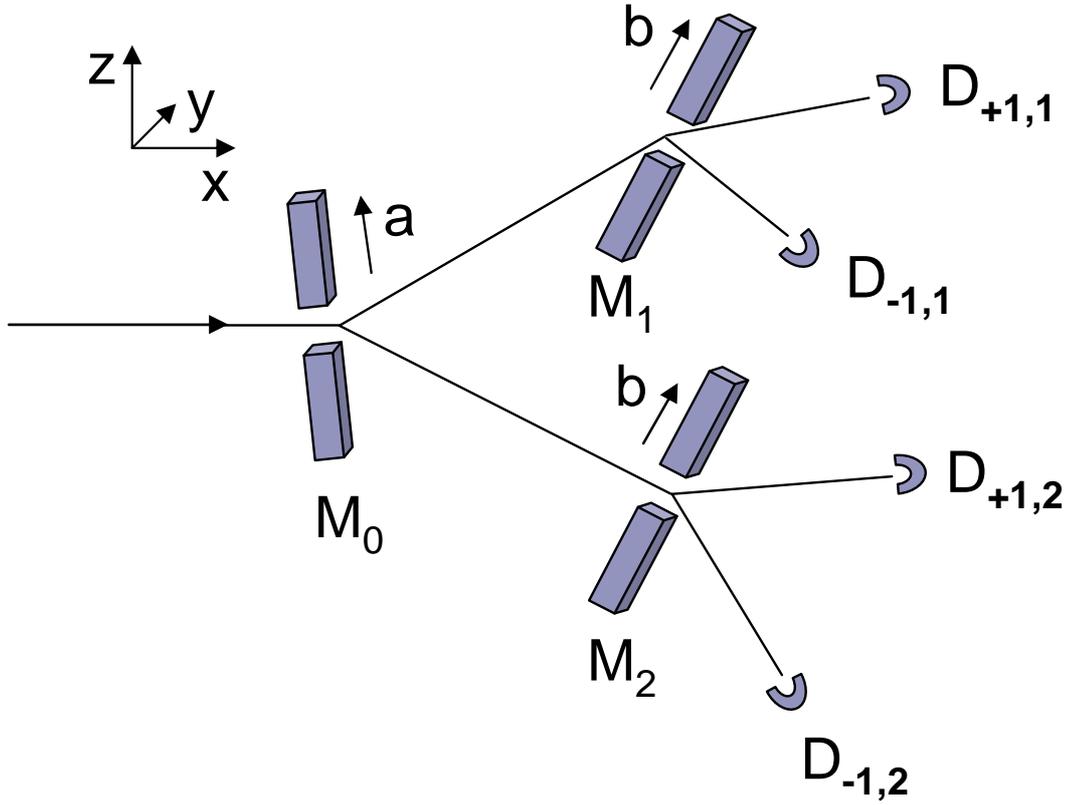}
\caption{Conceptual layout of a filtering type experiment.
Spin-1/2 particles pass through a Stern-Gerlach magnet $M_0$
that projects the spin onto either the
${\mathbf a}$ direction or the $-{\mathbf a}$ direction.
In case of the former (latter) projection, the particle
is directed to the Stern-Gerlach magnet $M_1$ ($M_2$).
$M_1$ and $M_2$ are assumed to be identical and
project the spin onto either the
${\mathbf b}$ direction or the $-{\mathbf b}$ direction.
A ``click'' of one of the four detectors
$D_{+1,1}$, $D_{-1,1}$, $D_{+1,2}$, and $D_{-1,2}$
signals the arrival of a particle.
}
\label{fig0}
\end{center}
\end{figure*}

In Fig.~\ref{fig0}, we show a schematic diagram of such an
experiment with two filtering stages, the generalization to
an arbitrary number of stages being trivial.
As we show below, the number $n$ of two-valued
variables that describe the result of the measurement of the
spin at each stage is equal to the number of filtering stages.
In other words, for each spin-1/2 particle passing through
a filtering apparatus with $n$ stages,
the experiment yields an $n$-tuple of two-valued variables.
In order to obtain the averages that quantum theory predicts,
we obviously have to repeat the single-spin experiment
using identical preparation.

Spin-1/2 particles enter the Stern-Gerlach magnet $M_0$, with its
magnetic field along direction ${\mathbf a}$. $M_0$ ``sends" each of
them either to Stern-Gerlach magnet $M_1$ or $M_2$. The magnets
$M_1$ and $M_2$, identical and both with their magnetic field along
direction ${\mathbf b}$, subdivide the particle stream once more and
finally, each of the particles is registered by one of the four
detectors $D_{+1,1}$, $D_{-1,1}$, $D_{+1,2}$, and $D_{-1,2}$.

We label the particles by a subscript $\alpha$. After the $\alpha$th
particle leaves $M_1$ or $M_2$, it will trigger one of the four
detectors (we assume ideal experiments, that is at any time one and
only one out of four detectors fires). We write $x_\alpha^{(i,j)}=1$
if the $\alpha$th particle was detected by detector $D_{i,j}$ and
$x_\alpha^{(i,j)}=0$ otherwise. Next, we define two new dichotomic
variables by
\begin{eqnarray}
S_{1,\alpha}&=&
\left( x_\alpha^{(+1,1)}+x_\alpha^{(-1,1)}\right)
-
\left( x_\alpha^{(+1,2)}+x_\alpha^{(-1,2)}\right)
,
\nonumber \\
S_{2,\alpha}&=&
\left( x_\alpha^{(+1,1)}+x_\alpha^{(+1,2)}\right)
-
\left( x_\alpha^{(-1,1)}+x_\alpha^{(-1,2)}\right)
.
\label{fil0}
\end{eqnarray}
If $S_{1,\alpha}=\pm1$, the spin has been projected on the
$\pm{\mathbf a}$ direction. Likewise, if $S_{2,\alpha}=\pm1$, the
spin has been projected on the $\pm{\mathbf b}$ direction.

We now describe this experiment by quantum theory. It is a
straightforward exercise (see pages 172 and 250 in
Ref.~\onlinecite{BALL03}) to show that the projection operators
$M(S_1,{\mathbf a}$) are given by
\begin{eqnarray}
M(S_1,{\mathbf a})&=&
\frac{\openone+S_1
\sigma\cdot\mathbf{a}}{2}
,
\label{fil1}
\end{eqnarray}
where we have omitted the spin subscript to make absolutely clear
that in this subsection, we consider measurements on one and the
same particle only. Of course,
the projection operators for the second stage follow the expression
of Eq.~(\ref{fil1}) with the unit vector $\mathbf{a}$ replaced by
$\mathbf{b}$ and $S_1$ replaced by $S_2$.

Assume now that the system is prepared in the state with the density
matrix
\begin{eqnarray}
\rho^{(1)}&=&
\frac{\openone+\sigma\cdot\mathbf{x}}{2}
,
\label{fil2}
\end{eqnarray}
where the vector $\mathbf{x}$ ($\Vert\mathbf{x}\Vert\le1$) fully
determines the state but is not specified further. Then, according
to quantum theory, the probability that we observe a given pair
$(S_1,S_2)$ is given by~\cite{BALL03}
\begin{eqnarray}
P^{(2)}(S_1,S_2)&=&
\mathbf{Tr} \rho^{(1)} M(S_1,{\mathbf a})M(S_2,{\mathbf b})M(S_1,{\mathbf a})
\nonumber \\
&=&
\frac{1+S_1 \mathbf{x}\cdot\mathbf{a}+S_2 \mathbf{x}\cdot\mathbf{a}\;\mathbf{a}\cdot\mathbf{b}
+S_1 S_2\mathbf{a}\cdot\mathbf{b}}{4}
.
\label{fil3}
\end{eqnarray}
Note that $[M(S_1,{\mathbf a}),M(S_2,{\mathbf b})]\not=0$ unless
$\mathbf{a}=\pm\mathbf{b}$, $[\rho,M(S_1,{\mathbf a})]\not=0$ unless
$\mathbf{x}=\pm\mathbf{a}$, and that $[\rho,M(S_2,{\mathbf
b})]\not=0$ unless $\mathbf{x}=\pm\mathbf{b}$. Thus, for virtually
all cases of interest, none of the operators in Eq.~(\ref{fil3})
commute, yet quantum theory yields the probability
$P^{(2)}(S_1,S_2)$ in all these cases. Note that except for an
inconsequential sign and independent
of the state of the system $\rho^{(1)}$, the two-spin correlation of
this filtering-type experiment on one particle
is the same as the two-spin correlation of
the EPRB filtering-type experiment considered by Bell~\cite{BELL93}.

The filtering-type experiment shown in Fig.~\ref{fig0} can be
extended to include $n$ successive measurements on the same particle.
As an example, we add one more stage. Imagine that we then replace
the four detectors in Fig.~\ref{fig0} by four identical
Stern-Gerlach magnets with their fields along the ${\mathbf c}$
direction followed by an array of eight detectors $D_{+1,1}$,
$D_{-1,1}$, $D_{+1,2}$, $D_{-1,2}$, $D_{+1,3}$, $D_{-1,3}$,
$D_{+1,4}$, and $D_{-1,4}$ (numbered from top to bottom, diagram not
shown). The path that a particle has followed is then uniquely
determined by the three dichotomic variables
\begin{eqnarray}
S_{1,\alpha}&=&
x_\alpha^{(+1,1)}+x_\alpha^{(-1,1)}
+
x_\alpha^{(+1,2)}+x_\alpha^{(-1,2)}
\nonumber \\
&&-
x_\alpha^{(+1,3)}-x_\alpha^{(-1,3)}
-
x_\alpha^{(+1,4)}-x_\alpha^{(-1,4)}
,
\nonumber \\
S_{2,\alpha}&=&
x_\alpha^{(+1,1)}+x_\alpha^{(-1,1)}
+
x_\alpha^{(+1,3)}+x_\alpha^{(-1,3)}
\nonumber \\
&&-
x_\alpha^{(+1,2)}-x_\alpha^{(-1,2)}
-
x_\alpha^{(+1,4)}-x_\alpha^{(-1,4)}
,
\nonumber \\
S_{3,\alpha}&=&
x_\alpha^{(+1,1)}+x_\alpha^{(+1,2)}
+
x_\alpha^{(+1,3)}+x_\alpha^{(+1,4)}
\nonumber \\
&&-
x_\alpha^{(-1,1)}-x_\alpha^{(-1,2)}
-
x_\alpha^{(-1,3)}-x_\alpha^{(-1,4)}
.
\nonumber \\
\label{fil4}
\end{eqnarray}
Then, according to quantum theory, the probability that we observe
the given triple $(S_1,S_2,S_3)$ is~\cite{BALL03}
\begin{widetext}
\begin{eqnarray}
P^{(3)}(S_1,S_2,S_3)&=&
\mathbf{Tr} \rho^{(1)} M(S_1,{\mathbf a})M(S_2,{\mathbf b})M(S_3,{\mathbf c})M(S_2,{\mathbf b})M(S_1,{\mathbf a})
\nonumber \\
&=&
\frac{1+S_1 \mathbf{x}\cdot\mathbf{a}+S_2 \mathbf{x}\cdot\mathbf{a}\;\mathbf{a}\cdot\mathbf{b}
+S_3 \mathbf{x}\cdot\mathbf{a}\;\mathbf{a}\cdot\mathbf{b}\;\mathbf{b}\cdot\mathbf{c}
+
S_1 S_2\mathbf{a}\cdot\mathbf{b}
+S_1 S_3\mathbf{a}\cdot\mathbf{b}\;\mathbf{b}\cdot\mathbf{c}
+S_2 S_3\mathbf{b}\cdot\mathbf{c}
+
S_1S_2 S_3 \mathbf{x}\cdot\mathbf{a}\;\mathbf{b}\cdot\mathbf{c}
}{8}
,
\label{fil5}
\end{eqnarray}
\end{widetext}
demonstrating that also for three actual  measurements on the same
particle, quantum
theory yields a well defined probability distribution.

Summarizing: For filtering-type experiments such as the one depicted in Fig.~\ref{fig0}
and the ones analyzed in Sections~\ref{Leggett} and \ref{EPR},
quantum theory guarantees the existence of probabilities $P^{(n)}(S_1,\ldots,S_n)$
even though the quantum theoretical description of the $n$ measurements
involves operators that may not commute.

\subsection{EBBI for quantum phenomena}\label{EBBI}

In the two previous subsections, we have shown that
for the type of experiments that we consider in this paper,
quantum theory guarantees the existence of
non-negative functions $P^{(n)}(S_1,\ldots,S_n)$
of $n$ dichotomic variables, for any value of $n$.
Without loss of generality, we may write
\begin{eqnarray}
P^{(1)}(S_1)&=&\frac{1+S_1E^{(1)}}{2}
\label{qt4}
\label{qt7}
\\
P^{(2)}(S_1,S_2)&=&
\frac{1+S_1E^{(2)}_1+S_2E^{(2)}_2+S_1S_2E^{(2)}}{4}\
,
\label{qt5}
\label{qt8}
\\
P^{(3)}(S_1,S_2,S_3)&=&
\frac{1+S_1E^{(3)}_1+S_2E^{(3)}_2+S_3E^{(3)}_3}{8}
\nonumber \\
&&+\frac{S_1S_2E^{(3)}_{12}+S_1S_3E^{(3)}_{13}
+S_2S_3E^{(3)}_{23}}{8}
\nonumber \\
&&+\frac{S_1S_2S_3E^{(3)}}{8}
.
\label{qt6}
\label{qt9}
\end{eqnarray}
We are now in the position to apply the results of Section~\ref{functions}
and state:
A quantum mechanical system that describes an experiment which measures
\begin{enumerate}
\item{
singles of a two-valued variable cannot violate the inequality
\begin{eqnarray}
|E^{(1)}|&\le&1
.
\label{qt10}
\end{eqnarray}
}
\item{
pairs of two-valued variables cannot violate the inequalities
\begin{eqnarray}
|E^{(2)}_i|\le 1\,,\,
|E^{(2)}|\le 1\,,\,
|E^{(2)}_1\pm E^{(2)}_2|\le 1 \pm E^{(2)}
.
\label{qt11}
\end{eqnarray}
}
\item{
triples of two-valued variables cannot violate Boole's inequalities
\begin{eqnarray}
|E^{(3)}_{ij}\pm E^{(3)}_{ik}|&\le&1\pm E^{(3)}_{jk}\quad,\quad
,
\label{BOOLE2}
\end{eqnarray}
}
\end{enumerate}
for $i=1,2$ and $(i,j,k)=(1,2,3),(3,1,2),(2,3,1)$.
It is important to note that inequalities Eq.~(\ref{BOOLE2}) follow
directly from the fact that the expression
Eq.~(\ref{rho3}) is non negative: No additional assumptions need be
invoked in order to prove the inequalities Eq.~(\ref{BOOLE2}). We
emphasize that Eq.~(\ref{BOOLE2}) can never be violated by a quantum
system that describes a triple of two-valued dynamical variables.
Notice that the derivation of the above results does not depend in
any way on a particular ``interpretation'' of quantum theory: We
have made use of the commonly accepted mathematical framework of
quantum theory only. The derivation of inequalities
Eqs.~(\ref{qt10}) -- (\ref{BOOLE2}) does not make reference to
metaphysical concepts: It is the mathematical structure of quantum
theory that imposes inequalities Eqs.~(\ref{qt10}) --
(\ref{BOOLE2}).

For the examples of quantum systems treated in Sections~\ref{Leggett} and \ref{EPR}
there is no need to deploy the full machinery of the density matrix
formalism as the states of these systems are described by pure states.
We briefly recapitulate how the description in terms of pure states fits
into the general density-matrix formalism.

The quantum system is said to be in a pure state if and only if $\rho=\rho^2$, see Ref.~\onlinecite{BALL03}
For a pure state the density matrix takes the form
\begin{eqnarray}
\rho&=&|\Psi\rangle\langle \Psi|
,
\label{rho0}
\end{eqnarray}
in which case $|\Psi\rangle$ is called the state vector or wave
function. Therefore, the expressions Eqs.~(\ref{qt4}) -- (\ref{qt6})
do not change and the inequalities Eqs.~(\ref{qt10}) --
(\ref{BOOLE2}) have to be satisfied.

For a system of $n$ spin-1/2 objects in a pure state, the state
vector $|\Psi\rangle$ can be expanded into the complete, orthonormal
set of many-body basis states $\{|S_1\ldots S_n\rangle\,|$ $ S_1=\pm1,\ldots,S_n=\pm1\}$. We have
\begin{eqnarray}
|\Psi\rangle
&=&
\sum_{\{S\}} c(S_1,\ldots,S_n)|S_1\ldots S_n\rangle
,
\label{rho1}
\end{eqnarray}
where $c(S_1,\ldots,S_n)$ are, in general, complex coefficients and the sum is over
the $2^n$ possible values of the $n$-tuple of eigenvalues $(S_1,\ldots,S_n)$.
For instance, the state vector of two spin-1/2 objects in the singlet state is
\begin{equation}
|Singlet\rangle=
\frac{
|+1,-1\rangle
-
|-1,+1\rangle
}{\sqrt{2}}
=\frac{
|\uparrow\downarrow\rangle
-
|\downarrow\uparrow\rangle
}{\sqrt{2}}
,
\label{singlet}
\end{equation}
such that $c(+1,-1)=-c(-1,+1)=2^{-1/2}$ and $c(+1,+1)=-c(-1,-1)=0$.

\subsection{Example}\label{example}

It may seem that the derivation of the inequalities
Eqs.~(\ref{qt10}) -- (\ref{BOOLE2}) depends on our choice that the
up and down states of the spins are eigenvectors of the
$z$-components of the spin operators. This is not the case. Let us
assume that the observation of, say,  spin one is not along the
$z$-direction but along some direction specified by a unit vector
$\mathbf{a}$. The corresponding matrix would then be
$\sigma_1\cdot\mathbf{a}$, not $\sigma_1^z$. This change has no
effect on the proof that leads to Eq.~(\ref{BOOLE2}) except, and
this is very important, we should keep track of the fact that the
measurement on spin one is performed along the direction
$\mathbf{a}$. Usually, this should be clear from the context but if
not, it is necessary to include the directions of measurement in the
notation of the probabilities by writing $P^{(1)}(S_1|\mathbf{a})$
instead of $P^{(1)}(S_1)$ etc.

As an illustration, let us consider a system of two spin-1/2 objects.
For such a system there are only three essentially different
averages of dynamical variables namely
$\langle\sigma_1\cdot{\mathbf a}\rangle$,
$\langle \sigma_2\cdot{\mathbf b}\rangle$, and
$\langle\sigma_1\cdot{\mathbf a}\,\sigma_2\cdot{\mathbf b}\rangle$
where ${\mathbf a}$ and
${\mathbf b}$ are unit vectors. Knowing these averages for ${\mathbf
a},{\mathbf b}=(1,0,0),(0,1,0),(0,0,1)$ suffices to completely
determine the state of the quantum system, that is $\rho^{(2)}$. In
the simplest version of the EPRB experiments, the two spins are
measured in three different directions ${\mathbf a}$, ${\mathbf b}$,
and ${\mathbf c}$. Accordingly, we obtain the probabilities
\begin{widetext}
\begin{eqnarray}
P^{(2)}(S_1,S_2|{\mathbf a}{\mathbf b})&=&
\frac{1+S_1 \langle \sigma_1\cdot{\mathbf a}\rangle+S_2\langle \sigma_2\cdot{\mathbf b}\rangle
+S_1S_2 \langle\sigma_1\cdot{\mathbf a}\,\sigma_2\cdot{\mathbf b}\rangle}{4}
,\nonumber \\
{\widehat P}^{(2)}(S_1,S_3|{\mathbf a}{\mathbf c})&=&
\frac{1+S_1 \langle \sigma_1\cdot{\mathbf a}\rangle+S_3\langle \sigma_2\cdot{\mathbf c}\rangle
+S_1S_3 \langle\sigma_1\cdot{\mathbf a}\,\sigma_2\cdot{\mathbf c}\rangle}{4}
,\nonumber \\
{\widetilde P}^{(2)}(S_2,S_3|{\mathbf b}{\mathbf c})&=&
\frac{1+S_2 \langle \sigma_1\cdot{\mathbf b}\rangle+S_3\langle \sigma_2\cdot{\mathbf c}\rangle
+S_2S_3 \langle\sigma_1\cdot{\mathbf b}\,\sigma_2\cdot{\mathbf c}\rangle}{4}
.
\label{exa0}
\end{eqnarray}
Let us assume that $\langle \sigma_1\cdot{\mathbf b}\rangle=\langle
\sigma_2\cdot{\mathbf b}\rangle$, which is the case for the quantum
theoretical description of the EPRB experiment. Then, from Theorem IV
we conclude that all the inequalities
\begin{eqnarray}
|\langle\sigma_1\cdot{\mathbf a}\,\sigma_2\cdot{\mathbf b}\rangle
\pm
\langle\sigma_1\cdot{\mathbf a}\,\sigma_2\cdot{\mathbf c}\rangle|&\le&
1\pm \langle\sigma_1\cdot{\mathbf b}\,\sigma_2\cdot{\mathbf c}\rangle
,\nonumber \\
|\langle\sigma_1\cdot{\mathbf a}\,\sigma_2\cdot{\mathbf b}\rangle
\pm
\langle\sigma_1\cdot{\mathbf b}\,\sigma_2\cdot{\mathbf c}\rangle|&\le&
1\pm \langle\sigma_1\cdot{\mathbf a}\,\sigma_2\cdot{\mathbf c}\rangle
,\nonumber \\
|\langle\sigma_1\cdot{\mathbf a}\,\sigma_2\cdot{\mathbf c}\rangle
\pm
\langle\sigma_1\cdot{\mathbf b}\,\sigma_2\cdot{\mathbf c}\rangle|&\le&
1\pm \langle\sigma_1\cdot{\mathbf a}\,\sigma_2\cdot{\mathbf b}\rangle
,
\label{exa1}
\end{eqnarray}
are satisfied if and only if there exists a
probability $P^{(3)}(S_1,S_2,S_3|{\mathbf a}{\mathbf b}{\mathbf c})$
that returns the probabilities Eq.~(\ref{exa0}) as marginals.

Anticipating the general discussion of Section~\ref{commut}, we show
now that non-commutation of the matrices $\sigma_1\cdot{\mathbf
a}\,\sigma_2\cdot{\mathbf b}$, $\sigma_1\cdot{\mathbf
a}\,\sigma_2\cdot{\mathbf c}$, and $\sigma_1\cdot{\mathbf
b}\,\sigma_2\cdot{\mathbf c}$ does not prohibit the existence of
$P^{(3)}(S_1,S_2,S_3|{\mathbf a}{\mathbf b}{\mathbf c})$ as a joint
probability. Assume therefore that $\sigma_1\cdot {\mathbf
a}\,\sigma_2\cdot {\mathbf b}$, $\sigma_1\cdot {\mathbf
a}\,\sigma_2\cdot {\mathbf c}$, and $\sigma_1\cdot {\mathbf
b}\,\sigma_2\cdot {\mathbf c}$ do not mutually commute and that the
inequalities Eq.~(\ref{exa1}) hold. Next, assume that $\langle
\sigma_1\cdot{\mathbf b}\rangle=\langle \sigma_2\cdot{\mathbf
b}\rangle$, which is indeed the case for the quantum theoretical
description of the EPRB experiment. Then, if $0\le
P^{(2)}(S_1,S_2|{\mathbf a}{\mathbf b})\le1$; $0\le {\widehat
P}^{(2)}(S_1,S_2|{\mathbf a}{\mathbf c})\le1$, and $0\le {\widetilde
P}^{(2)}(S_1,S_2|{\mathbf b}{\mathbf c})\le1$, we have
\begin{eqnarray}
P^{(3)}(S_1,S_2,S_3|{\mathbf a}{\mathbf b}{\mathbf c})
&=&
\frac{P^{(2)}(S_1,S_2|{\mathbf a}{\mathbf b})
+
{\widehat P}^{(2)}(S_1,S_3|{\mathbf a}{\mathbf c})
+
{\widetilde P}^{(2)}(S_2,S_3|{\mathbf b}{\mathbf c})
}{4}
\nonumber \\
&=&
\frac{1+S_1 \langle \sigma_1\cdot{\mathbf a}\rangle+S_2\langle \sigma_2\cdot{\mathbf b}\rangle
+S_3\langle \sigma_2\cdot{\mathbf c}\rangle+S_1S_2 \langle\sigma_1\cdot{\mathbf a}\,\sigma_2\cdot{\mathbf b}\rangle}{8}
\nonumber \\
&&+\frac{
S_1S_3 \langle\sigma_1\cdot{\mathbf a}\,\sigma_2\cdot{\mathbf c}\rangle
+S_2S_3 \langle\sigma_1\cdot{\mathbf b}\,\sigma_2\cdot{\mathbf c}\rangle
}{8}
.
\label{exa0a}
\end{eqnarray}
\end{widetext}
Theorem IV, Eq.~(\ref{five3}) shows that
$P^{(3)}(S_1,S_2,S_3|{\mathbf a}{\mathbf b}{\mathbf c})$ as given by
Eq.~(\ref{exa0a}) represents the well-defined probability to observe
a given triple $(S_1,S_2,S_3)$, even though the operators that are
being measured, do not commute. The necessary  condition
for $P^{(3)}(S_1,S_2,S_3|{\mathbf a}{\mathbf b}{\mathbf c})$ to
exist as a probability is that the EBBI are satisfied, independent
of the presence of non-commuting operators in the theory (for a more
extensive discussion, see Section~\ref{commut}).

\subsection{A trap to avoid III: Separable states}\label{sepa}

Separable (product) states are special in that the state of the
system is determined by the states of the individual,
distinguishable subsystems. In this subsection, we study this aspect in
its full generality, simply because nothing is gained by limiting
the discussion to spin-1/2 systems.

Let us consider a composite quantum system that consists of two
identical subsystems. The Hilbert space ${\cal H}$ of the composite
quantum system is the direct product of the Hilbert spaces ${\cal
H}_i$ of the subsystems, that is ${\cal H}={\cal H}_1\otimes{\cal
H}_2$~\cite{BALL03}. The subsystems are assumed to be in the state
represented by the density matrices $\rho^{(1)}_1(\lambda)$ and
$\rho^{(1)}_2(\lambda)$, respectively. The variable $\lambda$ is an
element of a set that does not need to be defined in detail. In the
following, to simplify the notation, it is implicit that matrices
with a subscript $i$ act on the Hilbert space ${\cal H}_i$ and are
unit matrices with respect to the Hilbert space ${\cal H}_{3-i}$. We
denote by $\mathbf{Tr}_i$ the trace over the subspace of the $i$th
subsystem. Next, we define the matrix
\begin{eqnarray}
\rho^{(2)}&=&
\int \rho^{(1)}_1(\lambda) \rho^{(1)}_2(\lambda)\mu(\lambda) d\lambda
,
\label{prod0}
\end{eqnarray}
where $\mu(\lambda)$ is a probability density, that is a non negative
function, which satisfies $\int \mu(\lambda)\,d\lambda =1$ (compare with Eq.~(\ref{bell0})).
Using the properties of the trace,
$\mathbf{Tr} \rho^{(1)}_1(\lambda)\rho^{(1)}_2(\lambda)=
\mathbf{Tr}_1 \rho^{(1)}_1(\lambda)\mathbf{Tr}_2\rho^{(1)}_2(\lambda)=1$
and the fact that $\rho^{(2)}$ is a sum of non negative
matrices, it follows that Eq.~(\ref{prod0}) is a density matrix for the system
consisting of subsystems one and two.
Density matrices of the form Eq.~(\ref{prod0}) are called separable.

Notice that expression Eq.~(\ref{prod0}) is not the most general
state of a system consisting of two subsystems: Any convex
combination of $\rho^{(1)}_1(\lambda) \rho^{(1)}_2(\lambda')$
qualifies as a density matrix  but, as will become clear from the
derivation that follows, for this general class of states one cannot
prove EBBI. The difference between states of the form
Eq.~(\ref{prod0}) and a general state is similar to the difference
between functions of triples and three functions of pairs discussed
in Sections~\ref{Boole} and \ref{games}. Indeed, the state
Eq.~(\ref{prod0}) of a composite systems of two identical subsystems
can be recovered from the state
\begin{eqnarray}
\rho^{(3)}&=&
\int \rho^{(1)}_1(\lambda) \rho^{(1)}_2(\lambda)\rho^{(1)}_3(\lambda)\mu(\lambda) d\lambda
,
\label{prod10}
\end{eqnarray}
of a composite system of three identical subsystems by performing the trace operation over
one of the three subsystems. For a general state, this construction fails.

Let there be three dynamical variables for subsystem $i=1,2$,
represented by the matrices $A_i$, $B_i$, and $C_i$.
In analogy with the Boole inequalities, we wish to derive
inequalities for sums and differences of the correlations
\begin{eqnarray}
\langle A_1B_2 \rangle&=& \mathbf{Tr}\rho^{(2)}  A_1B_2
\nonumber \\
&=&
\int \mathbf{Tr}_1\rho^{(1)}_1(\lambda)A_1\mathbf{Tr}_2 \rho^{(1)}_2(\lambda)B_2\mu(\lambda) d\lambda
\nonumber \\
&\equiv&
\int \langle A_1 \rangle_\lambda\langle B_2 \rangle_\lambda\mu(\lambda) d\lambda
,
\nonumber \\
\langle A_1C_2 \rangle&=& \mathbf{Tr}\rho^{(2)}  A_1C_2
\nonumber \\
&=&
\int \mathbf{Tr}_1\rho^{(1)}_1(\lambda) A_1\mathbf{Tr}_2\rho^{(1)}_2(\lambda)C_2\mu(\lambda) d\lambda
\nonumber \\
&\equiv&
\int \langle A_1 \rangle_\lambda\langle C_2 \rangle_\lambda\mu(\lambda) d\lambda
,
\nonumber \\
\langle B_1C_2 \rangle&=& \mathbf{Tr} \rho^{(2)} B_1C_2
\nonumber \\
&=&
\int \mathbf{Tr}_1\rho^{(1)}_1(\lambda)B_1\mathbf{Tr}_2 \rho^{(1)}_2(\lambda)C_2\mu(\lambda) d\lambda
\nonumber \\
&\equiv&
\int \langle B_1 \rangle_\lambda\langle C_2 \rangle_\lambda\mu(\lambda) d\lambda
.
\label{prod2}
\end{eqnarray}
As long as we confine ourselves to finite-dimensional Hilbert spaces
(as we do here), we may, without loss of generality, assume that
$A_i$, $B_i$, and $C_i$ are normalized such that the eigenvalues of
these matrices are in the interval $[-1,1]$. Then, from Postulate I
it follows that $|\langle A_i \rangle_\lambda|\le1$, $|\langle B_i
\rangle_\lambda|\le1$, and $|\langle C_i \rangle_\lambda|\le1$ for
all $\lambda$. From the algebraic identity $(1\pm xy)^2=(x \pm
y)^2+(1-x^2)(1-y^2)$ it follows that $|a\pm b|\le 1\pm ab$ for real
numbers $a$ and $b$ with $|a|\le1$ and $|b|\le1$. Then, it
immediately follows that $|ac\pm bc|\le 1\pm ab$ for real numbers
$a$, $b$, and $c$ such that $|a|\le1$, $|b|\le1$, and $|c|\le1$.
Combining all these results we find
\begin{eqnarray}
\left|\langle A_1B_2 \rangle\pm \langle A_1C_2 \rangle\right|
&\le&
\int \left|
\langle A_1 \rangle_\lambda\langle B_2 \rangle_\lambda
\pm
\langle A_1 \rangle_\lambda\langle C_2 \rangle_\lambda
\right|
\mu(\lambda) d\lambda
\nonumber \\
&\le&
\int \left( 1\pm
\langle B_2 \rangle_\lambda\langle C_2 \rangle_\lambda
\right)
\mu(\lambda) d\lambda
.
\label{prod3}
\end{eqnarray}
We can turn inequality Eq.~(\ref{prod3}) into a Boole-Bell inequality
if we assume that $\langle B_1 \rangle_\lambda=\langle B_2\rangle_\lambda$ for all $\lambda$,
which is the case if the two subsystems are identical.
Indeed, then Eq.~(\ref{prod3}) becomes
\begin{eqnarray}
\left|\langle A_1B_2 \rangle\pm \langle A_1C_2 \rangle\right|
&\le&
\int \left(1\pm
\langle B_1 \rangle_\lambda\langle C_2 \rangle_\lambda
\right)
\mu(\lambda) d\lambda
\nonumber \\
&\le&
1\pm
\langle B_1C_2 \rangle
,
\label{prod4}
\end{eqnarray}
and by permutation of the symbols $A$, $B$, and $C$, all other Boole-like inequalities follow.

We can now ask the question what conclusion one can draw if, for
some specific model, we find that inequality Eq.~(\ref{prod4}) is
violated. Disregarding technical conditions such as the requirements
on the spectral range of the matrices $A_i$, $B_i$, and $C_i$, the
only logically correct conclusion is that the density matrix
$\rho^{(2)}$ of the composite system cannot be represented by a
state of the form Eq.~(\ref{prod0}). In other words, a necessary
condition that a quantum system consisting of two identical,
distinguishable systems is represented by the separable state
Eq.~(\ref{prod0}) is that the inequalities Eq.~(\ref{prod4}) are not
violated. Although this is a nontrivial statement about the state of
the composite system no other conclusion can be drawn from the
violation of Eq.~(\ref{prod4}).

We emphasize that it is not legitimate to replace the quantum
theoretical expectations that appear in Eq.~(\ref{prod4}) by certain
empirical data, simply because Eq.~(\ref{prod4}) has been derived
within the mathematical framework of quantum theory, not for sets of
data collected, grouped and characterized by experimenters. The
latter can be tested against the original Boole inequalities only
and the conclusions that follow from their violation have no bearing
on the quantum theoretical model which as shown in
Section~\ref{QuantumTheory}, can never violate the EBBI
Eq.~(\ref{BOOLE2})~\cite{RAED07a,RAED07c}.

Although the derivation of Eq.~(\ref{prod4}) may seem to be
unrelated to the derivations of EBBI of the preceding
sections, this is not the case. Indeed, as mentioned earlier, the
system of two identical subsystems can be trivially embedded in a
system of three identical subsystems by constructing the density
matrix of the latter according to Eq.~(\ref{prod10}). If we now
limit ourselves to subsystems that have two states only, it is a
simple exercise to show that
\begin{equation}
P^{(3)}(S_1,S_2,S_3)
=
\int P^{(1)}(S_1|\lambda)P^{(1)}(S_2|\lambda)P^{(1)}(S_3|\lambda)
\mu(\lambda) d\lambda
,
\label{prod11}
\end{equation}
which is formally identical to Eq.~(\ref{bell3})
and hence, Theorems II and IV of Section~\ref{functions} apply.

Summarizing:
For a composite quantum system consisting of two
identical subsystems $i=1,2$ and described by a separable state,
correlations of three dynamical variables represented by
finite, normalized Hermitian matrices $A_i$, $B_i$, and $C_i$,
obey the Boole-like inequality Eq.~(\ref{prod4}).
As the (non-)commutativity of the
three matrices $A_i$, $B_i$, and $C_i$ does not enter the conditions
required to prove inequality Eq.~(\ref{prod4}), it would
be a logical fallacy to relate the apparent
violation of Eq.~(\ref{prod4}) to the non-commutativity of the three
matrices $A_i$, $B_i$, and $C_i$.

\subsection{Non-commuting operators, common probability spaces and EBBI}\label{commut}

It is well known that the involvement of non-commuting operators in
quantum problems may prohibit the use of one common (Kolmogorov)
probability space~\cite{FINE82,BALL03,HESS08} for these problems.
In essence, the point is this: If $A$ and $B$ are Hermitian matrices, they are
diagonalizable~\cite{BELL97}. If they commute ($[A,B]=0$), there
exists a unitary transformation that simultaneously diagonalizes
$A$, $B$, and $AB$~\cite{BELL97}. Therefore if $[A,B]=0$, then
according to Postulate II, the dynamical variables that are
represented by $A$, $B$ and $AB$ can simultaneously assume one of
their possible values. In this case, it becomes meaningful to speak
about the observation of events corresponding to $A$, $B$, and $AB$
and the product rule, one of the cornerstones of Kolmogorov's
axiomatic framework of probability theory is
satisfied~\cite{BALL03}. However, if  $[A,B]\not=0$, it is no longer
possible to simultaneously attribute eigenvalues to $A$, $B$ and
$AB$: Any attempt to assign numbers to the probabilities that appear
in the product rule fails~\cite{BALL03}. In this case, the dynamical
variables cannot be defined on one common Kolmogorov probability
space. However, for a given state of the quantum system, the
probability distributions corresponding to each of the dynamical
variables may be interrelated~\cite{BALL03}. The most important
consequence of such interrelation is the Heisenberg uncertainty
principle for the position and momentum of a particle~\cite{BALL03}.
We now show that the Heisenberg uncertainty principle, when applied to the
EPRB experiment, does not impose any relation between
probability distributions corresponding to different measurements.

If $X$, $Y$ and $Z=i[X,Y]$ are matrices, application of the Schwarz inequality yields~\cite{BALL03}
\begin{eqnarray}
\langle X^2-\langle X \rangle^2\rangle
\langle Y^2-\langle Y \rangle^2\rangle
&\ge&\frac{1}{4} |\langle Z\rangle|^2
,
\label{uncer0}
\end{eqnarray}
where the average of $X$ is defined by
$\langle X\rangle= \mathbf{Tr}\rho X$,
$\rho$ denoting the density matrix that describes the state of the quantum system.
If $X$ and $Y$ represent the coordinate and momentum operators, respectively,
Eq.~(\ref{uncer0}) reduces to the Heisenberg uncertainty relation in its original form.

In the standard EPRB experiment, described in Section~\ref{example}, we perform
three experiments, each experiment yielding a pair of two-valued variables
for the pairs of setting $({\mathbf a},{\mathbf b})$, $({\mathbf a},{\mathbf c})$,
and $({\mathbf b},{\mathbf c})$.
Using $\sigma_j\cdot{\mathbf x}\sigma_j \cdot {\mathbf
y}= {\mathbf x}\cdot{\mathbf y}+i({\mathbf x}\times{\mathbf y})\cdot
\sigma_j$ for $j=1,2$, it follows that
\begin{eqnarray}
\left[\sigma_1\cdot {\mathbf a}\,\sigma_2\cdot {\mathbf b},\sigma_1\cdot {\mathbf a}\,\sigma_2\cdot {\mathbf c}\right]
&=&
2i ({\mathbf b}\times{\mathbf c})\cdot \sigma_2
,
\nonumber \\
\left[\sigma_1\cdot{\mathbf a}\,\sigma_2\cdot{\mathbf b},\sigma_1\cdot{\mathbf b}\,\sigma_2\cdot{\mathbf c}\right]
&=&
2i ({\mathbf a}\times{\mathbf b})\cdot \sigma_1 +2i({\mathbf b}\times{\mathbf c})\cdot \sigma_2
,
\nonumber \\
\left[\sigma_1\cdot{\mathbf a}\,\sigma_2\cdot{\mathbf c},\sigma_1\cdot{\mathbf b}\,\sigma_2\cdot{\mathbf c}\right]
&=&
2i ({\mathbf a}\times{\mathbf b})\cdot \sigma_1
.
\label{comm0}
\end{eqnarray}
From Eq.~(\ref{comm0}), it follows that if ${\mathbf
a}\times{\mathbf b}\not=0$, ${\mathbf a}\times{\mathbf c}\not=0$,
and ${\mathbf b}\times{\mathbf c}\not=0$, none of the commutators in
Eq.~(\ref{comm0}) vanish. Suppose that ${\mathbf a}\times{\mathbf
b}=0$. Then ${\mathbf a}$ and ${\mathbf b}$ are (anti-) parallel and
of the two experiments that yield $\sigma_1\cdot{\mathbf
a}\,\sigma_2\cdot{\mathbf c}$ and $\sigma_1\cdot{\mathbf
b}\,\sigma_2\cdot{\mathbf c}$, one is redundant. The same holds for
the other cases in which two directions of measurement are
(anti-)parallel. Clearly, the condition for the three experiments to
be fundamentally distinct is that none of the commutators in
Eq.~(\ref{comm0}) vanishes. In other words, if one or two of the
commutators in Eq.~(\ref{comm0}) vanish, the experiment is
completely described by at most two dichotomic variables and hence
there exists no EBBI (see Section~\ref{EBBI}).

\begin{widetext}
Combining inequality Eq.~(\ref{uncer0}) and Eq.~(\ref{comm0}) we find
\begin{eqnarray}
\left(1-\langle\sigma_1\cdot\mathbf{a}\;\sigma_2\cdot\mathbf{b}\rangle^2\right)
\left(1-\langle\sigma_1\cdot\mathbf{a}\;\sigma_2\cdot\mathbf{c}\rangle^2\right)
&\ge&\left|(\mathbf{b}\times\mathbf{c})\cdot\langle\sigma_2\rangle\right|^2
,\nonumber \\
\left(1-\langle\sigma_1\cdot\mathbf{a}\;\sigma_2\cdot\mathbf{b}\rangle^2\right)
\left(1-\langle\sigma_1\cdot\mathbf{b}\;\sigma_2\cdot\mathbf{c}\rangle^2\right)
&\ge&\left|(\mathbf{a}\times\mathbf{b})\cdot\langle\sigma_1\rangle
+ (\mathbf{b}\times\mathbf{c})\cdot\langle\sigma_2\rangle\right|^2
,\nonumber \\
\left(1-\langle\sigma_1\cdot\mathbf{a}\;\sigma_2\cdot\mathbf{c}\rangle^2\right)
\left(1-\langle\sigma_1\cdot\mathbf{b}\;\sigma_2\cdot\mathbf{c}\rangle^2\right)
&\ge&\left|(\mathbf{a}\times\mathbf{b})\cdot\langle\sigma_1\rangle\right|^2
.
\label{uncer5}
\end{eqnarray}
\end{widetext}
As the EPRB experiment is described by a system in the singlet state
we have $\langle \sigma_1\rangle=\langle \sigma_2\rangle=0$ and hence
\begin{eqnarray}
\left(1-\langle\sigma_1\cdot\mathbf{a}\;\sigma_2\cdot\mathbf{b}\rangle^2\right)
\left(1-\langle\sigma_1\cdot\mathbf{a}\;\sigma_2\cdot\mathbf{c}\rangle^2\right)
\ge 0
,\nonumber \\
\left(1-\langle\sigma_1\cdot\mathbf{a}\;\sigma_2\cdot\mathbf{b}\rangle^2\right)
\left(1-\langle\sigma_1\cdot\mathbf{b}\;\sigma_2\cdot\mathbf{c}\rangle^2\right)
\ge 0
,\nonumber \\
\left(1-\langle\sigma_1\cdot\mathbf{a}\;\sigma_2\cdot\mathbf{c}\rangle^2\right)
\left(1-\langle\sigma_1\cdot\mathbf{b}\;\sigma_2\cdot\mathbf{c}\rangle^2\right)
\ge 0
.
\label{uncer7}
\end{eqnarray}
Clearly, none of these inequalities imposes any condition on or
any relation between the probability distributions for measuring
$\langle\sigma_1\cdot\mathbf{a}\;\sigma_2\cdot\mathbf{b}\rangle$,
$\langle\sigma_1\cdot\mathbf{a}\;\sigma_2\cdot\mathbf{c}\rangle$,
or
$\langle\sigma_1\cdot\mathbf{b}\;\sigma_2\cdot\mathbf{c}\rangle$,
also in the case where the operators involved do not commute.
Obviously, the fact that the operators in the quantum theoretical
description of the EPRB experiment do not commute does not impose
interrelations between the probability distributions for
measuring the eigenvalues of these operators.

We further address the question to what extent the non-commutativity
of the matrices that appear in the quantum theoretical description
of EPRB-like experiments (see Section~\ref{example})
leads to testable consequences.
The discussion that follows equally holds for all other
quantum systems considered in this paper.

We return to our derivation of the EBBI and exclude redundant
experiments (implying that none of the commutators in
Eq.~(\ref{comm0}) vanishes).
If the EBBI are satisfied, quantum
theory guarantees that $P^{(3)}(S1,S2,S3)$ exists while if the EBBI
are violated it does not. But in both cases, the matrices
$\sigma_1\cdot {\mathbf a}\,\sigma_2\cdot {\mathbf b}$,
$\sigma_1\cdot {\mathbf a}\,\sigma_2\cdot {\mathbf c}$, and
$\sigma_1\cdot {\mathbf b}\,\sigma_2\cdot {\mathbf c}$, never
mutually commute, independent of whether or not the EBBI are
satisfied. The logical implication is that the condition that these
matrices do not mutually commute is a superfluous condition
for the apparent violation of the EBBI. The apparent
violation of the EBBI does imply that $P^{(3)}(S1,S2,S3)$ does not
exist as a probability. However, it would be a logical fallacy
to directly relate this non-existence of a joint probability to a
general statement that the presence of non-commuting operators in
the theory prohibits the existence of a common probability
space~\cite{HESS08}.

Summarizing: We have shown that apparent violations of the EBBI
cannot be attributed to the non-commutativity of the (products of)
spin operators, the expectation values of which appear in the EBBI.
A more general, much stronger, indication that non-commutativity is
actually irrelevant for the apparent violations
of the EBBI is that these violations are also found for genuine
``classical'' models (see Section~\ref{apparent}), both in the case of
data and for ``factorizable'' probabilistic models.
Evidently, in the realm of these classical models, non-commutativity is
neither necessary nor sufficient for violations of EBBI
nor is commutativity necessary or sufficient to guarantee the validity of EBBI.

\section{Application to quantum flux tunneling}\label{Leggett}

\begin{figure*}[t]
\begin{center}
\includegraphics[width=14cm]{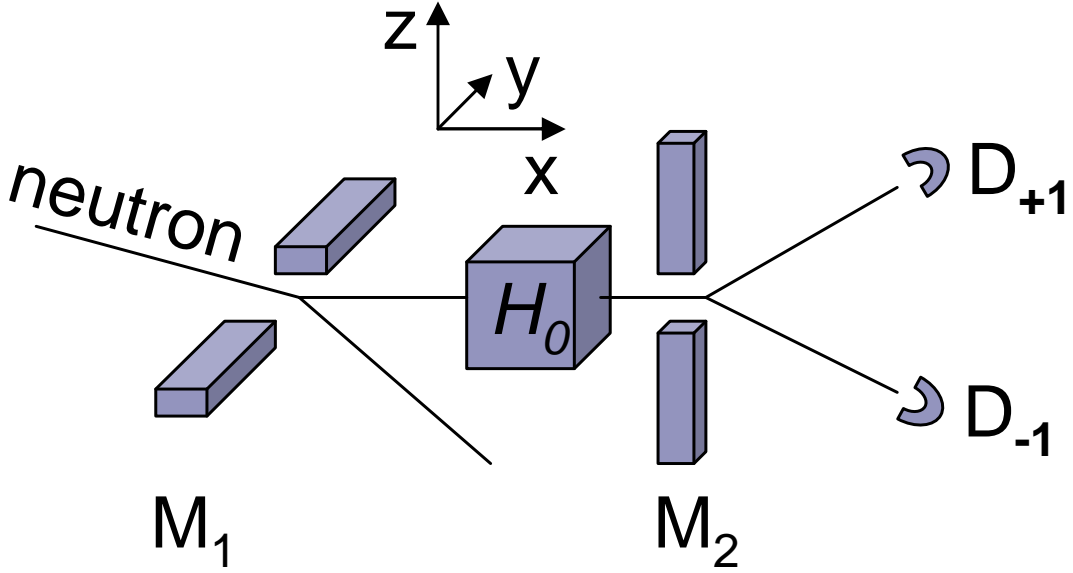}
\caption{Conceptual layout of an experiment to measure the magnetic flux through a SQUID.
A neutron passes through a Stern-Gerlach magnet ($M_1$) that aligns the magnetic moment
of the neutron along the $y$-direction, interacts with the magnetic moment of the system described by a Hamiltonian $H_0$,
and passes through another Stern-Gerlach magnet ($M_2$) that deflects the neutron according to the
projection of its magnetic moment on the $z$-direction.
The detectors $D_{+1}$ and $D_{-1}$ signal the
arrival of a neutron with spin up and spin down respectively.
}
\label{neutrons}
\end{center}
\end{figure*}

In an idealized picture, the flux trapped in a SQUID may be viewed as a prototype two-state system,
the macroscopic flux tunneling between the two states.
Leggett and Garg have described an experiment to detect signatures of the tunneling process by measuring the state
of the flux as a function of the time differences between measurements~\cite{LEGG85}.
To illustrate how the general theory applies to this problem,
we adopt the quantum mechanical model proposed by Ballentine~\cite{BALL87a}.
In this model, one neutron at a time is being propelled through the SQUID
and the state of the flux is inferred by measuring correlations
of the spin of the neutrons as a function of the time differences between successive neutrons~\cite{BALL87a}.

A schematic diagram of this experiment is shown in
Fig.~\ref{neutrons}. At time $t_0$, we prepare the system, that is
the SQUID, in spin state $|\phi_0\rangle$. At fixed times $t_0\le
t_1\le t_2\le t_3$, we shoot three neutrons one after each other
through the system, let the neutron spin interact with the magnetic
moment of the system, and detect the spin of the neutrons when they
no longer interact with the system. We repeat this procedure many
times and count the number of neutrons with spin up and spin down.
Then, we repeat the whole procedure, choosing again $t_1$, $t_2$ and
$t_3$, and study the counts as a function of $t_1-t_0$, $t_2-t_1$,
and $t_3-t_2$.

At $t=t_0$, the initial state (after preparation) of the system+neutrons is given by
\begin{eqnarray}
|\Psi(t_0)\rangle&=&|\phi_0\phi_1\phi_2\phi_3\rangle
,
\label{three0}
\end{eqnarray}
where
$|\phi_j\rangle$ with $j=1,2,3$
represents the state of the spin of the $j$th neutron.
Obviously, the system described by Eq.~(\ref{three0}) is initially in a product state,
which is equivalent to the (rather obvious) statement that in the initial state
there are no correlations between the four objects.
According to quantum theory, we have
\begin{equation}
P^{(3)}(S_1,S_2,S_3|t_3,t_2,t_1,\Psi(t_0))= |\langle S_1,S_2,S_3|\Psi(t_3,t_2,t_1)\rangle|^2
,
\label{three1}
\end{equation}
where $|\Psi(t_3,t_2,t_1)\rangle$ denotes the state of the
system+neutrons at the time that the third neutron has triggered one
of the detectors. In Eq.~(\ref{three1}) we have included $\Psi(t_0)$
into the list of conditions on the probability even though
$\Psi(t_0)$ is not an element of Boolean logic.
However, the condition $\Psi(t_0)$ in Eq.~(\ref{three1}) should be interpreted
operationally: At $t_0$, the system has been prepared in a
particular manner such that its state is represented by
$|\Psi(t_0)\rangle$~\cite{BALL03}.

The numerical quantities accessible through measurement are the clicks of the detector.
For each run of the experiment, there are three of these clicks (we assume 100\%
detection efficiency, no loss of neutrons etc.),
which we denote by the triples $(S_{1,\alpha},S_{2,\alpha},S_{3,\alpha})$.
From $M$ repetitions with the same $t_1$, $t_2$, and $t_3$, we compute
the empirical averages and correlations
\begin{eqnarray}
\langle S_i\rangle_3&=&\frac{1}{M}\sum_{\alpha=1}^M S_{i,\alpha}\quad,\quad i=1,2,3,
\nonumber \\
\langle S_iS_j\rangle_3&=&\frac{1}{M}\sum_{\alpha=1}^M S_{i,\alpha}S_{j,\alpha}
\quad,\quad (i,j)=(1,2),(1,3),(2,3),
\nonumber \\
\langle S_1S_2S_3\rangle_3&=&\frac{1}{M}\sum_{\alpha=1}^M S_{1,\alpha}S_{2,\alpha} S_{3,\alpha}
,
\label{three3}
\end{eqnarray}
where the subscript 3 in $\langle \cdot \rangle_3$ refers to the three observations that are made in each run of the experiment.
Assuming that quantum theory describes this experiment,
we expect to find that
\begin{eqnarray}
\langle S_i\rangle_3&\rightarrow&
E^{(3)}_{i}\quad,\quad i=1,2,3,
\nonumber \\
\langle S_iS_j\rangle_3&\rightarrow&
E^{(3)}_{ij}
\quad,\quad (i,j)=(1,2),(1,3),(2,3),
\nonumber \\
\langle S_1S_2S_3\rangle_3&\rightarrow&
E^{(3)}
,
\label{three4}
\end{eqnarray}
where the notation $A\rightarrow B$ means that
as $M\rightarrow\infty$, $A=B$ with probability one.

From Sections~\ref{Boole} and \ref{QuantumTheory}, we know that it
is mathematically impossible to violate the inequalities
\begin{eqnarray}
|\langle S_iS_j\rangle_3\pm \langle S_iS_k\rangle_3|&\le&1\pm \langle S_jS_k\rangle_3
,
\label{three5a}
\label{BOOLE3}
\\
|E^{(3)}_{ij}\pm E^{(3)}_{ik}|&\le&1\pm E^{(3)}_{jk}
.
\label{three5b}
\label{BOOLE4}
\end{eqnarray}
with $(i,j,k)=(1,2,3),(3,1,2),(2,3,1)$.
If the real experiment would show a violation of the Boole inequalities
Eq.~(\ref{three5a}), this can only imply that we
have made one or more mistakes in elementary arithmetic. Indeed,
this experiment complies with the condition that lead to
Eq.~(\ref{three5a}), namely that each instance yields a triple of
two-valued numbers $(S_{1,\alpha},S_{2,\alpha},S_{3,\alpha})$.

From a violation of Eq.~(\ref{three5b}) we can only deduce that the
specific quantum mechanical model calculation that yields the
expression of $E^{(3)}_{ij}$ needs to be revised. Indeed, we have
shown in Section \ref{QuantumTheory} that Eq.~(\ref{three5b}) must
be satisfied in general.

It is instructive to scrutinize the arguments claimed
in Ref.~\onlinecite{LEGG85} that lead to the wrong conclusion that
the above quantum mechanical system can violate Eq.~(\ref{three5b}).
Ref.~\onlinecite{LEGG85} starts with ``macroscopic realism": A
macroscopic system with two macroscopically distinct states
available to it will at all times be in one or the other of these
states.
Then, the crucial and incorrect assumption is made that
macroscopic realism implies the existence of
consistent joint probabilities $p_{12}(S_1,S_2)$, $p_{13}(S_1,S_3)$,
$p_{23}(S_2,S_3)$, and $p(S_1,S_2,S_3)$
that obey~\cite{LEGG85}
\begin{eqnarray}
p_{12}(S_1,S_2)&=&\sum_{S_3=\pm1}p(S_1,S_2,S_3),
\nonumber \\
p_{13}(S_1,S_3)&=&\sum_{S_2=\pm1}p(S_1,S_2,S_3),
\nonumber \\
p_{23}(S_2,S_3)&=&\sum_{S_1=\pm1}p(S_1,S_2,S_3).
\label{three5e}
\end{eqnarray}
Macroscopic realism does not imply Eq.~(\ref{three5e}) as should be
clear by now. However, together with the additional grouping into
triples (\CNTUH), it most definitely does. Then because the
measurements are performed on groups of three neutrons, we may
indeed follow Ref.~\onlinecite{LEGG85} and define the correlation
functions $K^{(3)}_{ij}$ by
\begin{eqnarray}
K^{(3)}_{ij}&=&\sum_{S_1=\pm1}\sum_{S_2=\pm1}\sum_{S_3=\pm1}S_iS_jp(S_1,S_2,S_3),
\\
&=&
\sum_{S_i=\pm1}\sum_{S_j=\pm1}S_iS_jp_{ij}(S_i,S_j),
\label{three5c}
\end{eqnarray}
for $(i,j)=(1,2),(1,3),(2,3)$
where the latter expression follows from the requirement of consistency.
As we have seen in Section~\ref{QuantumTheory}, the fact that $p(S_1,S_2,S_3)$ exists as a probability
is sufficient to prove that
\begin{eqnarray}
|K^{(3)}_{ij}\pm K^{(3)}_{ik}|&\le& 1\pm K^{(3)}_{jk}
,
\label{three5d}
\end{eqnarray}
for $(i,j,k)=(1,2,3),(3,1,2),(2,3,1)$,
containing the Leggett-Garg inequality~\cite{LEGG85} as a
particular case. Now, because there exists a joint probability for
triples, the EBBI and consequently also the Leggett-Garg inequality,
cannot be violated. However, in Ref.~\onlinecite{LEGG85} a
contradiction is predicted because it is assumed, without
justification, that $K^{(3)}_{ij}=P(t_j-t_i)$ with $P(t)\approx
e^{-\gamma|t|}\cos\omega t$, an expression obtained from a quantum
mechanical calculation of a correlation function that involves two
measurements only. This is inconsistent: Inequality
Eq.~(\ref{three5d}) has been derived from a probability distribution
that involves three, not only two, measurements. If the numerical
values of $K^{(3)}_{ij}$ as determined from experiments involving
two measurements lead to violations of inequality
Eq.~(\ref{three5d}), the only correct action is to reject the
assumption that these are the values of $K^{(3)}_{ij}$ that will be
observed in an experiment that performs three measurements. As we
have seen over and over again by now: In general one cannot deduce
inequalities such as Eq.~(\ref{three5d}) if experiment or theory
deal with pairs of two-valued variables only.

\subsection{Concrete example}
We adopt the specific model analyzed by Ballentine~\cite{BALL87a} to
illustrate how the line of thought adopted in
Ref.~\onlinecite{LEGG85} yields conclusions that are in conflict
with the EBBI, that is with elementary arithmetic. The
Hamiltonian of the system (the SQUID) is defined by
\begin{eqnarray}
H_0&=&\omega\sigma^x_0
.
\label{sm0}
\end{eqnarray}
This Hamiltonian describes a spin-1/2 object that is tunneling between the spin-up and spin-down state
with an angular frequency $\omega$.
During the time $\tau$ that the system interacts with the $j$th neutron, the Hamiltonian changes to
\begin{eqnarray}
H_j&=&\omega\sigma^x_0+\alpha\sigma^z_0\sigma^x_j
.
\label{sm2}
\end{eqnarray}
At time $t_0$, we prepare the system in the state with spin up, that is $|\phi_0\rangle=|\uparrow\rangle$
and we prepare neutrons such that their spins are aligned along the positive $y$-direction.
Thus, the initial state of the $j$th neutron is
\begin{eqnarray}
|\phi_j\rangle&=&\frac{1}{\sqrt{2}}(|\uparrow\rangle+i|\downarrow\rangle)
.
\label{three0a}
\end{eqnarray}

Following Ref.~\onlinecite{BALL87a}, we consider the limiting case
in which the interaction time $\tau\rightarrow0$ and the coupling
constant $\alpha\rightarrow\infty$ such that $\alpha\tau =\pi/4$.
For this choice of parameters, the correlation between the system
and neutron spin is maximal~\cite{BALL87a}. In this case, the wave
function after the three neutrons have interacted with the
system reads
\begin{eqnarray}
|\Psi(\Delta t_3,\Delta t_2,\Delta t_1)\rangle
=
\cos\omega\Delta t_3
\cos\omega\Delta t_2\cos\omega\Delta t_1
&&
|\uparrow\uparrow\uparrow\uparrow \rangle
\nonumber \\
-
\cos\omega\Delta t_3
\cos\omega\Delta t_2\sin\omega\Delta t_1
&&
|\downarrow\downarrow\downarrow\downarrow \rangle
\nonumber \\
+i
\cos\omega\Delta t_3
\sin\omega\Delta t_2\cos\omega\Delta t_1
&&
|\downarrow\uparrow\downarrow\downarrow \rangle
\nonumber \\
-i
\cos\omega\Delta t_3
\sin\omega\Delta t_2\sin\omega\Delta t_1
&&
|\uparrow\downarrow\uparrow\uparrow \rangle
\nonumber \\
+
\sin\omega\Delta t_3
\cos\omega\Delta t_2\cos\omega\Delta t_1
&&
|\downarrow\uparrow\uparrow\downarrow \rangle
\nonumber \\
+
\sin\omega\Delta t_3
\cos\omega\Delta t_2\sin\omega\Delta t_1
&&
|\uparrow\downarrow\downarrow\uparrow \rangle
\nonumber \\
-i
\sin\omega\Delta t_3
\sin\omega\Delta t_2\cos\omega\Delta t_1
&&
|\uparrow\uparrow\downarrow\uparrow\rangle
\nonumber \\
-i
\sin\omega\Delta t_3
\sin\omega\Delta t_2\sin\omega\Delta t_1
&&
|\downarrow\downarrow\uparrow\downarrow \rangle
,
\nonumber \\
\label{sm13}
\end{eqnarray}
where $\Delta t_i=t_i-t_{i-1}$.
For general $\Delta t_i$, Eq.~(\ref{sm13}) represents a highly entangled, four-spin state.
A straightforward calculation yields
\begin{eqnarray}
E^{(3)}_{12}&=&\cos2\omega\Delta t_2 ,
\nonumber \\
E^{(3)}_{13}&=&\cos2\omega\Delta t_3\cos2\omega\Delta t_2,
\nonumber \\
E^{(3)}_{23}&=&\cos2\omega\Delta t_3
,
\label{sm14}
\end{eqnarray}
where we omit the expressions of averages
that are not relevant for testing the inequalities.
Substituting the expressions Eq.~(\ref{sm14}) in the inequalities Eq.~(\ref{three5b}),
one finds that the latter are always satisfied, as expected on general grounds.
On the other hand, if we consider experiments in which we collect pairs instead of triples,
quantum theory yields
\begin{eqnarray}
E^{(2)}&=&\cos2\omega(t_2-t_1),
\nonumber \\
{\widehat E}^{(2)}&=&\cos2\omega(t_3-t_1),
\nonumber \\
{\widetilde E}^{(2)}&=&\cos2\omega(t_3-t_2).
\label{sm15}
\end{eqnarray}
Obviously, for this model $E^{(3)}_{12}=E^{(2)}$
and $E^{(3)}_{23}={\widetilde E}^{(2)}$ but $E^{(3)}_{13}\not={\widehat E}^{(2)}$.
Should we now make the mistake to assume that
$E^{(3)}_{12}=E^{(2)}=\cos2\omega(t_2-t_1)$, $E^{(3)}_{23}={\widetilde E}^{(2)}=\cos2\omega(t_3-t_2)$ and
$E^{(3)}_{13}={\widehat E}^{(2)}=\cos2\omega(t_3-t_1)$
and substitute these expressions into the inequalities Eq.~(\ref{three5b}), we would find
that the latter can be violated.
However, it is clear that the only conclusion that one can draw from this
violation is that the assumption
$E^{(3)}_{12}=E^{(2)}$, $E^{(3)}_{23}={\widetilde E}^{(2)}$, $E^{(3)}_{13}={\widehat E}^{(2)}$
is wrong:
Although the system that describes the two-neutron measurement
can quite naturally be embedded in a system that describes the three-neutron measurement,
this embedding is nontrivial in the sense that $E^{(3)}_{13}\not={\widehat E}^{(2)}$.

\subsection{Summary}
It is not legitimate to substitute the expressions of $E^{(2)}$,
${\widehat E}^{(2)}$, ${\widetilde E}^{(2)}$, as obtained from a
quantum theoretical description of an experiment that involves pairs
only, into inequalities that have been derived from a quantum
theoretical description of an experiment that involves triples of
variables. As shown in Section~\ref{QuantumTheory}, quantum theory
does not provide inequalities that put bounds on ${\widetilde
E}^{(2)}$ in terms of $E^{(2)}$ and ${\widehat E}^{(2)}$. The
derivation of the EBBI requires a system with at least
three different two-valued variables.

\section{Application to Einstein-Podolsky-Rosen-Bohm (EPRB) experiments}\label{EPR}
\subsection{Original EPRB experiment}

\begin{figure*}[t]
\begin{center}
\includegraphics[width=14cm]{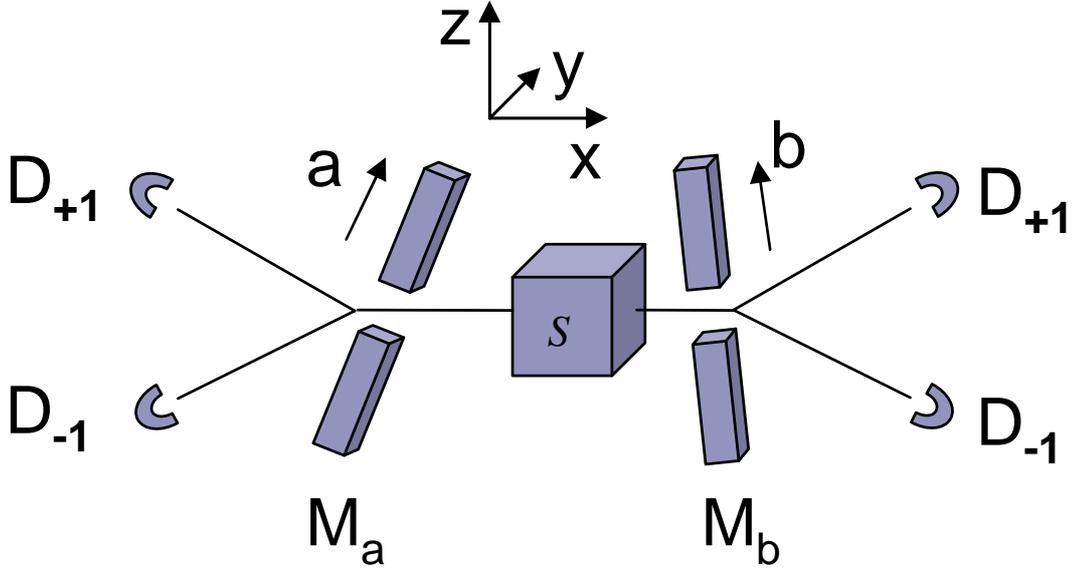}
\caption{Schematic diagram of the Einstein-Podolsky-Rosen-Bohm
(EPRB) thought experiment. The source $S$ produces pairs of spin-1/2
particles. The particle going to the left
(right) passes through a Stern-Gerlach magnet $M_a$ ($M_b$) that
directs the particle to either detector $D_{+1}$ or $D_{-1}$,
depending on whether its spin after passing the magnet is parallel
or anti-parallel to the direction $\mathbf{a}$ ($\mathbf{b}$). If
the detector $D_{+1}$ at the left (right) of the source fires, we
set $S_1=+1$ ($S_2=+1$), otherwise we set $S_1=-1$ ($S_2=-1$). In
this idealized experiment, each pair produced by the source
generates a pair of signals $(S_1=\pm1,S_2=\pm1)$. } \label{epr2}
\end{center}
\end{figure*}

In Fig.~\ref{epr2}, we show a schematic diagram of the
Einstein-Podolsky-Rosen thought experiment~\cite{EPR35} in the form
proposed by Bohm~\cite{BOHM51}. In the quantum mechanical
description of this experiment, it is assumed that the system
consists of two spin-1/2 objects. According to the axioms of quantum
theory~\cite{BALL03}, repeated measurements on the system described
by the normalized state vector $|\Psi\rangle$ yield
statistical estimates for the single-particle expectation values
$E^{(2)}_1=\langle \Psi|\mathbf{\sigma}_1\cdot\mathbf{a}|\Psi \rangle$,
$E^{(2)}_2=\langle \Psi|\mathbf{\sigma}_2\cdot\mathbf{b}|\Psi \rangle$
and for the two-particle correlation
$E^{(2)}=\langle \Psi|\mathbf{\sigma}_1\cdot\mathbf{a}\; \mathbf{\sigma}_2\cdot\mathbf{b} |\Psi\rangle$
where $\mathbf{a}$ and $\mathbf{b}$ are unit vectors.

For a quantum system of two spin-1/2 objects, we can derive an
inequality as follows. We consider two additional
experiments that yield
${\widehat E}^{(2)}=\langle \Psi|\mathbf{\sigma}_1\cdot\mathbf{a}\;
\mathbf{\sigma}_2\cdot\mathbf{c} |\Psi\rangle$ and ${\widetilde
E}^{(2)}=\langle \Psi|\mathbf{\sigma}_1\cdot\mathbf{b}\;
\mathbf{\sigma}_2\cdot\mathbf{c} |\Psi\rangle$ where $\mathbf{c}$ is
also a unit vector. Using the Schwartz inequality $|\langle\Psi
|X|\Psi\rangle|^2\le\langle \Psi|X^\dagger X|\Psi\rangle$ with
$X=X^\dagger=\sigma_1\cdot\mathbf{a}\;\sigma_2\cdot\mathbf{b} \pm
\sigma_1\cdot\mathbf{a}\;\sigma_2\cdot\mathbf{c} $ we find
$X^\dagger
X=2+2\mathbf{b}\cdot\mathbf{c}$ and hence
\begin{eqnarray}
\left| E^{(2)}\pm {\widehat E}^{(2)}\right|^2&\le& 2(1\pm\mathbf{b}\cdot\mathbf{c})
.
\label{aepr0}
\end{eqnarray}
Note that in essence, the proof of inequality Eq.~(\ref{aepr0}) follows from
the Schwartz inequality which in turn follows from
the assumption that the inner product on the Hilbert space is non negative.

If the system is in the singlet state Eq.~(\ref{singlet})
we have $E^{(2)}_1=E^{(2)}_2=0$,
$E^{(2)}=-\mathbf{a}\cdot\mathbf{b}$,
${\widehat E}^{(2)}=-\mathbf{a}\cdot\mathbf{c}$, and
${\widetilde E}^{(2)}=-\mathbf{b}\cdot\mathbf{c}$.
Substituting these expressions in Eq.~(\ref{aepr0}) yields
\begin{eqnarray}
\left| E^{(2)}\pm {\widehat E}^{(2)}\right|^2&=&|\mathbf{a}\cdot(\mathbf{b}\pm\mathbf{c})|^2
\nonumber \\
&=&(\mathbf{b}\pm\mathbf{c})^2\cos^2\theta_\pm
=2(1\pm\mathbf{b}\cdot\mathbf{c})\cos^2\theta_\pm
\nonumber \\
&\le&2(1\pm\mathbf{b}\cdot\mathbf{c})
,
\label{aepr2}
\end{eqnarray}
where $\theta_\pm$ denotes the angle between the vectors
$\mathbf{a}$ and $\mathbf{b}\pm\mathbf{c}$. Thus, from Eqs.~(\ref{aepr0}) and (\ref{aepr2})
we conclude that a quantum system in the singlet state satisfies Eq.~(\ref{aepr0})
with equality if $\mathbf{a}$ lies in the plane formed by $\mathbf{b}$ and $\mathbf{c}$.

\subsection{Summary}
The inequality Eq.~(\ref{aepr0}) has been derived for a quantum system consisting of two spin-1/2 objects.
If some numerical values of the correlations would lead to a violation of this inequality
this would merely indicate that the calculation that yields these numerical values is wrong.

It is well-known that if we read the superscript $(2)$ as $(3)$ and
substitute the expressions $E^{(2)}=-\mathbf{a}\cdot\mathbf{b}$,
${\widehat E}^{(2)}=-\mathbf{a}\cdot\mathbf{c}$, and ${\widetilde
E}^{(2)}=-\mathbf{b}\cdot\mathbf{c}$ into EBBI
Eq.~(\ref{BOOLE2}) then, for a range of choices of $\mathbf{a}$,
$\mathbf{b}$ and $\mathbf{c}$, at least one of the inequalities
Eq.~(\ref{BOOLE2}) is not satisfied~\cite{BELL93}. However, in
contrast to the far-reaching conclusions that many researchers have
drawn from this apparent violation, from the viewpoint of quantum
theory, the only logically correct conclusion one can draw is that
it is not allowed to read the superscript $(2)$ as $(3)$.
Alternatively, we may adopt the hypothesis that
the system is described by a density matrix of the form
Eq.~(\ref{prod0}). Then the observation that the singlet state may
lead to a violation of the inequality Eq.~(\ref{prod4}) merely
implies that this hypothesis is false.

\subsection{Extended EPRB experiment}\label{ExtEPR}

\begin{figure*}[t]
\begin{center}
\includegraphics[width=14cm]{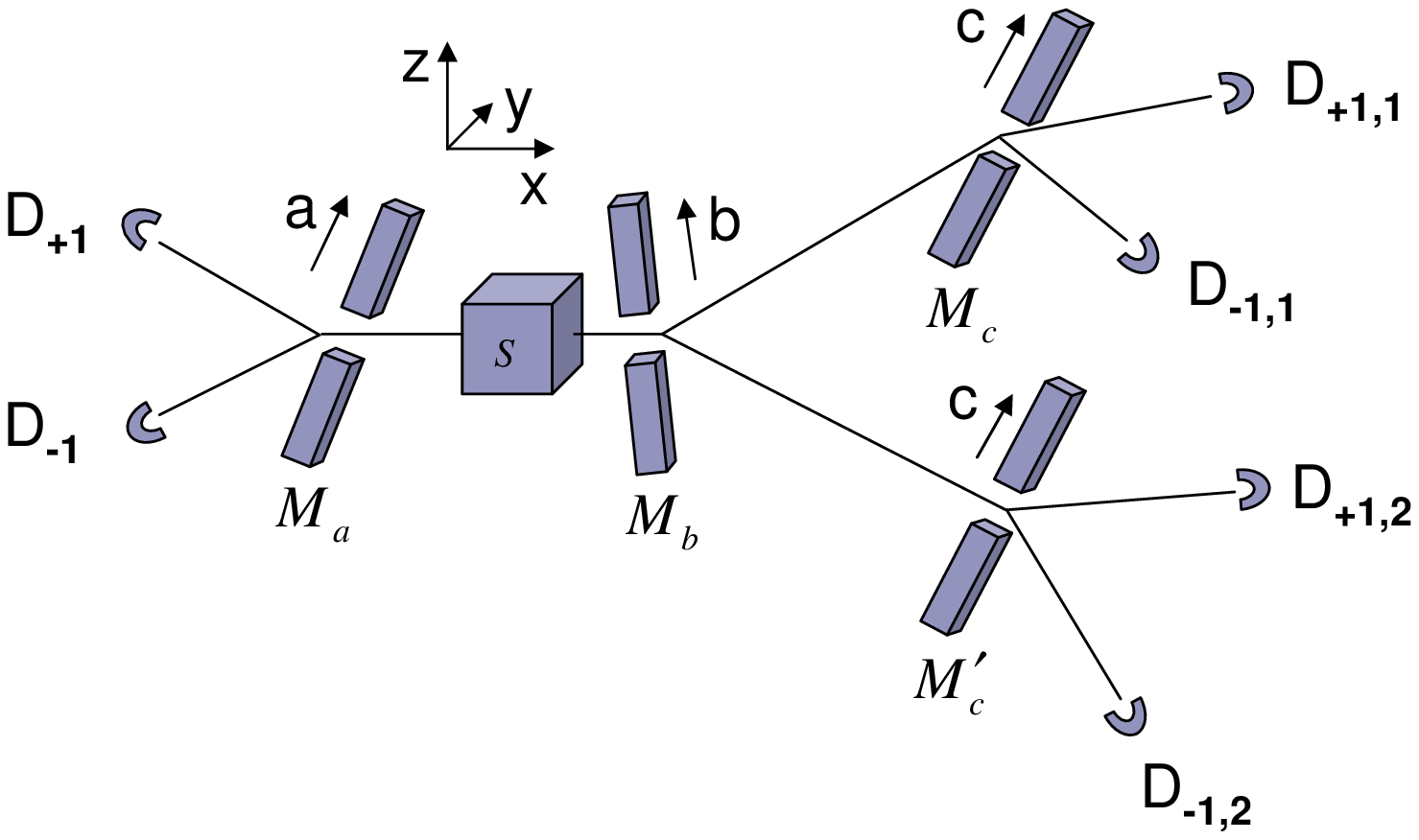}
\caption{ Same as Fig.~\ref{epr2} except that the detectors at
the right are replaced by two Stern-Gerlach magnets and four
detectors. The two additional Stern-Gerlach magnets $M_c$ and $M'_c$
are both assumed to be identical, $\mathbf{c}$ being the direction
of their magnetic fields. The detectors at the left yield the signal
$S_1=\pm1$. If detectors $D_{+1,1}$ or $D_{-1,1}$ fire, we set
$S_2=+1$, otherwise we set $S_2=-1$. If detectors $D_{+1,1}$ or
$D_{+1,2}$ fire, we set $S_3=+1$, otherwise we set $S_3=-1$. In this
idealized experiment, each pair produced by the source generates a
triple of signals $(S_1=\pm1,S_2=\pm1,S_3=\pm1)$. Note that the pair
$(S_1=\pm1,S_2=\pm1)$ expected from this experiment is the same as
the one that would be expected if one performs the experiment shown
in Fig.~\ref{epr2}. } \label{epr3}
\end{center}
\end{figure*}

In the original EPRB thought experiment, one only measures pairs of
two-valued variables. This fact has been used by many researchers to
(correctly) question the applicability of Bell's inequalities to
experimental data. However, there exists a straightforward extension
of the original EPRB experiment~\cite{SICA99} that allows us to
properly define the probability distribution of three two-valued
variables. We show below that this experiment (which is as
realizable as the original EPRB experiment) as well as its quantum theoretical
description can never lead to a violation of the EBBI.

The arrangement of this extended EPRB experiment is shown in
Fig.~\ref{epr3}. The key point of this experiment is that the
variable $S_2$, which in the original EPRB experiment is obtained by
measuring the spin as the particle leaves the Stern-Gerlach
apparatus $M_\mathbf{b}$ characterized by the unit vector $\mathbf{b}$, can be
retrieved from the data collected by the detectors $D_{+1,1}$,
$D_{-1,1}$, $D_{+1,2}$, and $D_{-1,2}$. At the same time, these four
detectors yield the value of a variable corresponding to $S_3$.

Thus, for each emitted pair labeled $\alpha$, this experiment yields
a triple ($S_{1,\alpha}$ , $S_{2,\alpha}$ , $S_{3,\alpha}$), which as Boole showed,
can never lead to a violation of Eq.~(\ref{BOOLE0}). Obviously, from
the construction of this experiment alone, one can expect that there
is some kind of correlation between $S_{2,\alpha}$ and
$S_{3,\alpha}$. Note that although the source emits pairs of
particles only, in this extended version of the EPRB experiment there
are six detectors and eight, not four, possible
outcomes.

What is left is to show explicitly that the quantum theoretical
results for the experiment shown in Fig.~\ref{epr3} satisfy EBBI Eq.~(\ref{BOOLE4}).
This demonstration is mainly for
pedagogical purposes. Indeed, from the general theory of
Section~\ref{QuantumTheory}, we already know that a quantum theory
for a system of three two-valued variables cannot violate
Eq.~(\ref{BOOLE4}). For simplicity of presentation, we consider
the case that $\mathbf{a}$, $\mathbf{b}$ and ${\mathbf c}$ lie in
the same plane (which is the case most readily realized in
experiments that use the photon polarization) and that the system is
in the singlet state Eq.~(\ref{singlet}). To fix the notation, we
put the vectors $\mathbf{a}$, $\mathbf{b}$ and $\mathbf{c}$ into the
$xz$-plane.

A Stern-Gerlach device of which the magnetic field makes an angle
$\theta$ with respect to the $z$-axis (by our convention the axis of
spin quantization) transforms the spin part of state vector
$v_\uparrow|\uparrow\rangle+v_\downarrow|\downarrow\rangle$ into
$w_\uparrow|\uparrow\rangle+w_\downarrow|\downarrow\rangle$ where
\begin{eqnarray}
\left(
\begin{array}{c}
w_\uparrow \\
w_\downarrow
\end{array}
\right)
&=&
\left(
\begin{array}{rr}
\cos\theta/2&\sin\theta/2 \\
-\sin\theta/2&\cos\theta/2
\end{array}
\right)
\left(
\begin{array}{c}
v_\uparrow \\
v_\downarrow
\end{array}
\right)
.
\label{aepr10}
\end{eqnarray}
Hence, after the particle passes through a Stern-Gerlach magnet
the eigenstates of the spin read
\begin{eqnarray}
|\uparrow_u\rangle&=&\cos\frac{\theta_u}{2}|\uparrow\rangle+\sin\frac{\theta_u}{2}|\downarrow\rangle
,
\\
|\downarrow_u\rangle&=&-\sin\frac{\theta_u}{2}|\uparrow\rangle+\cos\frac{\theta_u}{2}|\downarrow\rangle
,
\label{aepr11}
\end{eqnarray}
where $u=a,b,c$ and $\theta_u$ characterizes the direction of the field in the Stern-Gerlach magnet $M_u$.

\begin{figure*}[t]
\begin{center}
\includegraphics[width=16cm]{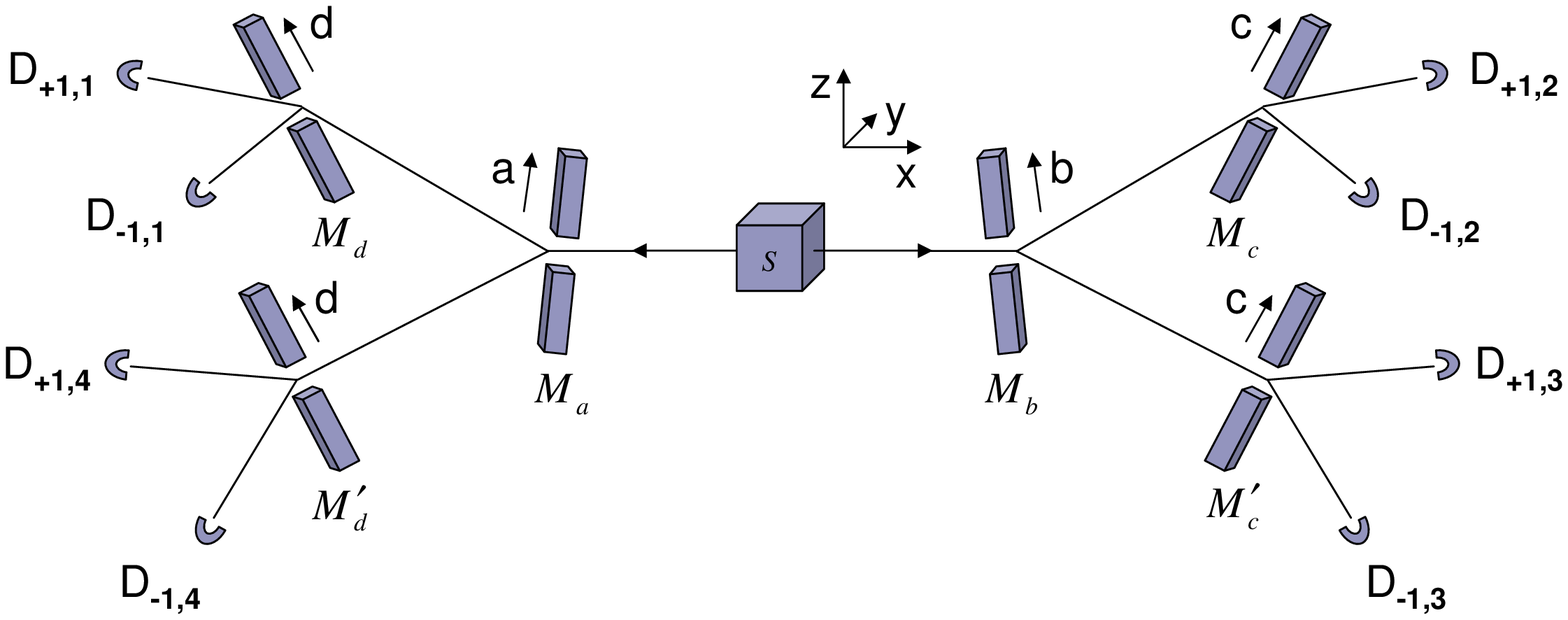}
\caption{Same as Fig.~\ref{epr3} except that the detectors at
the left are replaced by two Stern-Gerlach magnets and four
detectors. The two additional Stern-Gerlach magnets $M_d$ and $M_d$
are both assumed to be identical, $\mathbf{d}$ being the direction
of their magnetic fields.
If detectors $D_{+1,1}$ or $D_{-1,1}$ fire, we set $S_1=+1$,
whereas if $D_{+1,4}$ or $D_{-1,4}$ fire we set $S_1=-1$.
If detectors $D_{+1,1}$ or $D_{+1,4}$ fire, we set $S_4=+1$,
whereas if $D_{-1,1}$ or $D_{-1,4}$ fire we set $S_4=-1$.
Similary,
If detectors $D_{+1,2}$ or $D_{-1,2}$ fire, we set $S_2=+1$,
whereas if $D_{+1,3}$ or $D_{-1,3}$ fire we set $S_2=-1$.
If detectors $D_{+1,2}$ or $D_{+1,3}$ fire, we set $S_3=+1$,
whereas if $D_{-1,2}$ or $D_{-1,3}$ fire we set $S_3=-1$.
In this idealized experiment, each pair produced by the source generates a
quadruples of signals $(S_1=\pm1,S_2=\pm1,S_3=\pm1,S_4=\pm1)$.
Note that the pair $(S_1=\pm1,S_2=\pm1)$ expected from this experiment is the same as
the one that would be expected if one performs the experiment shown in Fig.~\ref{epr2}.
}
\label{epr4}
\end{center}
\end{figure*}

As an example, we calculate the probability that
detectors $D_{+1}$ and $D_{+1,1}$ fire.
This can only happen if the Stern-Gerlach magnet $M_b$
with orientation $\mathbf{b}$ directs the particle
to the Stern-Gerlach magnet $M_c$.
We assign the value $S_2=+1$ ($S_2=-1$) to
the path in which the particle has its spin (anti-)parallel to $\mathbf{b}$ .
According to quantum theory,
when the particles follow the paths corresponding to
$(S_1=+1,S_2=+1,S_3=+1)$ (see Fig.~\ref{epr3}),
the state vector 
of the two spins reads
\begin{widetext}
\begin{eqnarray}
\Phi(S_1=+1,S_2=+1,S_3=+1)
&=&
\frac{1}{\sqrt{2}}
|\uparrow_a\uparrow_c\rangle
\langle\uparrow_a\uparrow_c|\uparrow_a\uparrow_b\rangle
\langle\uparrow_a\uparrow_b|
\left(
|\uparrow\downarrow\rangle
-|\downarrow\uparrow\rangle
\right)
\nonumber \\
&=&
\frac{1}{\sqrt{2}}
\cos\frac{\theta_c-\theta_b}{2}
\sin\frac{\theta_b-\theta_a}{2}
|\uparrow_a\uparrow_c\rangle
.
\label{aepr12}
\end{eqnarray}
%
It is not difficult to see that in general,
\begin{eqnarray}
\Phi(S_1,S_2,S_3)
&=&
|S_1 S_3\rangle
\frac{(1+S_1S_2)s_{ba}+S_2(1-S_1S_2)c_{ba}}{2\sqrt{2}}
\frac{(1+S_2S_3)c_{cb}+S_2(1-S_2S_3)s_{cb}}{2}
,
\label{aepr15}
\end{eqnarray}
where $s_{uu'}=\sin(\theta_u-\theta_{u'})/2$
and $c_{uu'}=\cos(\theta_u-\theta_{u'})/2$.
Therefore, the probability to observe the triple $(S_1,S_2,S_3)$ is given by
\begin{eqnarray}
P(S_1,S_2,S_3)
&=&
\frac{1
- S_1S_2\cos(\theta_b-\theta_a)
- S_1S_3\cos(\theta_b-\theta_a)\cos(\theta_c-\theta_b)
+
S_2S_3\cos(\theta_c-\theta_b)}{8}
,
\label{aepr16}
\end{eqnarray}
From Eq.~(\ref{qt6}) and Eq.~(\ref{aepr16}) it follows that
\begin{eqnarray}
E^{(3)}_{12}&=&-\cos(\theta_b-\theta_a),
\nonumber \\
E^{(3)}_{13}&=&-\cos(\theta_b-\theta_a)\cos(\theta_c-\theta_b),
\nonumber \\
E^{(3)}_{23}&=&\cos(\theta_c-\theta_b)
,
\label{aepr17}
\end{eqnarray}
which in essence, are the same expressions as Eq.~(\ref{sm14}). As
in the case of flux tunneling, we see that $E^{(3)}_{12}=E^{(2)}$
but $E^{(3)}_{23}=-{\widetilde E}^{(2)}$ and
$E^{(3)}_{13}\not={\widehat E}^{(2)}$, where $E^{(2)}$, ${\widehat
E}^{(2)}$ and ${\widetilde E}^{(2)}$ are calculated for the original
EPRB thought experiment (see previous subsection). As expected from
the general theory of Section~\ref{QuantumTheory}, the expressions
Eq.~(\ref{aepr17}) always satisfy the EBBI Eq.~(\ref{BOOLE2}).
As a consistency check, we also compute the two-variable correlations using the
formalism of Section~\ref{filter}.
For a quantum system of two spin-1/2 particles in the singlet state,
the probability to observe the triple $(S_1,S_2,S_3)$ is given by
\begin{eqnarray}
P^{(3)}(S_1,S_2,S_3)&=&
\mathbf{Tr} \rho^{(2)}
M(S_1,{\mathbf a})M(S_2,{\mathbf b})
M(S_3,{\mathbf c})M(S_2,{\mathbf b})M(S_1,{\mathbf a})
\nonumber \\&=&
\frac{1-{\mathbf a}\cdot{\mathbf b}\,S_1S_2-{\mathbf a}\cdot{\mathbf b}\,{\mathbf b}\cdot{\mathbf c}\,S_1S_3
+{\mathbf b}\cdot{\mathbf c}\,S_2S_3}{8}
,
\label{chsh0a}
\end{eqnarray}
from which Eq.~(\ref{aepr17}) can be obtained
if the vectors $\mathbf{a}$, $\mathbf{b}$ and $\mathbf{c}$ are chosen to lie in the $xz$-plane.
Recall (see Section~\ref{filter}) that the spin-1/2 operators that measure $S_2$ and $S_3$ do not
necessarily commute.

For completeness, we discuss an extended EPRB experiment~\cite{SICA99} that
could be used to check the violation of the CHSH inequality.
The diagram of the experiment is presented in Fig.~\ref{epr4}
and is a logical extension of Fig.~\ref{epr3}.
According to quantum theory (see Section~\ref{filter}),
the probability to observe the quadruple $(S_1,S_2,S_3,S_4)$ is given by
\begin{eqnarray}
P^{(4)}(S_1,S_2,S_3,S_4)&=&
\mathbf{Tr} \rho^{(2)}
M(S_1,{\mathbf a})M(S_4,{\mathbf d})M(S_2,{\mathbf b})
M(S_3,{\mathbf c})M(S_2,{\mathbf b})M(S_4,{\mathbf d})M(S_1,{\mathbf a})
,
\label{chsh0}
\end{eqnarray}
\end{widetext}
disposing of the folklore that quantum theory cannot yield a joint probability distribution for all
possible measurements if, as in this example, non commuting operators are involved (see Section~\ref{filter}).
From Eq.~(\ref{chsh0}), it is straightforward to compute all two-particle correlations.
For a quantum system of two spin-1/2 particles in the singlet state we find
$E^{(4)}_{12}=-{\mathbf a}\cdot{\mathbf b}$,
$E^{(4)}_{13}=-({\mathbf a}\cdot{\mathbf b})({\mathbf b}\cdot{\mathbf c})$,
$E^{(4)}_{14}={\mathbf a}\cdot{\mathbf d}$,
$E^{(4)}_{23}={\mathbf b}\cdot{\mathbf c}$,
$E^{(4)}_{24}=-({\mathbf a}\cdot{\mathbf b})({\mathbf a}\cdot{\mathbf d})$,
and
$E^{(4)}_{34}=-({\mathbf a}\cdot{\mathbf b})({\mathbf a}\cdot{\mathbf d})({\mathbf b}\cdot{\mathbf c})$.
As expected from the general theory, CHSH inequalities such as
\begin{eqnarray}
|E^{(4)}_{12}-E^{(4)}_{13}+E^{(4)}_{24}+E^{(4)}_{34}|\le 2
,
\label{chsh1}
\end{eqnarray}
cannot be violated for the EPRB experiment depicted in Fig.~\ref{epr4}.

\section{Apparent violations of extended Boole-Bell inequalities in actual experiments}\label{apparent}

After these rather lengthy explanations, it is desirable to
illustrate the major aspects using actual experiments as an example.
We present three distinctly different but logically related
possibilities of violating Boole-Bell inequalities. The first
example is a simple, realistic every-day experiment
involving doctors who perform allergy tests on patients.
The second example shows how a innocent looking modification
of Bell's model of the EPRB experiment can lead to violations
of the EBBI while obeying the same local realism criteria
as Bell's model.
The third example relates to EPRB experiments
as they are performed in the laboratory
and is of a different nature than the first two.
It deals with space-time by attaching special importance to the time
synchronization of the two-particle measurements. Together these
examples represent an infinitude of possibilities to explain
apparent violations of Boole-Bell inequalities in an Einstein local
way.

\subsection{Games with symptoms and patients: From Boole to Bell}
\label{games}

As already mentioned, the early definitions of probability by Boole were
related to a one-to-one correspondence that Boole established
between actual experiments and idealizations of them through
elements of logic with two possible outcomes. His view gave the
concept of probability precision in its relation to sets of
experiments and this precision is expressed by Boole's discussion of
probabilities as related to possible experience.
These discussions can be best explained by an example
that has its origins in the works of
Boole and relates to the work of Bell inasmuch as
it can be used as a counterexample to Bell's conclusions related to
non-locality~\cite{KARL09}.

Consider an allergy to alcohol that strikes persons in different ways
depending on circumstances such as place of birth and place of
diagnosis etc..
Assume that we deal with patients that are born in
Austria (subscript $\bf a$), in Brazil (subscript $\bf b$) and in Canada (subscript $\bf c$).
Assume further that doctors are gathering information about the allergy
in the three cities Lille, Lyon and Paris, all in France.
The doctors are careful and perform the investigations on randomly chosen
but identical dates.
The patients are denoted by the symbol
$A_{\bf o}^l(n)$ where ${\bf o} = {\bf a}, {\bf b}, {\bf c}$ depending on
the birthplace of the patient, $l = 1, 2, 3$ depending on where the
doctor gathered information, $l$ designating Lille, $2$ Lyon and $3$
Paris respectively, and $n = 1, 2, 3,\ldots,N$ denotes just a given
random day of the examination.
Note that eventually the doctors could also label with the time and date of observation,
the type of weather or any other label that the doctors think to be relevant for the outcome of their observations.

The doctors perform the same alcohol allergy test on the persons visiting their office.
The test consists of serving the persons a glass of wine diluted with water from the tap.
When a person is allergic he or she gets a pimply red rash that disappears within one hour after drinking the diluted glass of wine.
When the person shows an allergic reaction the doctor assigns a value $A_{\bf o}^l(n)=+1$ to the person and otherwise
$A_{\bf o}^l(n)=-1$.

Assume that on even days the tap water contains no additives in Lille, iron in Lyon and chlorine in Paris.
On odd days the tap water contains fluorine and iron in Lille, chlorine and fluorine in Lyon and fluorine and iron in Paris.
This information is not known to the doctors performing the examinations,
hence they assume that they are performing identical allergy tests.
Also not known to the doctors is that persons born in Austria are allergic to alcohol, not allergic to chlorine or iron,
and also not allergic if alcohol and fluorine are present at the same time.
Persons born in Brazil are allergic to alcohol, not allergic to fluorine or chlorine,
and also not allergic if alcohol and iron are both present.
Persons born in Canada are allergic to fluorine only.
In Table~\ref{table7}, we list the results of all possible examinations.

\begin{table}[t]
    \caption{
The absence or presence of the additives fluorine (F), chlorine (Cl), and iron (Fe)
in tap water of Lille ($l=1$), Lyon ($l=2$), and Paris ($l=3$),
are indicated by -- or X, respectively.
The results of the allergy tests of patients
born in Austria, ($A_{\bf a}^l$), Brasil ($A_{\bf b}^l$), and Cananda ($A_{\bf c}^l$)
are indicated by $+1$ (allergic) and $-1$ (not allergic), respectively.}
  \begin{center}
\begin{tabular}{l|ccc|ccc}
\multicolumn{1}{c}{}&\multicolumn{3}{|c|}{Even days}&\multicolumn{3}{c}{Odd days}\\
\hline
$l$&1&2&3&1&2&3\\
\hline
F&-&-&-& X&X&X\\
Cl&-&-&X& -&X&-\\
Fe&-&X&-& X&-&X\\
\hline
$A_{\bf a}^l$&$+1$&$+1$&$+1$&$-1$&$-1$&$-1$\\
$A_{\bf b}^l$&$+1$&$-1$&$+1$&$-1$&$+1$&$-1$\\
$A_{\bf c}^l$&$-1$&$-1$&$-1$&$+1$&$+1$&$+1$\\
\end{tabular}
    \label{table7}
  \end{center}
\end{table}

The first variation of this investigation of the alcohol allergy is
performed as follows. The doctor in Lille examines only patients of
type $\bf a$, the doctor in Lyon only of type $\bf b$ and the
doctor in Paris only of type $\bf c$. On any given day of examination (of
precisely one patient for each doctor and day) they write down their
diagnosis and then, after many exams, concatenate the results and
form the following sum of pair-products of exam outcomes at a given
date described by $n$:
\begin{eqnarray}
\Gamma(w,n)
&=&
A_{\bf a}^1(w,n)A_{\bf b}^2(w,n) + A_{\bf a}^1(w,n)A_{\bf c}^3(w,n)
\nonumber \\&&
+ A_{\bf b}^2(w,n)A_{\bf c}^3(w,n)
,
\label{hla23n1}
\end{eqnarray}
where the variable $w$ denotes the fact that a glass of wine diluted with water from the tap was served to make the allergy test.
Boole noted now that
\begin{equation}
\Gamma(w,n) \geq -1
,
\label{hla23n2}
\end{equation}
which can be found by inserting all possible values for the patient
outcomes summed in Eq.~(\ref{hla23n1}). For the average (denoted by
$\langle . \rangle$) over all examinations we have then also:
\begin{equation}
\Gamma (w)= \langle\Gamma(w,n)\rangle=\frac{1}{N}\sum_{n=1}^N \Gamma(w,n)
\geq -1
.
\label{hla23n3}
\end{equation}
This equation gives conditions for the product averages and
therefore for the frequencies of the occurrence of certain values
of $A_{\bf a}^1(w,n), A_{\bf b}^2(w,n)$ etc.
These latter frequencies must therefore obey these conditions.
Thus we obtain rules or non-trivial
inequalities for the frequencies of occurrence of the patients
symptoms. Boole calls these rules ``conditions
of possible experience". In case of a violation, Boole states that
then the ``evidence is contradictory''.

As mentioned earlier, in the opinion of the authors,
the term ``possible experience'' introduced by Boole
is somewhat of a misnomer.
The experimental outcomes have been
determined from an experimental procedure in a scientific way and
are therefore possible.
What may not be possible is the one-to-one correspondence
of Boole's logical elements or variables to the
experimental outcomes that the scientist or statistician has chosen.

In this first example, we may indeed regard the various $A_{\bf o}^l(w,n) = \pm1$ with given
indices as the elements of Boole's logic to which the actual
experiments can be mapped. As shown by Boole, this is a sufficient
condition for the inequality of Eq.~(\ref{hla23n3}) to be valid. We
may in this case also omit all the indices except for those
designating the birth place and still will obtain a valid equation
that never can be violated:
\begin{equation}
\langle A_{\bf a}(w)A_{\bf b}(w)\rangle + \langle A_{\bf a}(w)A_{\bf c}(w)\rangle + \langle A_{\bf
b}(w)A_{\bf c}(w)\rangle \geq -1
.
\label{hla23n3b}
\end{equation}
The reason is simply that three arbitrary dichotomic variables i.e.
variables that assume only two values ($\pm 1$ in our case) must
always fulfill Eq.~(\ref{hla23n3b}) no matter what their logical
connection to experiments is because we deduce the three products of
Eq.~(\ref{hla23n3b}) from sequences of each three measurement
outcomes.
Note that Eq.~(\ref{hla23n3b}) contains six factors with
each birthplace appearing twice and representing then the
identical result. We will now discuss a slightly
modified experiment that is much more general and contains six
measurement results for the six factors.

In this second variation of the investigation,
we let only two doctors, one in Lille and one in Lyon perform
the examinations. The doctor in Lille examines randomly all patients of
types $\bf a$ and $\bf b$ and the one in Lyon all of type $\bf b$ and
$\bf c$ each one patient at a randomly chosen date. The doctors are
convinced that neither the date of examination nor the location
(Lille or Lyon) has any influence and therefore denote the patients
only by their place of birth.
After a lengthy period of examination they find
\begin{eqnarray}
\Gamma (w)&=& \langle A_{\bf a}(w)A_{\bf b}(w)\rangle + \langle A_{\bf a}(w)A_{\bf c}(w)\rangle
\nonumber \\&&
+ \langle A_{\bf b}(w)A_{\bf c}(w)\rangle = -3 .
\label{hla23n5}
\end{eqnarray}
They further notice that the single outcomes of $A_{\bf a}(w), A_{\bf
b}(w)$ and $A_{\bf c}(w)$ are randomly equal to $\pm 1$. This latter fact
completely baffles them. How can the single outcomes be entirely
random while the products are not random at all and how can a Boole
inequality be violated hinting that we are not dealing with a
possible experience? After lengthy discussions they conclude that
there must be some influence at a distance going on and the outcomes
depend on the exams in both Lille and Lyon such that a single
outcome manifests itself randomly in one city and that the outcome
in the other city is then always of opposite sign.

However, there are
also other ways that remove the cyclicity, ways that do not need to
take recourse to influences at a distance. In this example, although not known
by the doctors beforehand, we have a
time and a city dependence of the allergy as described above.
Obviously for
measurements on random dates we have the outcome that $A_{\bf
a}(w), A_{\bf b}(w)$ and $A_{\bf c}(w)$ are randomly equal to $\pm 1$ while
at the same time $\Gamma(w,n) = -3$ and therefore $\Gamma (w)= -3$. We
need no deviation from conventional thinking to arrive at this
result because now, in order to deal with Boole's elements of logic,
we need to add the coordinates of the cities to obtain
$\Gamma (w)= \langle A_{\bf a}^1 (w)A_{\bf b}^2(w)\rangle + \langle A_{\bf a}^1 (w)A_{\bf c}^2(w)\rangle +
\langle A_{\bf b}^1 (w)A_{\bf c}^2(w)\rangle \geq -3$
and the inequality is of the trivial kind because the cyclicity is
removed. The date index does not matter for the products since both
signs are reversed on even and odd days leaving the products unchanged.
Including the city labels the doctors realize that
$A_{\bf b}^1(w,n)=-A_{\bf b}^2(w,n)$, totally against their expectations.
Contacting the water delivering company can however resolve this mistery.

We note that in connection with EPR experiments and questions
relating to interpretations of quantum mechanics,
Eq.~(\ref{hla23n3}) is of the Bell-type.
It is often claimed that a
violation of such inequalities implies that either realism or
Einstein locality should be abandoned. As we saw in our
counterexample which is both Einstein local and realistic in the
common sense of the word, it is the one to one correspondence of the
variables to the logical elements of Boole that matters when we
determine a possible experience, but not necessarily the choice
between realism and Einstein locality.

Realism plays a role in the arguments of Bell and followers because
they introduce a variable $\lambda$ representing an element of
reality and then write
\begin{eqnarray}
\Gamma (\lambda)&=& \langle A_{\bf a}(\lambda) A_{\bf b}(\lambda)\rangle + \langle A_{\bf
a}(\lambda) A_{\bf c}(\lambda)\rangle
\nonumber \\&&
+ \langle A_{\bf b}(\lambda) A_{\bf c}(\lambda)\rangle \geq -1
.
\label{hla23n7}
\end{eqnarray}
Because no $\lambda$ exists that would lead to a violation except a
$\lambda$ that depends on the index pairs ($\bf a$, $\bf b$), ($\bf
a$, $\bf c$) and ($\bf b$, $\bf c$) the simplistic conclusion is
that either elements of reality do not exist or they are non-local.
The mistake here is that Bell and followers insist from the start
that the same element of reality occurs for the three different
experiments with three different setting pairs. This assumption
implies the existence of the combinatorial-topological cyclicity
that in turn implies the validity of a non-trivial inequality but
has no physical basis. Why should the elements of reality not all be
different? Why should they, for example not include the time of
measurement? There is
furthermore no reason why there should be no parameter of the
equipment involved. Thus the equipment could involve time and
setting dependent parameters such as $\lambda_{\bf a}(t),
\lambda_{\bf b}(t), \lambda_{\bf c}(t)$ and the functions $A$ might
depend on these parameters as
well~\cite{FINE82,HESS01,RAED06c,ZHAO08,KHRE09}.

We note that although this example violates the Bell-type inequality Eq.~(\ref{hla23n3})
it does not violate the CHSH inequality.

\subsection{Factorizable model}\label{factor}

\begin{figure*}[t]
\begin{center}
\includegraphics[width=14cm]{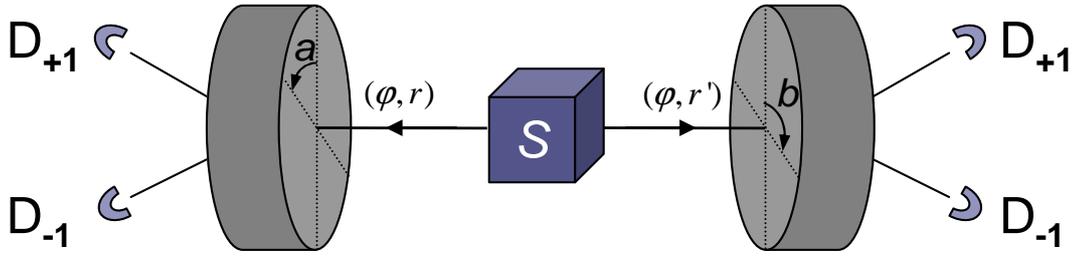}
\caption{
Schematic diagram of a factorizable model for the EPRB experiment.
The properties of the particle going to the  left (right) are
represented by an angle $\varphi$ and a number
$-1\le r \le +1$ ($-1\le r' \le +1$).
The source $S$ emits these particles with a random,
uniformly distributed angle $\varphi$ and
with $(r,r')$ distributed according
to the density $\mu(r,r')$ (see text).
Based on the setting $a$ ($b$) and $(\varphi,r)$ ($(\varphi,r')$)
the gray cilinders direct the particles to one of the detectors
$D_{\pm1}$ where they generate a ``click''
depending on the choice of $\mu(r,r')$.
This locally causal, factorizable model can violate
the Bell inequalities $|E^{(2)}(a,b)\pm E^{(2)}(a,c)|\le 1 - E^{(2)}(b,c)$.
}
\label{fig5}
\end{center}
\end{figure*}

The models that we consider in this subsection do not pretend to account for
the correlations of two spin-1/2 particles in the singlet state but
provide further illustrations of the ideas presented above.

Imagine the standard EPRB setup
with a source emitting two particles carrying the variables $(\varphi,r)$ and $(\varphi,r')$, where
$0\le\varphi\le2\pi$ and $-1\le r,r'\le1$, see Fig.~\ref{fig5}. The source imposes some relation between the variables $r$ and $r'$, as
explained later. One particle flies to a station with the detector in orientation $a$ and
the other particle flies to another station with the detector in orientation $b$.
The detection process and the correlation between the events in both stations are
defined by the probabilities
\begin{eqnarray}
P^{(1)}(S|a \varphi r)&=&
\Theta\left[ S\left(\cos(\varphi-a)-r\right) \right]
,
\nonumber \\
P^{(1)}(S'|b \varphi r')&=&
\Theta\left[ S'\left(\cos(\varphi-b)-r'\right) \right]
,
\nonumber \\
P^{(2)}(S,S'|a b)&=&\frac{1}{2\pi}\int_0^{2\pi}\,d\varphi \int_{-1}^{+1}\,dr\int_{-1}^{+1}\,dr' P^{(1)}(S|a \varphi r)
\nonumber \\
&&\hbox to 1cm{\hfill}\times P^{(1)}(S'|b \varphi r')\mu(r,r')
,
\label{hid0}
\label{e104}
\end{eqnarray}
respectively. Here $\Theta(.)$ is the unit step function and $\mu(r,r')$ is a probability density.

We consider three choices for $\mu(r,r')$, namely $\mu(r,r')=1/4$,
$\mu(r,r')=\delta(r-r')/2$, and $\mu(r,r')=\delta(r+r')/2$.
These three models are local realist, hidden variable models~\cite{BELL93}.
For any of these three choices, we have
\begin{eqnarray}
P^{(1)}(+1|a\varphi)&=&\int_{-1}^{+1}\,dr \int_{-1}^{+1}\,dr' P^{(1)}(S|a \varphi r)\mu(r,r')
\nonumber \\
&=&\int_{-1}^{+1}\,dr \int_{-1}^{+1}\,dr' P^{(1)}(S|a \varphi r')\mu(r,r')
\nonumber \\
&=&\cos^2\frac{a-\varphi}{2}
,
\label{hid0a}
\end{eqnarray}
hence all three models reproduce Malus law for the single-particle probabilities.

For $\mu(r,r')=1/4$ we find
\begin{eqnarray}
E^{(2)}(a,b)=-\frac{1}{2}\cos(a-b)
,
\label{hid1a}
\end{eqnarray}
while for $\mu(r,r')=\delta(r-r')/2$ we have
\begin{eqnarray}
E^{(2)}(a,b)=1-\frac{4}{\pi}|\sin\frac{a-b}{2}|
.
\label{hid1}
\end{eqnarray}
It follows that $|E^{(2)}(a,b)\pm E^{(2)}(a,c)|\le 1\pm E^{(2)}(b,c)$,
with the $E^{(2)}$'s given by Eq.~(\ref{hid1a}) or Eq.~(\ref{hid1}),
is always satisfied, independent of the choice of $a$, $b$, and $c$.
If we write $f^{(2)}(S,S')=P^{(2)}(S,S'|a b)$, ${\widehat f}^{(2)}(S,S')=P^{(2)}(S,S'|a c)$, and
${\widetilde f}^{(2)}(S,S')=P^{(2)}(S,S'|b c)$ (see Eq.~(\ref{bell0})),
it follows from Section~\ref{Bell} that there exists a common probability distribution
for all possible experiments and hence the EBBI cannot be violated.

However, for $\mu(r,r')=\delta(r+r')/2$ we have
\begin{eqnarray}
E^{(2)}(a,b)=\frac{4}{\pi}|\cos\frac{a-b}{2}|-1
,
\label{hid2}
\end{eqnarray}
If we substitute expression Eq.~(\ref{hid2}) in
$|E^{(2)}(a,b)\pm E^{(2)}(a,c)|\le 1\pm E^{(2)}(b,c)$,
we find that this inequality may be violated
(for $b=a+2\pi$ and $c=a+\pi$ for instance).

This is not a surprise: If $\mu(r,r')=\delta(r+r')/2$ then
\begin{eqnarray}
P^{(2)}(S,S'|a b)&=&\frac{1}{4\pi}\int_0^{2\pi}\,d\varphi \int_{-1}^{+1}\,dr P^{(1)}(S|a \varphi [+r])
\nonumber \\
&&\hbox to 1cm{\hfill}\times P^{(1)}(S'|b \varphi [-r]))
,
\label{hid3}
\label{e107}
\end{eqnarray}
cannot be brought in the form
\begin{eqnarray}
P^{(2)}(S,S'|a b)&=&\int \,d\lambda  P^{(1)}(S|a \lambda)P^{(1)}(S'|b \lambda)
,
\label{hid4}
\end{eqnarray}
for all possible values of $a$ and $b$, hence the derivation of the Bell inequality stops here.
Although Eq.~(\ref{hid3}) has the same factorizable structure as the local
hidden variable models considered by Bell, the fact that it cannot
be brought into the form Eq.~(\ref{hid4}) illustrates, once again,
the importance of having the common label ``$\lambda$'' appear in
all factors for the derivation of the Bell inequality to hold true.

To relate the model to actual experiments, one needs to relate $(\varphi, r)$ to some elements of reality.
Bell assumes identical triples of elements of reality for the left and right going particles
but in fact, this assumption lacks a physical, let alone a logical, basis.
By considering $\mu(r,r')=\delta(r+r')/2$, we avoid this assumption and
find violations of the EBBI.
It is of interest to note that if we substitute Eq.~(\ref{hid2}) into the CHSH inequality~\cite{CLAU69,BELL93}
\begin{equation}
-2 \le E^{(2)}(a,b)-E^{(2)}(a,c)+E^{(2)}(d,b)+E^{(2)}(d,c)\le 2
,
\label{hid5}
\end{equation}
we find that it is always satisfied.

Summarizing: The local realist model with $\mu(r,r')=\delta(r+r')/2$ provides
an example of a factorizable model that violates
the Bell inequality but satisfies the CHSH inequality.
Nevertheless, we have constructed a local realist, factorizable model that violates the EBBI.
Hence neither local realism nor factorability are necessary conditions for the EBBI to hold.

\subsection{EPR-Bohm experiments and measurement time synchronization}\label{RealEPR}

To the best of our knowledge, all real EPRB experiments that have been performed
up to date employ an operational procedure to decide whether
the two detection events correspond to either the observation of
one two-particle system or (exclusive) to the observation of two single-particle systems.
In EPRB experiments, this decision is taken on the basis of coincidence
in time~\cite{KOCH67,FREE72,ASPE82b,TAPS94,TITT98,WEIH98,ROWE01,FATA04,SAKA06}.
The set of data that is collected in these real laboratory experiments
can be written as
%
\begin{eqnarray}
\Lambda^{(2)}&=&\left\{ ({\bf d}_{1,\alpha},{\bf d}_{2,\alpha})|\alpha =1,\ldots ,M \right\}
\nonumber \\&=&
\left\{ (S_{1,\alpha},t_{1,\alpha},{\bf a}_{1,\alpha},S_{2,\alpha},t_{2,\alpha},{\bf a}_{2,\alpha})
|\alpha =1,\ldots ,M \right\},
\nonumber \\
\label{Lambda}
\end{eqnarray}
%
where ${\bf d}_{i,\alpha}=(S_{i,\alpha},t_{i,\alpha},{\bf a}_{i,\alpha})$ and
$S_{i,\alpha}=\pm1$ is a dichotomic variable that indicates which of the two detectors in station $i=1,2$ detected the
particle (photon, proton, ...),  $t_{i,\alpha}$ is the time at which the detector in station $i=1,2$ fired, and
${\bf a}_{i,\alpha}$ denotes a vector of numbers that specifies the instrument settings at station $i=1,2$.
For instance, in the experiment of Weihs et al.~\cite{WEIH98}, the ${\bf a}_{i,\alpha}$'s may contain the rotations of the
photon polarization induced by the electro-optic modulators.
In Eq.~(\ref{Lambda}) (first line), we have made explicit that the data is collected in pairs, each
pair consisting of several variables, some of which are not dichotomic.
The second line of Eq.~(\ref{Lambda}) gives another view of the same data, namely as 6-tuples of real-valued numbers.
Recalling that the dichotomic character of the variables was essential for the derivation of the Boole inequalities,
it is unlikely that similar inequalities hold for the raw data Eq.~(\ref{Lambda}), for an exception see Ref.~\onlinecite{LARS04}.
Therefore, if the desire is to make contact with the Boole inequalities,
some further processing of the data is required.

It is quite natural to identify coincidences by comparing the time differences
$\{ t_{1,\alpha}-t_{2,\alpha} \vert \alpha =1,\ldots ,M \}$ with a time window $W$
and this is indeed what is being done in EPRB experiments~\cite{KOCH67,FREE72,ASPE82b,TAPS94,TITT98,WEIH98,ROWE01,FATA04,SAKA06}.
Note however that the aim of these experiments is to use a value
of $W$ that is as small as technically feasible whereas the time differences
become irrelevant in the limit $W\rightarrow\infty$ only.
Furthermore, to obtain a data set that
consists of pairs only, the events are selected such that
$\mathbf{a}_{1,\alpha}=\mathbf{a}_1$ and $\mathbf{a}_{2,\alpha}=\mathbf{a}_2$
where $(\mathbf{a}_1,\mathbf{a}_2)$ is one particular pair of instrument settings.
Accordingly, the reduced data set becomes
\begin{widetext}
\begin{eqnarray}
\Lambda'^{(2)}(\mathbf{a}_1,\mathbf{a}_2)&=&
\left\{ (S_{1,\alpha},S_{2,\alpha})|
\mathbf{a}_{1,\alpha}=\mathbf{a}_1,\mathbf{a}_{2,\alpha}=\mathbf{a}_2,
\vert t_{1,\alpha} -t_{2,\alpha}|\le W,
\alpha =1,\ldots ,M \right\}
.
\label{Cxy}
\end{eqnarray}
\end{widetext}

We are now in the position to apply the results of the earlier sections.
Let us consider the case where there are three pairs originating
from experiments with different
instrument settings, namely
$(\mathbf{a}_1,\mathbf{a}_2)=(\mathbf{a},\mathbf{b})$,
$(\mathbf{a}_1,\mathbf{a}_2)=(\mathbf{a},\mathbf{c})$, and
$(\mathbf{a}_1,\mathbf{a}_2)=(\mathbf{b},\mathbf{c})$.
The three pairs of instrument settings yield the data sets
$\Upsilon^{(2)}=\Lambda'^{(2)}(\mathbf{a},\mathbf{b})$,
${\widehat \Upsilon}^{(2)}=\Lambda'^{(2)}(\mathbf{a},\mathbf{c})$,
and
${\widetilde \Upsilon}^{(2)}=\Lambda'^{(2)}(\mathbf{b},\mathbf{c})$ but,
as we have seen several times, there are no
Boole inequalities Eq.~(\ref{BOOLE0}) for the corresponding
pair correlations unless we make the hypotheses that
there is an underlying process of triples that gives rise to the data.
Should we therefore find that the pair correlations violate
the Boole inequalities Eq.~(\ref{BOOLE0}), the only logically valid
conclusion is that the named hypothesis is false.

We have shown in a series of
papers~\cite{RAED06c,RAED07a,RAED07b,RAED07c,ZHAO08} that it is
possible to construct models, that is algorithms, that are locally
causal in Einstein's sense, generate the data set Eq.~(\ref{Lambda})
and reproduce exactly the correlation that is characteristic for a
quantum system in the singlet state. These algorithms can be viewed
as concrete realizations of Fine's synchronization
model~\cite{FINE82}.
According to Bell's theorem, such models do not exist. This apparent
paradox is resolved by the work presented in this paper: There
exists no Bell inequality for triples of pairs, there are only EBBI
for pairs extracted from triples.

\section{Summary and Conclusions}\label{conclusions}

The central result of this paper is that the necessary conditions
and the proof of the inequalities of Boole for $n$-tuples of
two-valued data (see Section~\ref{Boole}) can be generalized to real
non negative functions of two-valued variables (see
Section~\ref{functions}) and to quantum theory of two-valued
dynamical variables (see Section~\ref{QuantumTheory}). The resulting
inequalities, that we refer to as
extended Boole-Bell inequalities (EBBI) for reasons explained in the
Introduction and in Section~\ref{functions}, have the same form as
those of Boole and Bell. Equally central is the fact that these EBBI
express arithmetic relations between numbers that can never be
violated by a mathematically correct treatment of the problem: These
inequalities derive from the rules of arithmetic and the
non negativity of some functions only. A violation of these
inequalities is at odds with the commonly accepted rules of
arithmetic or, in the case of quantum theory, with the commonly
accepted postulates of quantum theory.

Applied to specific examples, the main conclusions of the present work are:

\begin{itemize}
\item{In the original Einstein-Podolsky-Rosen-Bohm (EPRB) thought experiment,
one collects the three data sets
$\Upsilon^{(2)}=\{(S_{1,\alpha},S_{2,\alpha})|\alpha=1,\ldots,M\}$,
${\widehat \Upsilon}^{(2)}=\{({\widehat S}_{1,\alpha},{\widehat S}_{2,\alpha})|i=1,\ldots,M\}$,
and
${\widetilde \Upsilon}^{(2)}=\{({\widetilde S}_{1,\alpha},{\widetilde S}_{2,\alpha})|\alpha=1,\ldots,M\}$.
From these data sets, one extracts
the correlations $F^{(2)}$, ${\widehat F}^{(2)}$, and ${\widetilde F}^{(2)}$.
Then, Bell and followers assume that it is legitimate to
substitute $F^{(2)}$ for $F^{(3)}_{ij}$, ${\widehat F}^{(2)}$ for
$F^{(3)}_{ik}$, and ${\widetilde F}^{(2)}$ for $F^{(3)}_{jk}$ into
the Boole inequalities $|F^{(3)}_{ij}\pm F^{(3)}_{ik}|\le1\pm
F^{(3)}_{jk}$ for $(i,j,k)=(1,2,3),(3,1,2),(2,3,1)$, which does hold
for triples $(S_{1,\alpha},S_{2,\alpha},S_{3,\alpha})$, but not
necessarily for pairs of two-valued data. Therefore, if it then
turns out that a data set leads to a violation of Boole's
inequalities, the only conclusion that one can draw is that the data
set does not satisfy the conditions necessary to prove the Boole
inequalities, namely that three data sets of pairs can be extracted
from a single data set of triples (see Section~\ref{Boole}).
}
\item{A violation of
the EBBI cannot be attributed to influences at a
distance. The only possible way that a violation could arise is if
grouping is performed in pairs (see Section~\ref{games})}.
\item{In the original EPRB thought experiment, one can measure pairs of data only,
making it de-facto impossible to use Boole's inequalities properly.
This obstacle is removed in the extended EPRB thought experiment
discussed in Section~\ref{ExtEPR}.
In this extended EPRB experiment, one can
measure both pairs and triples and consequently, it is impossible for the data to
violate Boole's inequalities.
This statement is generally true: It does not depend on whether the
internal dynamics of the apparatuses induces some correlations among
different triples or that there are influences at a distance.
The fact that this experiment yields triples of two-valued numbers
is sufficient to guarantee that Boole's inequalities cannot be violated.
}
\item{The rigorous quantum theoretical treatment of a quantum flux tunneling problem
(see Section~\ref{Leggett}) and the EPR-Bohm experiment (see
Section~\ref{EPR}) provide explicit examples that quantum theory can
never give rise to violations of the EBBI.
}
\end{itemize}

\section*{Acknowledgement}
We thank K. De Raedt, F. Jin, S. Miyashita, S. Yuan and S. Zhao for extensive discussions.

\bibliography{../../epr}   

\end{document}